\documentclass[reqno, 11pt]{amsart}

\usepackage{amsfonts,amssymb,amsbsy,amsmath,amsthm,enumerate}
\usepackage{txfonts}
\usepackage{eucal}
\usepackage{dsfont}
\usepackage{mathrsfs}

\usepackage{bbm}

\usepackage{stackrel}

\usepackage[small,nohug,
]{diagrams}
\diagramstyle[labelstyle=\scriptstyle]
\newarrow{Corresponds}<--->

\usepackage[dvips]{graphicx}
\usepackage{color}

\newtheorem{theorem}{Theorem}[section]
\newtheorem{proposition}[theorem]{Proposition}
\newtheorem{lemma}[theorem]{Lemma}
\newtheorem{corollary}[theorem]{Corollary}
\newtheorem{assumption}[theorem]{Assumption}
\newtheorem{definition}[theorem]{Definition}
\newtheorem{remark}[theorem]{Remark}

\newtheorem{example}[theorem]{Example}

\numberwithin{equation}{section}

\numberwithin{figure}{section}
\numberwithin{table}{section}

\newcommand\beq{\begin{equation}}
\newcommand{\bea}{\begin{eqnarray}}
\newcommand{\eea}{\end{eqnarray}}
\newcommand{\beas}{\begin{eqnarray*}}
\newcommand{\eeas}{\end{eqnarray*}}
\newcommand{\beql}{\begin{equation} \label}
\newcommand{\eeq}{\end{equation}}

\newcommand{\R}{\mathbb R}
\newcommand{\N}{\mathbb N}
\newcommand{\C}{\mathbb C}                           %NUMBER SETS
\newcommand{\Q}{\mathbb Q}
\newcommand{\Z}{\mathbb Z}
\newcommand{\T}{\mathbb T}

\newcommand{\s}[1]{\CMcal{#1}}
\newcommand{\f}[1]{\mathcal{#1}}                  %FONTS
\newcommand{\bb}[1]{\mathscr{#1}}
\newcommand{\rr}[1]{\mathfrak{#1}}
\newcommand{\n}[1]{\mathds {#1}}
\newcommand{\expo}[1]{{\rm e}^{#1}}                 %SPECIAL FUNCTION
\newcommand{\dd}{\,{\rm d}}
\newcommand{\ii}{\,{\rm i}\,}
\newcommand{\ncint}{\mathrel{{\ooalign{$\int$\cr\kern+.07em\raise.15ex\hbox{$\pmb{\scriptstyle-}$}\cr}}}}           \newcommand{\ncpartial}{\mathrel{{\ooalign{$\partial$\cr\kern+.29em\raise.79ex\hbox{$\pmb{\scriptstyle-}$}\cr}}}}

\newcommand{\virg}[1]{\lq\lq#1\rq\rq}                %TIPOGRAPHIC
\newcommand{\ie}{{\sl i.\,e.\,}}
\newcommand{\eg}{{\sl e.\,g.\,}}
\newcommand{\cf}{{\sl cf.\,}}

\begin{document}
%------------------------------------------------------------------------------------------------------------------------%
%                                                                        heading
%------------------------------------------------------------------------------------------------------------------------%

\title[Chiral vector bundles]{Chiral vector bundles
}

%-----%

\author[G. De~Nittis]{Giuseppe De Nittis}
\address[De~Nittis]{Facultad de Matem\'aticas \& Instituto de F\'{\i}sica,
Pontificia Universidad Cat\'olica,
Santiago, Chile}
\email{gidenittis@mat.uc.cl}

\author[K. Gomi]{Kiyonori Gomi}
\address[Gomi]{Department of Mathematical Sciences, Shinshu University,  Nagano, Japan}
\email{kgomi@math.shinshu-u.ac.jp}



%-----%

\begin{abstract}
\vspace{-4mm}
This paper focuses on the study of a new category of vector bundles. The objects of this category, called \emph{chiral vector bundles}, are pairs given by a complex vector bundle along with one of its automorphisms. 
We provide a classification for the \emph{homotopy} equivalence classes of these objects
based on the construction of a suitable classifying space. The computation of the cohomology of the latter allows us to introduce a proper set of  characteristic cohomology classes: some of those just reproduce the ordinary Chern classes but there are also new odd-dimensional classes which take care of the extra topological information introduced by the chiral structure. Chiral vector bundles provide a geometric model for topological quantum systems in class
 {\bf AIII}, namely for systems endowed with a (pseudo-)symmetry of chiral type. The classification of the chiral vector bundles over spheres and tori (explicitly computable up to dimension 4), recovers the commonly accepted classification for topological insulators of class {\bf AIII}  which is usually based on the K-group $K_1$. However, our classification turns out to be  even richer since it takes care also for  possible non-trivial Chern classes.
\end{abstract}

%-----%

\maketitle

\vspace{-5mm}
\tableofcontents

%--------------------%
\section{Introduction}\label{sect:intro}

The mathematical investigation of the nature of the topological phases of the matter is a recent hot topic in mathematical physics. Starting from the seminal work by  Kane and  Mele \cite{kane-mele-05}, where for the first time $\Z_2$ topological phases were predicted, a big effort has been devoted to the understanding of a general scheme for the complete classification of all possible phases  
of \emph{topological insulators} {\cite{altland-zirnbauer-97,schnyder-ryu-furusaki-ludwig-08,ryu-schnyder-furusaki-ludwig-10}}. The first considerable contribution to this program was  provided by Kitaev's \virg{Periodic Table} for topological insulators and superconductors \cite{kitaev-09}. This is a $K$-theory-based procedure which organizes into a systematic scheme, the \emph{stable} topological phases of symmetry-protected gapped free-fermion systems. {Even though the idea of using $K$-theory in condensed matter dates back to the 80s with the seminal paper by Bellissard \cite{bellissard-86},
after the Kitaev's  paper the interest in the subject exploded} and a certain amount of valuable foundational works devoted to the justification and the generalization of Kitaev's table appeared in recent years: \cite{freed-moore-13, thiang-14,kennedy-zirnbauer-15},{\cite{bourne-carey-rennie-16,grossmann-schulz-baldes-16,
prodan-schulz-baldes-14,prodan-schulz-baldes-book-16,
kellendonk-17,kubota-17}} just to mention few of them.

\medskip

A quite different approach to the study of  the topological phases of quantum systems
consists of focusing on the rigorous geometrical analysis
  of each single topological class. In fact, if from one side the various fundamental symmetries     
(or better \emph{pseudo}-symmetries as pointed out in \cite{kennedy-zirnbauer-15}) can be used to organize the different classes of topological insulators according to the \virg{Bott-clock} induced by the underlying Clifford structure, on the other side each class is described by its own geometric model. The latter is usually specified by  the set of symmetries defining the class itself along with  
a proper notion of \emph{ground state}, often identified with a Fermi projection in a gap (a different but equivalent point of view has been recently considered in \cite{kennedy-zirnbauer-15}). For periodic fermionic systems, or more generally for \emph{topological quantum systems} (in the sense of \cite{denittis-gomi-14,denittis-gomi-14-gen,denittis-gomi-15}), these geometric models are realizable as vector bundles enriched by the presence of extra structures. For the case of systems with an \emph{even} or \emph{odd} time-reversal symmetry (class {\bf AI} and {\bf AII}, respectively) {the underlying geometric theories have been studied in depth from different authors and using different techniques. For instance fermionic systems in dimension two and three have been classified in \cite{porta-graf-13,fiorenza-monaco-panati-14}  by means of obstruction techniques and in  \cite{carpentier-delplace-fruchart-gawedzki-15,carpentier-delplace-fruchart-gawedzki-tauber-15,monaco-tauber-17} by using the Wess-Zumino amplitudes. 
 In the more general setting of topological quantum systems, 
a general classification scheme for the vector bundle structures
underlying systems of type {\bf AI} and {\bf AII}
 has been developed in \cite{denittis-gomi-14,denittis-gomi-14-gen,denittis-gomi-17,denittis-gomi-18}. This scheme,  based on the construction of suitable characteristic classes, allows  some important achievements like} the possibility to access  the \emph{unstable} regime which is usually out of the domain of  K-theory (there is a recent interest in that \cite{kennedy-guggenheim-15,kennedy-zirnbauer-15}), or  the possibility to consider also systems adiabatically perturbed by external fields (meaning that one can replace the Brillouin torus or sphere with quite general topological spaces), or the identifications of proper cohomology classes capable of completely classifying the topological phases. 
An even more relevant aspect  is the possibility to describe the topological classification in terms of
differential-geometric invariants {(see \cite{denittis-gomi-15} for class {\bf AI} and   \cite{carpentier-delplace-fruchart-gawedzki-15,carpentier-delplace-fruchart-gawedzki-tauber-15,monaco-tauber-17} for class {\bf AII}).}

\medskip

This work is devoted to the study of a geometric model for the class {\bf AIII} or, said differently, for the class of quantum systems protected by a \emph{chiral} symmetry (in the jargon of topological insulators \cite{schnyder-ryu-furusaki-ludwig-08,hasan-kane-10}). There is a renewed recent interest for the study of the topology of these models (see \eg \cite{thiang-15} or \cite{prodan-schulz-baldes-14,prodan-schulz-baldes-book-16} for the case of random systems). On the one hand it is a common belief that systems in class {\bf AIII} are classified by the $K$-group $K_1$. On the other hand a closer look at these systems reveals ambiguities and deficiencies in the various definitions, as recently pointed out in \cite{thiang-15} (see also the introduction of \cite{thiang-14}). The main open questions seem to be summarized by the following list:
\begin{enumerate}
\item[$\s{Q}$.1)] Which is the class of transformations which preserve the  chiral phases? \vspace{1mm}
\item[$\s{Q}$.2)] The topological indices which label the various chiral phases are \emph{absolute} or \emph{relative}? \vspace{1mm}
\item[$\s{Q}$.3)] Is $K_1$ enough to completely  describe  the topology of chiral systems?
\end{enumerate}
It is our conviction that a complete answer to the previous questions (and to the observations raised in \cite{thiang-15}) needs a precise definition of a geometric model for the description of systems in class {\bf AIII}. In this work we will show how to construct such a model, 
a fact which is important by itself for a better understanding  of the geometry underlying chiral systems. Moreover, a rigorous study of the related topology will provide a finer classification for the class {\bf AIII}.

\medskip

From a mathematical point of view the geometric object of our interest is a new class of vector bundles that we decided to call \emph{chiral vector bundles}, or \emph{$\chi$-vector bundles} for short. These are formally  defined as follow:
\begin{definition}[Chiral vector bundle]\label{def:chi-VB}
A \emph{chiral vector bundle} over $X$ is a pair $(\bb{E},\Phi)$ where $\pi:\bb{E}\to X$ is a complex hermitian vector bundle
and $\Phi\in \rm{Aut}(\bb{E})$ is an automorphism of $\bb{E}$.
The \emph{rank} of the chiral vector bundle $(\bb{E},\Phi)$ is the complex dimension of the fibers of $\bb{E}$.
\end{definition}
\noindent
We point out that we are not only interested in the classification over spheres or tori (as usual in the business of topological insulators). Our requirements on the nature of the base space $X$ are quite weak. Hereinafter in this paper we will assume that:
 \begin{assumption}\label{ass:1}
$X$ is a compact and path-connected Hausdorff space with a CW-complex structure.
\end{assumption}
\noindent
The particular choice of the nomenclature is justified by the fact that $\chi$-vector bundles provide a
suitable model  for the study of the topology of ground states of topological quantum systems endowed with a {chiral symmetry}. This connection will be briefly
discussed at the end of this introduction and
rigorously established  in  Section \ref{sect:chi_bloch_bun}. 

\medskip

Chiral vector bundles over a topological space $X$ form a  category when endowed with a proper notion of morphisms. These technical aspects will be discussed in full  detail in Section \ref{sect:chi_bun}. The only important information for the aims of this introductory discussion is  that in this category the notion of isomorphism does not coincide with that of homotopy equivalence, a fact which makes  chiral vector bundles quite different from complex vector bundles. The pure notion of isomorphism turns out to be too strong to set up a reasonable classification theory of topological nature. For this reason one needs to weaken the notion of equivalence by looking at  homotopy (see Example \ref{ex:homot_iso_bis} for more details). With this precaution the problem of the classification of chiral phases can be translated in the problem of the enumeration of the set
$$
{{\rm Vec}}_\chi^m(X)\;:=\;\Big\{\text{equivalence classes of \emph{homotopy equivalent}\ } \text{rank}\ m\ \text{chiral vector bundles over}
\ X
\Big\}.
$$
For a precise definition of the notion of homotopy equivalence we refer to Definition \ref{def:eq_rel_chi_VB}. The virtue of this definition is that ${{\rm Vec}}_\chi^m(X)$ can be classified by homotopy classes of maps from $X$ to a classifying space $\n{B}^m_\chi$, \ie 
\begin{equation}\label{eq:intro_1}
{{\rm Vec}}_\chi^m(X)\;\simeq\; \big[X,\n{B}_\chi^m\big]\;.
\end{equation}
The isomorphism \eqref{eq:intro_1} is  established in Corollary \ref{cor:iso_pullback} while the space $\n{B}_\chi^m$ is described in Section \ref{sect:model_classifying_space}. Both homotopy and cohomology groups of $\n{B}_\chi^m$ can be explicitly determined 
(\cf Section \ref{sec:homot_B} and Section \ref{sec:cohom_B}, respectively)
and along with \eqref{eq:intro_1} this allows for a classification of ${{\rm Vec}}_\chi^m(X)$. For instance, for the classification  of the topological phases of {free}-fermion systems the base space can be identified with a \virg{Brillouin} sphere (see the discussion in \cite[Section 2]{denittis-gomi-14}) and so, after setting
 $X=\n{S}^d$ in \eqref{eq:intro_1} one obtains:
\begin{theorem}[Classification over spheres]\label{theo:class_spher}
Rank $m$ chiral vector bundles over spheres are classified by the following bijections
$$
{{\rm Vec}}_\chi^m\big(\n{S}^d\big)\;\simeq\;\pi_d\big(\n{B}_\chi^m\big)\;\simeq\; \pi_{d}\big(\n{U}(m)\big)\;\oplus\; \pi_{d-1}\big(\n{U}(m)\big)\;.
$$
\end{theorem}

\noindent
In the case of  low-dimensional spheres the result of the classification is displayed in Table \ref{tab:intro_01}.1 along with the standard classification for complex  vector bundles
(see Section \ref{sect:class_spher} for a more detailed discussion).
%
%\begin{center}
 \begin{table}[h]\label{tab:intro_01}
 \centering
 \begin{tabular}{|c|c||c|c|c|c|}
 \hline
 \multicolumn{6}{|c|}{ 
{\bf Free-fermion systems}}\\
\hline
VB  & AZC  & $d=1$ & $d=2$&$d=3$&$d=4$\\
\hline
 \hline
 \rule[-3mm]{0mm}{10mm}
 ${\rm Vec}_{\C}^m(\n{S}^d)$& {\bf A} & $0$ & $\Z$ & $0$ &   \begin{tabular}{cc}
 {${0}$}& {($m=1$)}\vspace{1mm}\\
  $\Z$& ($m\geqslant2$)\\
\end{tabular}   \\
\hline
 \rule[-3mm]{0mm}{13mm}
${\rm Vec}_{\chi}^{m}(\n{S}^d)$ & {\bf AIII} & $\boxed{\Z}$ & $\Z$ & \begin{tabular}{cc}
 {$0$}& {($m=1$)}\vspace{1mm}\\
 $\boxed{\Z}$& ($m\geqslant2$)\\
\end{tabular}& \begin{tabular}{cc}
 ${0}$& {($m=1$)}\vspace{1mm}\\
 {$\Z\oplus\boxed{\Z_2}$}& {($m=2$)}\vspace{1mm}\\
  $\Z$& ($m\geqslant3$)\\
\end{tabular}\\
\hline
\end{tabular}\vspace{2mm}
 \caption{\footnotesize The column VB lists the symbols for the sets
of equivalence classes of vector bundles in the \emph{complex} and \emph{chiral} category, respectively.
The related
 Altland-Zirnbauer-Cartan labels \cite{schnyder-ryu-furusaki-ludwig-08}  are displayed in column AZC.  
 In the complex case the classification is specified by the first Chern class $c_1$ in dimension $d=2$ and by the second Chern class $c_2$
 in dimension $d=4$ (\cf \cite[Section 4]{denittis-gomi-14}). In the chiral case the classification  is richer and in addition to the usual Chern classes one has new odd-dimensional characteristic classes called \emph{chiral classes} (\cf Section \ref{sec:charact_class}).
The extra invariants introduced by the chiral structure are displayed by the  \virg{boxed} entries. In dimension $d=1$ and $d=3$ the $\Z$ classification is provided by the first chiral class $w_1$ and by the second chiral class $w_2$, respectively.
In dimension $d=4$ there is an \emph{unstable} chiral invariant $\Z_2$ only in rank $m=2$ which is related to the homotopy group $\pi_4(\n{S}^3)$ via a Hopf's fibration (\cf Section \ref{sect:class_spher}). 
}
 \end{table}
 %\end{center}
%

\medskip

The content of Table \ref{tab:intro_01}.1 leads to an immediate observation: the topology of chiral vector bundles seems to be just richer than the topology of complex vector bundles. This is indeed true and not surprising at all! From its very definition it turns out that a chiral vector bundle is, in particular, a complex vector bundle  plus an extra structure. Then, there is a \virg{forgetting} map  which associates to a given $\chi$-bundle $(\bb{E},\Phi)$ just its underlying complex vector bundle $\bb{E}$.  This map is evidently compatible with the notions of isomorphisms of complex vector bundles and homotopy equivalence of  $\chi$-bundles (which is weaker than the isomorphism equivalence) and leads to a well-defined  morphism
\begin{equation}\label{eq:intro_2}
\jmath\;:\:{{\rm Vec}}_\C^m(X)\;\hookrightarrow \; {{\rm Vec}}_\chi^m(X)
\end{equation}
which is indeed an injection. In particular the image of ${{\rm Vec}}_\C^m(X)$ in ${{\rm Vec}}_\chi^m(X)$ under $\jmath$ is described by (homotopy) equivalence classes of the type $[(\bb{E},{\rm Id}_{\bb{E}})]$ which are classified, at least in low dimension, by the help of Chern classes. On the opposite side there is another interesting subclass  of $\chi$-bundles ${{\rm Vec}}_{\chi\ast}^m(X)\subset{{\rm Vec}}_\chi^m(X)$ which consists of homotopy equivalence  classes of the type $[(X\times\C^m,\Phi)]$ where the  underlying vector bundle is 
trivial. Since the automorphisms of a trivial product bundle  $X\times\C^m$ (endowed with a Hermitian metric) are described by maps
$\varphi:X\to\n{U}(m)$, one has that
\begin{equation}\label{eq:intro_3}
{{\rm Vec}}_{\chi\ast}^m(X)\;\simeq \; \big[X,\n{U}(m)\big]\;.
\end{equation}
This is the part of ${{\rm Vec}}_{\chi}^m(X)$ which is \virg{purely chiral} in the sense that $\chi$-bundles in ${{\rm Vec}}_{\chi\ast}^m(X)$ have vanishing Chern classes due to the triviality of the underlying vector bundle and hence they are classified only by \emph{pure} chiral invariants. This subset, which under stabilization can be classified by $K_1(X)$, describes the systems of class {\bf AIII} usually considered in the physical literature concerning topological insulators. However, at least in our opinion, this seems to be a strong physical restriction which is not at all necessary. As a matter of fact,
elements in  ${{\rm Vec}}_{\chi\ast}^m(X)$ describe chiral systems which posses a ground  state (or Fermi projection) with trivial Chern topology and  this is not at all the general situation {in the theory of band operators (see \eg Section \ref{sec:toy_mos})}. On the other side the full set ${{\rm Vec}}_\chi^m(X)$ contains also models of  chiral systems associated to ground  states with a non-trivial Chern charge and for this reason we believe that a deeper understanding of the chiral phase requires 
the study of the full set ${{\rm Vec}}_\chi^m(X)$ rather than of its \virg{Chern-trivial} subset ${{\rm Vec}}_{\chi\ast}^m(X)$. This interpretative aspect will be properly developed and deepened in Section \ref{sect:chi_bloch_bun}.

\medskip

The classification of chiral vector bundles becomes more complicated when one replaces spheres with a generic base space $X$. In this situation, by taking a cue from the similar problem for complex vector bundles, one finds  advantageous to construct suitable characteristic classes. As usual these classes can be defined as the pullback with respect to the classifying maps in 
\eqref{eq:intro_1} of the integral cohomology of the space $\n{B}_\chi^m$, the latter being completely computable.
In this way one can associate a set of  cohomology classes to each $\chi$-bundle (\cf Section \ref{sec:charact_class}) and, at least in low-dimension, these classes suffice for a complete characterization of the topology. More precisely, one can prove the following result.
\begin{theorem}[Classification via characteristic classes]\label{theo:intro1}
Let $X$ be as in Assumption \ref{ass:1} and let $d\in\N$ be  the maximal dimension of its cells.  
\begin{enumerate}
\item[(i)] The Picard group of chiral line bundles is classified by the isomorphism
$$
(w_1,c_1)\;:\;{{\rm Vec}}_\chi^1(X)\;\stackrel{\simeq}{\longrightarrow}\;H^1(X,\Z)\;\oplus\;H^2(X,\Z)
$$
induced by the first chiral class $w_1$ and the first Chern class $c_1$.\vspace{1.3mm}
\item[(ii)] In the case of a generic rank $m\geqslant2$ 
and for low dimensional base spaces
 $1\leqslant d \leqslant3$ there are bijections of sets
\begin{equation}\label{eq:class_low_dim_int}
(w_1,c_1,w_2)\;:\;{{\rm Vec}}_\chi^m\big(X\big)\;\stackrel{\simeq}{\longrightarrow}\;H^1(X,\Z)\;\oplus\;H^2(X,\Z)\;\oplus\;H^3(X,\Z)
\end{equation}
 induced by the first two chiral classes $w_1,w_2$ and the first Chern class $c_1$. \vspace{1.3mm}
 
\item[(iii)] If $d=4$ and in the \emph{stable} range $m\geqslant 3$ there are bijections of sets
\begin{equation}\label{eq:class_low_dim_int_d4}
(w_1,c_1,w_2,c_2)\;:\;{{\rm Vec}}_\chi^m\big(X\big)\;\stackrel{\simeq}{\longrightarrow}\;H^1(X,\Z)\;\oplus\;H^2(X,\Z)\;\oplus\;H^3(X,\Z)\;\oplus\;H^4(X,\Z)
\end{equation}
induced by the first two chiral classes $w_1,w_2$ and the first two Chern classes $c_1,c_2$. 
 \end{enumerate}
\end{theorem}
\noindent
Since $H^k(X,\Z)=0$ if $k>d$, equation \eqref{eq:class_low_dim_int} has to be decorated by the obvious constraints $c_1=w_2=0$ if $d=1$ and $w_2=0$ if $d=2$.
Item (i) is proved in Proposition \ref{propos:chi_lin_bun} while the proof of items (ii) and (iii) is described in Proposition \ref{propos:class_low_dim}.
The first interesting unstable case $d=4, m=2$ is not covered by the above theorem. {A general analysis of this case reserves considerable mathematical difficulties as  discussed in Section \ref{sect_tori_4}.} 
With the help of Theorem \ref{theo:intro1} one can for instance classify $\chi$-bundles over tori of low dimension. The latter provides the geometric model for the study of the topological phases for
systems of fermions which interact with a \emph{periodic} background (see the discussion in \cite[Section 2]{denittis-gomi-14}). In the case of   
\virg{Brillouin} tori 
 $X=\n{T}^d$ the classification is displayed in Table \ref{tab:intro_02}.2. 
 \begin{table}[h]\label{tab:intro_02}
 \centering
 \begin{tabular}{|c|c||c|c|c|}
 \hline
 \multicolumn{5}{|c|}{ 
{\bf Periodic fermion systems}}\\
\hline
VB  & AZC   & $d=2$&$d=3$&$d=4$\\
\hline
 \hline
 \rule[-2mm]{0mm}{10mm}
 ${\rm Vec}_{\C}^m(\n{T}^d)$& {\bf A}  & $\Z$ & $\Z^3$ &   
 \begin{tabular}{cc}
 {$\Z^6$}& {($m=1$)}\vspace{1mm}\\
  $\Z^6\oplus\Z$& ($m\geqslant2$)\\
\end{tabular}   \\
\hline
 \rule[-2mm]{0mm}{14mm}
${\rm Vec}_{\chi}^{m}(\n{T}^d)$ & {\bf AIII}  & $\boxed{\Z^2}\oplus\Z$ & \begin{tabular}{cc}
 $\boxed{\Z^3}\oplus\Z^3$& {($m=1$)}\vspace{1.4mm}\\
 $\boxed{\Z^3}\oplus\Z^3\oplus\boxed{\Z}$& ($m\geqslant2$)\\
\end{tabular}& \begin{tabular}{cc}
 $\boxed{\Z^4}\oplus\Z^6$& {($m=1$)}\vspace{1mm}\\
 $\boxed{\Z^4}\oplus\Z^6\oplus\boxed{\Z^4}\oplus\Z\oplus\boxed{\Z_2}$& {($m=2$)}\vspace{1mm}\\
  $\boxed{\Z^4}\oplus\Z^6\oplus\boxed{\Z^4}\oplus\Z$& ($m\geqslant3$)\vspace{1mm}\\
\end{tabular}
\\
\hline
\end{tabular}\vspace{2mm}
 \caption{\footnotesize 
Notations and abbreviations are the same already used in the caption of Table \ref{tab:intro_02}.1. The case $d=1$ is already covered in Table \ref{tab:intro_02}.1 since $\T^1=\n{S}^1$.
 In the complex case the classification is specified by the first Chern class $c_1$ up to dimension $d=3$ and by the
 first and second Chern classes $(c_1,c_2)\in \Z^6\oplus\Z$ in dimension $d=4$ (\cf \cite[Section 4]{denittis-gomi-14}). 
 The extra invariants introduced by the  chiral symmetry are listed by the  \virg{boxed} entries. 
The splitting in the direct sum of the groups 
is order by the sequence of characteristic classes $(w_1,c_1,w_2,c_2)$ modulo the obvious conditions of triviality given by
$H^k(\n{T}^d,\Z)=0$ if $k>d$.
{The \emph{unstable} case   $d=4, m=2$,  not covered by Theorem \ref{theo:intro1}, is discussed in Section \ref{sect:class_tori}.}
}
 \end{table}

{
In this work we are interested in a topological classification of chiral vector bundles valid for topological base spaces with a CW-complex
structure. However, it is worth to emphasize that when the base space is a differential manifold (with no torsion) one can identify the characteristic classes
 in Theorem \ref{theo:intro1} with suitable differential forms. We will refer to Remark \ref{rk_diff_aspect} for some more detail on this aspect.
}

\medskip

Before ending this introduction, let us briefly justify  our basic assumption that chiral vector bundles, as introduced in Definition \ref{def:chi-VB}, provide the proper geometric models for the topology of the ground state of quantum systems protected by a chiral symmetry. This aspect will be dealt with in depth in Section \ref{sect:chi_bloch_bun} and  completed with a comparison with the existing literature. Let us start with the following definition which generalizes the usual notion of (translational invariant) topological insulator in class {\bf AIII}.
\begin{definition}[Topological quantum systems with chiral symmetry\footnote{The setting described in Definition \ref{def:tsq_Chi} can be generalized to unbounded operator-valued maps $x\mapsto H(x)$ by requiring 
the continuity  of the \emph{resolvent map} $x\mapsto R_z(x):=\big(H(x)-z\n{1}\big)^{-1}\in\bb{K}(\s{H})$. 
 Another possible generalization is to replace the norm-topology with the open-compact topology as in \cite[Appendix D]{freed-moore-13}. However, these kinds of generalizations have no particular consequences  for the purposes of this work.}]
 \label{def:tsq_Chi}
Let $X$ be a locally compact, path-connected, Hausdorff space.
Let $\s{H}$ be a separable complex Hilbert space and denote by
 $\bb{K}(\s{H})$  the algebra of  compact operators on $\s{H}$, respectively. 
A \emph{topological quantum system} is a self-adjoint map
\begin{equation}\label{eq:tqsA1}
X\;\ni\;x\; \longmapsto\; H(x)=H(x)^*\;\in\;\bb{K}(\s{H})
\end{equation}
continuous with respect to the norm-topology of $\bb{K}(\s{H})$. Let $\sigma(H(x))=\{\lambda_j(x)\ |\ j\in\s{I}\subseteq\Z\}\subset\R$  be the sequence of eigenvalues of $H(x)$ ordered  according to  $\ldots\lambda_{-2}(x)\leqslant\lambda_{-1}(x)<0\leqslant\lambda_1(x)\leqslant\lambda_2(x)\leqslant \ldots$. The map $x\mapsto \lambda_j(x)$ (which is continuous by standard perturbative arguments \cite{kato-95}) is called \emph{$j$-th energy band}. 
An \emph{isolated family} of energy bands is any (finite) collection $\Omega:=\{\lambda_{j_1}(\cdot),\ldots,\lambda_{j_m}(\cdot)\}$
of energy bands such that 
\begin{equation}\label{eq:tqsA2}
\min_{x\in X}\ {\rm dist}\left(
\bigcup_{s=1}^m\{\lambda_{j_s}\}\;,\; \bigcup_{j\in\s{I}\setminus\{j_1,\ldots,j_m\}}\{\lambda_{j}\}
\right)\;=\;C_g\;>0.
\end{equation}
Inequality \eqref{eq:tqsA2} is usually called \emph{gap condition}. We say that the system is subjected to a \emph{chiral symmetry} if there is 
 a continuous unitary-valued map $x\mapsto \chi(x)$ on $X$ such that
\begin{equation}\label{eq:tqsA3}
\left\{
\begin{aligned}
\chi(x)\; H(x)\; \chi(x)^*\;&=\;-\;H(x)\\ 
\chi(x)^2\;&=\; \n{1}_\s{H}\;
\end{aligned}\;,\qquad\qquad x\in X\;
\right.
\end{equation}
where $\n{1}_\s{H}$ is the identity operator\footnote{We notice that the second condition in  \eqref{eq:tqsA3} can be replaced by the equivalent constraint $\chi'(x)^2=- \n{1}_\s{H}$ under the identification $\chi=\ii \chi'$.
}.
\end{definition}

\noindent 
{Before discussing the geometric implications of Definition \ref{def:tsq_Chi}
let us make a comment. In many of the most relevant physical applications related to electronic systems the operator $\chi$ acts only at  level of internal (spinorial) degrees of freedom and not on the  spatial degrees of freedom. In  these cases the dependence of $\chi$  by the coordinate (or \emph{quasi-momentum}) $x$ is trivial in the sense that the 
unitary-valued map $x\mapsto \chi(x)$ in  \eqref{eq:tqsA3} is a constant. 
On the other hand this is not the most general possible situation and it is possible to design models where the basic symmetries need to be \emph{twisted} by a phase which depends on the position (\eg in the spirit of the Zak's magnetic translations). Toy models of this type are discussed in Section \ref{sec:toy_mos}; See in particular Example \ref{Ex:hofs_C2}.   For this reason we chose to consider in this work, at least for a mathematical interest, the more general situation described in Definition \ref{def:tsq_Chi}.}

\medskip

A standard construction, which is described in some detail in Section \ref{sec:bloc_bund_top_phas}, 
associates to each  topological quantum system
\eqref{eq:tqsA1} with an {isolated family} $\Omega_-$ of $m$ \emph{negative} energy bands \eqref{eq:tqsA2} a complex vector bundle $\bb{E}_{\Omega_-}\to X$ of rank $m$ which is usually called \emph{Bloch-bundle}. Due to the chiral symmetry the system possesses also a twin family $\Omega_+$ of $m$ \emph{positive} bands which define a  \virg{twin} Bloch-bundle  $\bb{E}_{\Omega_+}\to X$. These two vector bundles are in general non-trivial
and the chiral symmetry induces an isomorphism $\tilde\Theta_\chi:\bb{E}_-\to \bb{E}_+$. 
These data are enough to univocally specify (up to isomorphisms) a \emph{Clifford vector bundle} over $X$ of type $(0,1)$ (\cf Section \ref{sect:chi_bloch_bun}). Objects of this type provide a geometric model for the construction of $K_1(X)$,
as showed by Karoubi in \cite[Chapter III, Section 4]{karoubi-97}. This is the point of view  adopted in \cite{freed-moore-13,thiang-14} for the classification of system in class {\bf AIII}. However, although on the one hand the K-theory introduced by Karoubi fits perfectly  with the need to explain the \virg{Bott-clock}, on the other hand this K-theory does not seem to take into account the topological content associated with the Chern classes of the underlying vector bundle. A more precise discussion on this delicate point is postponed to the end of Section \ref{sect:com_lit}. The link between topological quantum systems of type {\bf AIII}, or equivalently Clifford vector bundles, and chiral vector bundles depends on the choice of a \emph{reference} isomorphism between $\bb{E}_{\Omega_-}$ and $\bb{E}_{\Omega_+}$. In order to make a comparison with the existing literature
it is better to provide a different splitting of the total Bloch bundle
$\bb{E}_{\Omega_-}\oplus\bb{E}_{\Omega_+}\simeq \bb{E}_{-}\oplus\bb{E}_{+}$
where the subbundles $\bb{E}_{\pm}$ are defined by a symmetric gradation (\cf Lemma \ref{lemma:gradiation_splitting}) and one has isomorphisms $\bb{E}_{\Omega_\pm}\simeq\bb{E}_{\pm}$ (\cf Lemma \ref{lemm:iso_ener_bund}) as well as an induced chiral isomorphism $\Theta_\chi:\bb{E}_-\to \bb{E}_+$. Any {reference} isomorphism between $\bb{E}_{\Omega_-}$ and $\bb{E}_{\Omega_+}$ also induces an isomorphism
$h_{\rm ref}:\bb{E}_-\to \bb{E}_+$ and once  such a $h_{\rm ref}$ has been chosen, one can define a rank $m$ chiral vector bundle $(\bb{E},\Phi)$ just by setting  
$\bb{E}:=\bb{E}_-$ and $\Phi:=h_{\rm ref}^{-1}\circ\Theta_\chi$. In our opinion, it is exactly the topology of $(\bb{E},\Phi)$ which completely describes the topological phase of the original topological quantum system with chiral symmetry. The arbitrariness inherent in the choice of $h_{\rm ref}$ seems to be an inescapable aspect 
and explicitly \cite{thiang-14,thiang-15} or implicitly \cite{schnyder-ryu-furusaki-ludwig-08,ryu-schnyder-furusaki-ludwig-10,qi-zhang-11,budich-trauzettel-13,prodan-schulz-baldes-14} this problem seems to be common to all attempts to define properly
 the chiral invariants. We refer to  Section \ref{sect:com_lit} for a more precise discussion.

\medskip

 At the end of this long introduction, let us summarize our point of view about the description of the topology of systems with a chiral symmetry as follows:
\begin{assumption}[Topological phase for chiral topological quantum systems]\label{defi:top_phases}
Let $X$ be a topological space which fulfills Assumption \ref{ass:1} and $X\ni x\mapsto H(x)$
a topological quantum system with chiral symmetry  $X\ni x\mapsto \chi(x)$ as in Definition
\ref{def:tsq_Chi}. 
Assume that there exists a \emph{ground state} described by
a system $\Omega_-$ of $m$ strictly negative energy bands (\emph{zero energy gap} condition, \cf  Assumption \ref{ass:2}) which are separated from the rest of the spectrum in the sense of \eqref{eq:tqsA2}.
 With these data one can build a pair of \virg{twin} rank $m$ vector bundles $\bb{E}_\pm\to X$ and a \emph{chiral} isomorphism $\Theta_\chi:\bb{E}_-\to \bb{E}_+$.
Let $h_{\rm ref}:\bb{E}_-\to \bb{E}_+$ be a \emph{reference} isomorphism arbitrarily chosen.
We \emph{assume} that the \emph{topological phase} of the 
chiral  system, relatively to the reference map $h_{\rm ref}$, is detected by the equivalence class
$$
[(\bb{E},\Phi)]\;\in\;{{\rm Vec}}_\chi^m(X)
$$
where $\bb{E}:=\bb{E}_-$ and $\Phi:=h_{\rm ref}^{-1}\circ\Theta_\chi$.
\end{assumption}

\medskip

\noindent
The virtue of our Assumption \ref{defi:top_phases}  is that it allows us to answer,  from our own point of view, the questions $\s{Q}$.1) - $\s{Q}$.3) stated at the very beginning of this introduction.
\begin{enumerate}
\item[$\s{R}$.1)] The chiral phases are \emph{homotopy invariants} of the system. Isomorphisms (chiral isometries) are too strong 
and generic unitary transformations as in \cite{thiang-15} are too weak to set up a proper topological theory for chiral systems.\vspace{1mm}

\item[$\s{R}$.2)]  The notion of topological phase of a chiral system is \emph{not} absolute 
but depends on an arbitrary choice of a reference  isomorphism.
 Only the notion of \emph{relative} phase
between two different states of the system has an absolute interpretation. 
This is equivalent to establishing a priori the \virg{reference} trivial chiral phase. 
{In conclusion, to the best  of our understanding of the problem, we consider  the chiral phase as a label for the (topologically) distinct ways that the positive and negative energy states of a system can be related by a chiral (pseudo-)symmetry (\cf Example \ref{Ex:hofs_C2}).}\vspace{1mm}

\item[$\s{R}$.3)] In the classification of the topological phases of  chiral systems 
also the topology of the \emph{ground state}, described by the topology of the underlying Bloch-bundle, plays a role.
The K-group $K_1$ is not enough. For instance it does not contain information about the Chern classes.
\end{enumerate}
\medskip

\noindent 
{In our opinion $\s{R}$.2) is worth of a final comment. In the axiomatic theory of characteristic classes the fact that the topological invariants are defined only \emph{relatively} to a given choice seems to be the rule rather than the exception. For instance this is true also for the familiar Chern classes which can be defined from the   
Hirzebruch's system of axioms only up to a normalization given by fixing the Chern class of the tautological line bundle over $\C P^\infty$
\cite[Chapter 17, Section 3]{husemoller-94}.}

%-----------------%
\medskip

\noindent
{\bf Acknowledgements.} 
GD wishes to thank A. Verjovsky and G. Thiang for many stimulating discussions.
GD's research is supported
 by
the  grant \emph{Iniciaci\'{o}n en Investigaci\'{o}n 2015} - $\text{N}^{\text{o}}$ 11150143 funded  by FONDECYT.	 KG's research is supported by 
the JSPS KAKENHI Grant Number 15K04871.
\medskip
%-----------------%

%--------------------%
\section{Chiral vector bundles}\label{sect:chi_bun}

Before starting the formal study of  $\chi$-bundles, {let us introduce the basic notions.
Every complex vector bundle $\pi:\bb{E}\to X$  can be  always endowed  with} a hermitian fiber-metric $\rr{m}:\bb{E}\times_\pi \bb{E}\to \C$ which turns out to be unique up to isomorphisms (see \eg \cite[Chapter I, Theorem 8.7]{karoubi-97}). 
Here $\bb{E}\times_\pi \bb{E}$ denotes the set of pairs $(p_1,p_2)\in \bb{E}\times \bb{E}$ such that $\pi(p_1)=\pi(p_2)$. 
This means that without loss of generality we can restrict our attention to \emph{hermitian} complex vector bundles of rank $m\in\N$. The latter are characterized by their structure group (the group in which the transition functions take values) which is the unitary group $\n{U}(m)$. An automorphism  of a hermitian rank $m$ complex vector bundle $\pi:\bb{E}\to X$ is a vector bundle isomorphism
\beql{eq:diag1}
\begin{diagram}
\bb{E}&& \pile{\rCorresponds^{\Phi}
} &&\bb{E}\\
&\rdTo_\pi&& \ldTo_{\pi}& \\
&&X &&  \\
\end{diagram}\;
\eeq
 which is also an isometry with respect to the hermitian fiber-metric, \ie $\rr{m}\big(p_1,p_2\big)=\rr{m}\big(\Phi(p_1),\Phi(p_2)\big)$ for all $p_1,p_2\in\bb{E}\times_\pi \bb{E}$.
We denote with ${\rm Aut}(\bb{E})$ the automorphism group of $\bb{E}$. Let us  remark that automorphisms in ${\rm Aut}(\bb{E})$ can be locally identified with maps $X\supset\f{U}\to\n{U}(m)$ where $\f{U}$ is a trivializing open set for $\bb{E}$.

\medskip

\medskip

The construction of a coherent topological theory of chiral vector bundles requires the introduction of a proper notion of isomorphism. Quite naturally one can say that two rank $m$ $\chi$-bundles $(\bb{E}_1,\Phi_1)$ and $(\bb{E}_2,\Phi_2)$ over $X$ are \emph{isomorphic} if there exists a vector bundle isomorphism $f:\bb{E}_1\to \bb{E}_2$ which makes the following diagram 
\beql{eq:diag2}
\begin{diagram}
           &                     &\bb{E}_1 & \pile{\rCorresponds^{\Phi_1}} &\bb{E}_1       &         &\\
           &  \ldTo^{\pi_1}  &            &                                                          &                 &\rdTo^{\pi_1} &\\
X&                     & \dCorresponds_{f}  &                                                     & \dCorresponds^{f}   &        &X   \\
           &  \luTo_{\pi_2}  &            &                                                          &                 &\ruTo_{\pi_2} &\\
            &                      &\bb{E}_2 & \pile{\rCorresponds_{\Phi_2}} &\bb{E}_2   &
\end{diagram}
\eeq
commutative. We use the notation $(\bb{E}_1,\Phi_1)\approx(\bb{E}_2,\Phi_2)$ to say that
 $(\bb{E}_1,\Phi_1)$ and $(\bb{E}_2,\Phi_2)$ are isomorphic. The \emph{trivial} rank $m$ $\chi$-bundle is, by definition, the pair $(X\times\C^m,\rm{Id}_{X\times\C^m})$ where $X\times\C^m\to X$ is the standard product vector bundle (endowed with the first factor projection)   and $\rm{Id}_{X\times\C^m}$ is the identity map which fixes each point $(x,\rm{v})\in X\times\C^m$. A $\chi$-bundle $(\bb{E},\Phi)$ is said to be \emph{strongly trivial} if $(\bb{E},\Phi)\approx (X\times\C^m,\rm{Id}_{X\times\C^m})$.   The notion of isomorphism $\approx$ induces an equivalence relation and one can define in a usual way the set of isomorphism classes 
\begin{equation}\label{eq:iso_rel}
\widetilde{{\rm Vec}}_\chi^m(X)\;:=\;\Big\{\text{rank}\ m\ \text{chiral vector bundles over}
\ X
\Big\}\;\Big/\;\approx\;.
\end{equation}

\begin{remark}[Weak local triviality]\label{rk:week_LT}{\upshape
Chiral vector bundles are not \virg{genuine} locally trivial objects. Indeed, a notion of \emph{local triviality} for a $\chi$-bundle
 $(\bb{E},\Phi)$, which is also
coherent with the above definitions of isomorphism $\approx$ and strong triviality, should require  the existence of a cover $\{\f{U}\}$ of open subsets of $ X$ and subordinate trivializing maps  that establish isomorphisms $(\bb{E}|_{\f{U}},\Phi|_{\f{U}})\approx(\f{U}\times\C^m,\rm{Id}_{\f{U}\times\C^m})$. However,  if from one side the existence of isomorphisms $h_{\f{U}}:\bb{E}|_{\f{U}}\to \f{U}\times\C^m$ is guaranteed by the fact that the underlying complex vector bundle $\bb{E}\to X$ is locally trivial, from the other side
it is not generally possible to choose the $h_{\f{U}}$ in such a way that $h_{\f{U}}\circ\Phi|_{\f{U}}\circ h_{\f{U}}^{-1}=\rm{Id}_{\f{U}\times\C^m}$.  
As an example the reader can just consider the complex product line bundle $\n{S}^1\times \C\to \n{S}^1$ endowed with the endomorphism $\Phi_n(\theta,{\rm v})=(\theta,\expo{\ii n \theta}{\rm v})$ for all $(\theta,{\rm v})\in \n{S}^1\times \C$.
Since the structure group $\n{U}(1)$ is commutative there is no way to change  globally, or just locally, the endomorphism $\Phi_n$ into the identity map.
Nevertheless,
a closer look  at the structure of a $\chi$-bundle $(\bb{E},\Phi)$  shows that $\Phi$ can be locally identified with maps $\phi_{\f{U}}:{\f{U}}\to\n{U}(m)$ through the assignment 
 $h_{\f{U}}\circ\Phi|_{\f{U}}\circ h_{\f{U}}^{-1}={\rm Id}_{\f{U}}\times \phi_{\f{U}}$. This observation allows us to affirm that chiral vector bundles are locally trivial in a \emph{weak} sense, meaning that each $\chi$-bundle admits local isomorphisms  of the type $(\bb{E}|_{\f{U}},\Phi|_{\f{U}})\approx(\f{U}\times\C^m,{\rm Id}_{\f{U}}\times \phi_{\f{U}})$ which provide a local \virg{twisted} product structure.
}\hfill $\blacktriangleleft$
\end{remark}

\medskip 

Given a   vector bundle $\pi:\bb{E}\to X$ and a map $\varphi:Y\to X$ one can construct a new vector bundle $\varphi^*\bb{E}\to Y$ 
called the \emph{pullback} of $\bb{E}$ over $Y$ via $\varphi$. The fibers of $\varphi^*\bb{E}$ are explicitly described by
\cite[Chapter 2, Section 5]{husemoller-94}
\begin{equation}\label{eq:pullback1}
\varphi^*\bb{E}|_y\;:=\;\Big\{(y,p)\in Y\times\bb{E}\ \big |\ \varphi(y)=\pi(p)\Big\}\;\simeq\;\bb{E}|_{\varphi(y)}\;,\qquad\quad \text{for a fixed}\ \ y\in Y\;.
\end{equation}
When  $\varphi_1,\varphi_2:Y\to X$ are homotopic maps (and $Y$ is paracompact)  the pullbacks $\varphi_1^*\bb{E}$ and $\varphi_2^*\bb{E}$ turn out to be isomorphic in the category of complex vector bundles over $Y$ \cite[Chapter 3, Theorem 4.7]{husemoller-94}. One of the major consequences of this result is that one can reinterpret the notion of isomorphism between complex vector bundles in terms of homotopy equivalences. 
More precisely one has that $\bb{E}_1\to X$ and $\bb{E}_2\to X$ are isomorphic as complex vector bundles if and only if there is a vector bundle $\bb{E}\to X\times [0,1]$ such that $\bb{E}|_{X\times \{0\}}$ (resp. $\bb{E}|_{X\times \{1\}}$) is isomorphic to $\bb{E}_1$ (resp. $\bb{E}_2$).

 \begin{remark}\label{rk:iso_homot_CVB}{\upshape
There is a different homotopy characterization of the notion of isomorphism: Two complex vector bundles $\bb{E}_1\to X$ and $\bb{E}_2\to X$ turn out to be isomorphic (in the proper category)  if and only if there exists a map $\varphi:X\to X$ such that $\varphi$ is homotopic to ${\rm Id}_X$ and $\varphi^*\bb{E}_1$ is isomorphic to $\bb{E}_2$. This claim is ultimately a consequence of the fact that ${\rm Id}_X^*\bb{E}\simeq \bb{E}$ \cite[Chapter 2, Proposition 5.7]{husemoller-94}.
}\hfill $\blacktriangleleft$
\end{remark}

\medskip

Unfortunately, the situation is less simple in the context of chiral vector bundles. Of course, the notion of pullback is still well defined. Indeed if $(\bb{E},\Phi)$ is a $\chi$-bundle over $X$ and $\varphi:Y\to X$ is a map we can define the pullback $\chi$-bundle
$\varphi^*(\bb{E},\Phi):=(\varphi^*\bb{E},\varphi^*\Phi)$ over $Y$ where the underlying complex vector bundle $\varphi^*\bb{E}\to Y$ is described
fiberwise by \eqref{eq:pullback1} and the morphism $\varphi^*\Phi:\varphi^*\bb{E}\to \varphi^*\bb{E}$ is defined by
\begin{equation}\label{eq:pullback2}
\varphi^*\Phi\;:\;(y,p)\;\longmapsto\; (y,\Phi(p))\;, \qquad\quad \forall\ (y,p)\in \varphi^*\bb{E}\;.
\end{equation}
Since $\Phi$ is an automorphism of $\bb{E}$ the map  $\varphi^*\Phi$ is an isomorphism on each fiber $\varphi^*\bb{E}|_y$ and this is enough to affirm that $\varphi^*\Phi\in{\rm Aut}(\varphi^*\bb{E})$ \cite[Chapter 3, Theorem 2.5]{husemoller-94}.
Let us now consider two homotopic maps $\varphi_1,\varphi_2:Y\to X$ ($Y$  paracompact) and the related pullback $\chi$-bundles
$\varphi_1^*(\bb{E},\Phi)$ and $\varphi_2^*(\bb{E},\Phi)$ over $Y$. Since $\varphi_1^*\bb{E}$ and $\varphi_2^*\bb{E}$ are isomorphic as complex vector bundles there exists a map $f:\varphi_1^*\bb{E}\to \varphi_2^*\bb{E}$ which restricts to a linear isomorphism $f_y$ on each pair of fibers, \ie
$$
\bb{E}|_{\varphi_1(y)}\;{\simeq}\;\varphi_1^*\bb{E}|_y\;\stackrel{f_y}{\simeq}\; \varphi_2^*\bb{E}|_y\;{\simeq}\;\bb{E}|_{\varphi_2(y)}\;,\qquad\quad \forall\ y\in Y\;.
$$
Let $(y,p)\in \varphi_1^*\bb{E}|_y$ and $f(y,p)=(y,f_y(p))$ the corresponding point in  $\varphi_2^*\bb{E}|_y$
(with an abuse of notation, we are identifying $f_y$ with the linear isomorphism $f_y:\bb{E}|_{\varphi_1(y)}\to\bb{E}|_{\varphi_2(y)}$). A straightforward calculation shows that
$$
f\circ\varphi_1^*\Phi\;:\;(y,p)\;\to\; \big(y,f_y(\Phi(p))\big)\;,\qquad\qquad \varphi_2^*\Phi\circ f\;:\;(y,p)\;=\;\big(y, \Phi(f_y(p))\big)\;.
$$
Thus, the condition $f\circ\varphi_1^*\Phi=\varphi_2^*\Phi\circ f$ for a $\chi$-bundle isomorphism between
$\varphi_1^*(\bb{E},\Phi)$ and $\varphi_2^*(\bb{E},\Phi)$ in the sense of \eqref{eq:diag2}
necessitates that  validity of
\begin{equation}\label{eq:pullback3}
f_y\;\circ\;\Phi|_{\bb{E}|_{\varphi_1(y)}}\;=\;\Phi|_{\bb{E}|_{\varphi_2(y)}}\;\circ\; f_y\;,\qquad\quad \forall\ y\in Y\;.
\end{equation}
The set of equations \eqref{eq:pullback3} represents the obstruction for a homotopy between $\chi$-bundles to be an isomorphism.

\begin{example}\label{ex:homot_iso}{\upshape
 Let $(\n{S}^1\times\C,\Phi_n)$ be the $\chi$-line bundle  considered in Remark \ref{rk:week_LT}.
Given a map $\varphi:\n{S}^1\to \n{S}^1$ one has that $\varphi^*(\n{S}^1\times\C)$ is isomorphic to $\n{S}^1\times\C$ 
as  complex vector bundles. However, the pullback automorphism $\varphi^*\Phi_n$ acts on each point $(\theta,{\rm v})\in \n{S}^1\times\C$ as
$\varphi^*\Phi_n(\theta,{\rm v})=(\theta,\expo{\ii n \varphi(\theta)}{\rm v})$. Hence, if $\varphi_1,\varphi_2:\n{S}^1\to \n{S}^1$
are two distinct maps one has that $\varphi^*_1\Phi_n$ and $\varphi^*_2\Phi_n$ cannot be related by the equation \eqref{eq:pullback3}, even in the case one assumes $\varphi_1$ and $\varphi_2$ homotopic. This is just because the underlying structure group $\n{U}(1)$ is commutative.
On the other side it seems natural to believe that all the relevant topology of the $\chi$-line bundle $(\n{S}^1\times\C,\Phi_n)$ has to be contained in the map $\theta\mapsto \expo{\ii n \theta}$, which is topologically characterized by its \emph{winding number} $n$. In the  case the map $\varphi$ is homotopic to the identity map ${\rm Id}_{\n{S}^1}$ one has that also $\theta\mapsto \expo{\ii n \varphi(\theta)}$ is completely specified by the {winding number} $n$. Then,  the two $\chi$-line bundle $(\n{S}^1\times\C,\Phi_n)$ and $\varphi^*(\n{S}^1\times\C,\Phi_n)$ should be considered \virg{topologically equivalent} even though they are not isomorphic in the strong sense of  diagram \eqref{eq:diag2}.
 }\hfill $\blacktriangleleft$
\end{example}

\medskip

 Example \ref{ex:homot_iso} suggests that it is necessary  to relax the notion of equivalence of  $\chi$-bundles in \eqref{eq:iso_rel} in order to have a classification theory  which is preserved  under homotopy equivalences.

\begin{definition}[Equivalence of chiral vector bundles]\label{def:eq_rel_chi_VB}
Let $(\bb{E}_1,\Phi_1)$ and $(\bb{E}_2,\Phi_2)$ be two rank $m$ chiral vector bundles over the same base space $X$. We say that $(\bb{E}_1,\Phi_1)$ and $(\bb{E}_2,\Phi_2)$ are \emph{equivalent}, and we write $(\bb{E}_1,\Phi_1)\sim(\bb{E}_2,\Phi_2)$,  if and only if there is a rank $m$ chiral vector bundle
$(\bb{E},\Phi)$ over $X\times[0,1]$ such that
$$
(\bb{E}_1,\Phi_1)\;\approx \big(\bb{E}|_{X\times \{0\}},\Phi|_{X\times \{0\}}\big)\;\qquad\text{and}\qquad (\bb{E}_2,\Phi_2)\;\approx \big(\bb{E}|_{X\times \{1\}},\Phi|_{X\times \{1\}}\big)
$$
 where $\approx$ stands for the isomorphism relation between  $\chi$-bundles established by the diagram \eqref{eq:diag2}.
\end{definition} 

\medskip

\noindent
From Definition \ref{def:eq_rel_chi_VB} one immediately deduces that $\sim$ is an equivalence relation on the set of chiral vector bundles. This differs from the isomorphism relation $\approx$ and, in particular, it turns out to be \emph{weaker}. Indeed, just by choosing $\bb{E}=\bb{E}_1\times [0,1]$ and $\Phi=\Phi_1\times{\rm Id}_{[0,1]}$, one can verify that $(\bb{E}_1,\Phi_1)\approx(\bb{E}_2,\Phi_2)$ implies  $(\bb{E}_1,\Phi_1)\sim(\bb{E}_2,\Phi_2)$.
The set of equivalence classes of rank $m$ $\chi$-bundles with respect to $\sim$ will be denoted by
\begin{equation}\label{eq:equi_rel}
{{\rm Vec}}_\chi^m(X)\;:=\;\Big\{\text{rank}\ m\ \text{chiral vector bundles over}
\ X
\Big\}\;\Big/\;\sim\;.
\end{equation}
We will show in the rest of this work that ${{\rm Vec}}_\chi^m(X)$ admits a homotopic classification which can be characterized in term of proper characteristic classes, at least in low dimension.

\begin{example}\label{ex:homot_iso_bis}{\upshape
 Let us consider two $\chi$-line bundles $(\n{S}^1\times\C,\Phi_i)$, $i=1,2$, defined by
 $\Phi_i(\theta,{\rm v})=(\theta,\expo{\ii  \phi_i(\theta)}{\rm v})$ with  $\phi_1,\phi_2:\n{S}^1\to \n{S}^1$ two continuous maps. As showed in Example \ref{ex:homot_iso} the equality $\phi_1=\phi_2$ is a necessary and sufficient condition for the isomorphism (in the strong sense of  diagram \eqref{eq:diag2}) between the two  $\chi$-line bundles. On the other side the equivalence relation of Definition \ref{def:eq_rel_chi_VB} is less restrictive and requires only the homotopy equivalence between the maps $\phi_1$ and $\phi_2$. Indeed, let us assume the existence of a homotopy $\hat{\phi}:\n{S}^1\times[0,1]\to\n{S}^1$ such that $\hat{\phi}(\theta,0)=\phi_1(\theta)$
 and $\hat{\phi}(\theta,1)=\phi_2(\theta)$ for all $\theta\in\n{S}^1$. Then one can check that  $(\n{S}^1\times\C,\Phi_1)\sim(\n{S}^1\times\C,\Phi_2)$ just by looking at the $\chi$-line bundle $(\n{S}^1\times[0,1]\times\C,\Phi)$ over the base space $\n{S}^1\times[0,1]$ defined by the automorphism
 $\Phi(\theta,t,{\rm v})=(\theta,t,\expo{\ii  \hat{\phi}(\theta,t)}{\rm v})$. Summarizing,  one has that
 $$
\widetilde{{\rm Vec}}_\chi^1\big(\n{S}^1\big)\;\simeq\;{\rm Map}\big(\n{S}^1,\n{U}(1)\big)\;,\quad\quad {{\rm Vec}}_\chi^1\big(\n{S}^1\big)\;\simeq\;\pi_1\big(\n{S}^1\big)\;\simeq\;\Z\;,\quad\quad {{\rm Vec}}_\C^1\big(\n{S}^1\big)\;\simeq\;\pi_1\big(C\n{P}^\infty\big)\;\simeq\;0
 $$
 and this clearly indicates that the equivalence relation $\sim$ introduced in Definition \ref{def:eq_rel_chi_VB} provides a  proper structure  for a topological classification of chiral vector bundles. Conversely, the notion of \emph{strong} isomorphism given by \eqref{eq:diag2} results too strong (it distinguishes too much) while the notion of isomorphism in the complex category is too week (it does not distinguish anything).
 }\hfill $\blacktriangleleft$
\end{example}

The following result extends the homotopy property of complex vector bundles \cite[Chapter 3, Theorem 4.7]{husemoller-94} to the class of chiral vector bundles, the right notion of equivalence being the one introduced in Definition \ref{def:eq_rel_chi_VB}.

\begin{proposition}[Homotopy property]\label{prop:eq_rel_chi_VB}
Let $(\bb{E},\Phi)$  be a rank $m$ chiral vector bundle over the base space $X$
and $\varphi_1,\varphi_2:Y\to X$ two homotopic maps ($Y$ is assumed to be paracompact).
Then 
$\varphi_1^*(\bb{E},\Phi)\sim\varphi_2^*(\bb{E},\Phi)$ as chiral vector bundles over $Y$.
\end{proposition} 
\proof
 Let $\hat{\varphi}:Y\times[0,1]\to X$ be the homotopy between $\varphi_1$ and $\varphi_2$
  such that $\hat{\varphi}(y,0)=\varphi_1(y)$ and $\hat{\varphi}(y,1)=\varphi_2(y)$ for all $y\in Y$.
 Consider the vector bundle $\pi':\bb{E}'\to Y\times[0,1]$ with fibers 
$
{\pi'}^{-1}(y,t)=\hat{\varphi}_t^*\bb{E}
$
where we used the short notation $\hat{\varphi}_t:=\hat{\varphi}(\cdot,t)$ for $t\in[0,1]$. 
Each point in $\bb{E}'$ is identified by a triplet $(y,t,p)$ such that $\pi'(y,t,p)=(y,t)$ and $p\in\bb{E}|_{\hat{\varphi}_t(y)}$.
We can endow $\bb{E}'$ with a chiral structure $\Phi'$ as follows: $\Phi'(y,t,p)=(y,t,\Phi(p))$.
This is enough to show that $(\bb{E}|_{X\times \{0\}},\Phi|_{X\times \{0\}})\approx \varphi_1^*(\bb{E},\Phi)$
and $(\bb{E}|_{X\times \{1\}},\Phi|_{X\times \{1\}})\approx \varphi_2^*(\bb{E},\Phi)$.
\qed

\section{Homotopy classification for chiral vector bundles}

The celebrated \emph{Brown's representability theorem} \cite{brown-62}, \cite[Theorem 4E.1]{hatcher-02} may ensure abstractly (upon  a verification of few structural axioms)   the existence of a classifying
space $\n{B}_\chi^m$ for  chiral vector bundles of rank $m$ in the sense that
\begin{equation}\label{eq:class_space0}
{{\rm Vec}}_\chi^m(X)\;\simeq\; \left[X,\n{B}_\chi^m\right]
\end{equation}
where on the left-hand side there is the set of equivalence classes of  $\chi$-bundles in the sense of \eqref{eq:equi_rel} while  on the right-hand side there is the set of   homotopy classes of maps of the base space $X$ into  $\n{B}_\chi^m$.
The aim of this section is to provide a concrete geometric model for $\n{B}_\chi^m$.

\subsection{Geometric model for the classifying space}\label{sect:model_classifying_space}
{For each pair of integers $1\leqslant m\leqslant n$    
$$
G_m(\C^n)\simeq\n{U}(n)/\big(\n{U}(m)\times \n{U}(n-m)\big)
$$
denotes the \emph{Grassmann manifold} of $m$-dimensional (complex) subspaces of $\C^n$. 
 The inclusions $\C^n \subset \C^{n+1} \subset\ldots$ given by ${\rm v}\mapsto ({\rm v},0)$  yield inclusions $G_m(\C^n)\subset G_m(\C^{n+1})\subset\ldots$ and one can equip
$$
G_m(\C^\infty)\;:=\;\bigcup_{n=m}^{\infty}\;G_m(\C^n)\;,
$$
with the direct limit topology. The inclusions $\C^n \subset \C^{n+1} \subset\ldots$  also yield inclusions $\n{U}(n)\subset\n{U}(n+1)\subset\ldots$ and the space
$$
\n{U}(\infty)\;:=\;\bigcup_{n=1}^{\infty}\;\n{U}(n)\;,
$$
endowed with the direct limit topology, is sometimes called \emph{Palais unitary group} \cite{palais-65}.
The homotopy groups of the Palais group are described by the \emph{Bott periodicity} \cite{bott-58,bott-59}
$$
\pi_{k}\big(\n{U}(\infty)\big)\;=\;
\left\{
\begin{aligned}
&0&&\text{if}\quad k\ \ \text{even or}\ \ 0\\
&\Z&&\text{if}\quad k\ \ \text{odd}\;.\\
\end{aligned}
\right.
$$
}

\medskip

Let us introduce the following family of sets parametrized by $1\leqslant m\leqslant n$:
\begin{equation}\label{eq:class_space1}
\chi_m\big(\C^n\big)\;:=\;\Big\{(\Sigma,u)\in\ G_m(\C^n)\times\n{U}(n)\ \big|\ u(\Sigma)=\Sigma\Big\}\;.
\end{equation}
Each element in $\chi_m\big(\C^n\big)$ is a pair $(\Sigma,u)$ given by a subspace $\Sigma\subset\C^n$ of complex dimension $m$ and a unitary matrix $u\in \n{U}(n)$ which preserves $\Sigma$. 
 The inclusions $\C^n \subset \C^{n+1} \subset\ldots$  also yield the obvious inclusions $\chi_m\big(\C^n\big)\subset \chi_m\big(\C^{n+1}\big)\subset\ldots$ and this  suggests the following
\begin{definition}[Classifying space for $\chi$-bundles]\label{defi:class_space}
For each pair of integers $1\leqslant m\leqslant n$ let $\chi_m\big(\C^n\big)$ be the space given by \eqref{eq:class_space1}. The space 
$$
\n{B}_\chi^m\;:=\;\bigcup_{n=m}^{\infty}\;\chi_m\big(\C^n\big)\;,
$$
equipped with the direct limit topology, will be called the \emph{classifying space} for chiral vector bundles.
\end{definition}

\medskip

\noindent
In order to justify this name we need to prove that this space $\n{B}_\chi^m$ 
really provides a model for the realization of the isomorphism \eqref{eq:class_space0}.

\subsection{The chiral universal vector bundle} 
Each manifold $G_m(\C^n)$ is the base space of a {canonical} rank $m$ complex vector bundle $\bb{T}_m^n\to G_m(\C^n)$ with total space $\bb{T}_m^n$ given by all pairs $(\Sigma,{\rm v})$ with $\Sigma\in G_m(\C^n)$  and ${\rm v}$ a vector in $\Sigma$. The bundle projection  is defined by the natural restriction
 $(\Sigma,{\rm v})\mapsto\Sigma$. Now, when  $n$ tends to infinity, the same construction leads to 
 the \emph{tautological} $m$-plane bundle $\bb{T}_m^\infty\to G_m(\C^\infty)$. 
This vector bundle is the universal object which classifies  complex vector bundles in the sense that any rank $m$ complex vector bundle $\bb{E}\to X$ can be realized, up to isomorphisms, as the pullback of  $\bb{T}_m^\infty$ with respect to a \emph{classifying map} $\varphi : X \to G_m(\C^\infty)$, that is $\bb{E}\simeq \varphi^\ast \bb{T}_m^\infty$.
Since pullbacks of homotopic maps yield isomorphic complex vector bundles (\emph{homotopy property}), the classification of  $\bb{E}$ only depends on the homotopy class  of $\varphi$.  This leads to the fundamental result 
\beql{eq:class_compl_VB1}
{\rm Vec}^m_\C(X)\; \simeq\;\big[X , G_m(\C^\infty)\big]
\eeq
where in the right-hand side  there is the set of the equivalence classes of homotopic maps of $X$ into $G_m(\C^\infty)$ \cite{atiyah-67,milnor-stasheff-74,husemoller-94}.

\medskip

Based on the definition \eqref{eq:class_space1} one can consider the composition 
\begin{equation}\label{eq:class_space3}
\chi_m\big(\C^n\big)\;\stackrel{\imath}{\longrightarrow}\;G_m(\C^n)\;\times\;\n{U}(n)\;\stackrel{\rm pr_1}{\longrightarrow}\;G_m(\C^n)
\end{equation}
where $\imath$ is the inclusion given by $\chi_m\big(\C^n\big)\subset G_m(\C^n)\times\n{U}(n)$
and ${\rm pr_1}$ is the first factor projection. The pullback ${~^{\chi}\!\bb{T}_m^n}:=({\rm pr_1}\circ \imath)^*\bb{T}_m^n$ defines a rank $m$ vector bundle over $\chi_m\big(\C^n\big)$ that can be endowed  with a  natural automorphism which can be easily  constructed after 
a close inspection of 
the structure of the total space
\begin{equation}\label{eq:chi_taut1}
\begin{aligned}
{~^{\chi}\!\bb{T}_m^n}\;:&=\;\Big\{\Big((\Sigma, u),(\Sigma, {\rm v})\Big)\in \chi_m\big(\C^n\big)\times \bb{T}_m^n\ \big|\ u(\Sigma)=\Sigma,\ \ {\rm v}\in\Sigma \Big\}\\
&\simeq\;\Big\{(\Sigma, u, {\rm v})\in G_m(\C^n)\times\n{U}(n)\times \C^n\ \big|\ u(\Sigma)=\Sigma,\ \ {\rm v}\in\Sigma  \Big\}
\end{aligned}
\end{equation}
of the vector bundle $\pi:{~^{\chi}\!\bb{T}_m^n}\to \chi_m\big(\C^n\big)$. The second representation provides two important facts: first of all the bundle projection $\pi$
can be equivalently described as  $\pi(\Sigma, u, {\rm v})=(\Sigma, u)$; second, the vector bundle
${~^{\chi}\!\bb{T}_m^n}$ is identifiable with a vector subbundle of the product vector bundle $\big(G_m(\C^n)\times\n{U}(n)\big)\times\C^m$. On this product bundle, the usual action of $\n{U}(n)$ on $\C^n$ defines the automorphism
\begin{equation}\label{eq:uni_aut}
\Phi_m^n\;:\;(\Sigma, u,{\rm v})\;\longmapsto\;(\Sigma, u,u\cdot{\rm v})\;,\qquad\quad\forall\  (\Sigma, u,{\rm v})\;\in\;\Big(G_m(\C^n)\times\n{U}(n)\Big)\times\C^m\;.
\end{equation}
By construction this automorphism preserves the fibers of ${~^{\chi}\!\bb{T}_m^n}$. Hence,
 with a slight abuse of notation, 
equation \eqref{eq:uni_aut} defines an element $\Phi_m^n\in{\rm Aut}({~^{\chi}\!\bb{T}_m^n})$.

\medskip

Under the inclusions $\C^n \subset \C^{n+1} \subset\ldots$ given by ${\rm v}\mapsto ({\rm v},0)$
one has the related inclusions ${~^{\chi}\!\bb{T}_m^n}\subset {~^{\chi}\!\bb{T}_m^{n+1}} \subset\ldots$ which are compatible with the bundle projections and the automorphisms $\Phi_m^n$. The natural generalization of the idea of a universal vector bundle for $\chi$-bundles can be  described as follows:
\begin{definition}[Universal $\chi$-bundle]
For each pair of integers $1\leqslant m\leqslant n$ let $\pi : {~^{\chi}\!\bb{T}_m^n}\to \chi_m\big(\C^n\big)$ be the rank $m$ complex vector bundle \eqref{eq:chi_taut1} endowed with the automorphism $\Phi_m^n\in{\rm Aut}({~^{\chi}\!\bb{T}_m^n})$ defined by \eqref{eq:uni_aut}. The space 
$$
{~^{\chi}\!\bb{T}_m^\infty}\;:=\;\bigcup_{n=m}^{\infty}\;{~^{\chi}\!\bb{T}_m^n}\;,
$$
equipped with the direct limit topology, 
provides the total space for a rank $m$ complex vector bundle $\pi :{~^{\chi}\!\bb{T}_m^\infty}\to \n{B}_\chi^m$ over the classifying space introduced in Definition \ref{defi:class_space}. Moreover, the map $\Phi_m^\infty$ defined by $\Phi_m^\infty|_{{~^{\chi}\!\bb{T}_m^n}}=\Phi_m^n$ provides an automorphism $\Phi_m^\infty\in{\rm Aut}({~^{\chi}\!\bb{T}_m^\infty})$. The pair $({~^{\chi}\!\bb{T}_m^\infty}, \Phi_m^\infty)$ has the structure of a rank $m$ chiral vector bundle and it will be called the \emph{universal} $\chi$-bundle.
\end{definition}

\medskip

\noindent
Notice that $\n{B}_\chi^m$ is a subspace of $G_m(\C^\infty)\times\n{U}(\infty)$ and ${~^{\chi}\!\bb{T}_m^\infty}$ can be identified as the pullback of the universal bundle $\bb{T}_m^\infty$ for complex vector bundles under the mappings
\begin{equation}\label{eq:class_space33}
\n{B}_\chi^m\;\stackrel{\imath}{\longrightarrow}\;G_m(\C^\infty)\;\times\;\n{U}(\infty)\;\stackrel{\rm pr_1}{\longrightarrow}\;G_m(\C^\infty)\;.
\end{equation}

\subsection{The homotopy classification} 
The main aim of this section is the proof of the following result:
\begin{theorem}\label{teo:iso_pullback}
Let $X$ be a compact 
Hausdorff space. Every rank $m$
chiral vector bundle $(\bb{E},\Phi)$ over $X$ admits a map $\varphi:X\to\n{B}_\chi^m$ such
that  $(\bb{E},\Phi)$ is isomorphic (in the sense of the relation $\approx$ given in diagram \eqref{eq:diag2}) to
the pullback $\varphi^*({~^{\chi}\!\bb{T}_m^\infty},\Phi_m^\infty):=(\varphi^*{~^{\chi}\!\bb{T}_m^\infty},\varphi^*\Phi_m^\infty)$ of the universal $\chi$-bundle.
\end{theorem}

\medskip

\noindent
Before proving this result, we first present the main consequence of Theorem \ref{teo:iso_pullback}.
\begin{corollary}[Homotopy classification]\label{cor:iso_pullback}
Let $X$ be a compact 
Hausdorff space. Then there is a natural bijection 
\begin{equation*}
{{\rm Vec}}_\chi^m(X)\;\simeq\; \left[X,\n{B}_\chi^m\right]
\end{equation*}
where on the left-hand side there is the set of equivalence classes of  $\chi$-bundles in the sense of \eqref{eq:equi_rel} while  on the right-hand side there is the set of   homotopy classes of maps of the base space $X$ into the classifying space  $\n{B}_\chi^m$ described in Definition \ref{defi:class_space}.
\end{corollary}
\proof
Proposition \ref{prop:eq_rel_chi_VB} 
says that the map $\kappa: \big[X,\n{B}_\chi^m\big] \to {{\rm Vec}}_\chi^m(X)$
which associates to each homotopy class $[\varphi]$ the equivalence class $[\varphi^*({~^{\chi}\!\bb{T}_m^\infty},\Phi_m^\infty)]$ of $\chi$-bundles is well-defined and Theorem \ref{teo:iso_pullback} implies that $\kappa$ is surjective.
The injectivity of $\kappa$ is a consequence of Definition \ref{def:eq_rel_chi_VB} for the equivalence of $\chi$-bundles.  
Let $(\bb{E}_i,\Phi_i):= \varphi^*_i({~^{\chi}\!\bb{T}_m^\infty},\Phi_m^\infty)$, $i=1,2$ be two  $\chi$-bundles
obtained by pullbacking with respect to the maps $\varphi_i:X\to \n{B}_\chi^m$
and assume that
$(\bb{E}_1,\Phi_1)\sim (\bb{E}_2,\Phi_2)$. This implies the existence of a $\chi$-bundle
$(\bb{E},\Phi)$ over $X\times[0,1]$ such that
$
(\bb{E}_1,\Phi_1)\approx \big(\bb{E}|_{X\times \{0\}},\Phi|_{X\times \{0\}}\big)$ and $(\bb{E}_2,\Phi_2)\approx \big(\bb{E}|_{X\times \{1\}},\Phi|_{X\times \{1\}}\big)
$. Theorem \ref{teo:iso_pullback} assures that $(\bb{E},\Phi)$ can be identified, up to an isomorphism, with $\hat{\varphi}^*({~^{\chi}\!\bb{T}_m^\infty},\Phi_m^\infty)$ for some map $\hat{\varphi}:X\times[0,1]\to \n{B}_\chi^m$.
The isomorphisms $\varphi^*_1({~^{\chi}\!\bb{T}_m^\infty},\Phi_m^\infty)\approx \hat{\varphi}_0^*({~^{\chi}\!\bb{T}_m^\infty},\Phi_m^\infty)$ and $\varphi^*_2({~^{\chi}\!\bb{T}_m^\infty},\Phi_m^\infty)\approx \hat{\varphi}_1^*({~^{\chi}\!\bb{T}_m^\infty},\Phi_m^\infty)$ and the defintion of pullback imply that $\varphi_1=\hat{\varphi}_0:=\hat{\varphi}(\cdot,0)$ and 
$\varphi_2=\hat{\varphi}_1:=\hat{\varphi}(\cdot,1)$, namely $\hat{\varphi}$ is an homotopy between $\varphi_1$ and $\varphi_2$.
Hence, $\varphi_1$ and $\varphi_2$ define the same class in $\big[X,\n{B}_\chi^m\big]$ and $\kappa$ turns out to be injective.
\qed

\medskip
\medskip

\proof[Proof of Theorem \ref{teo:iso_pullback}]
Let $X$ be compact and  Hausdorff and $(\bb{E},\Phi)$ a $\chi$-bundle over $X$. Then,  there is a
positive integer $n$ and a map $\psi:X\to G_m(\C^n)$ such that 
$\bb{E}':=\psi^*\bb{T}_m^n$ is
isomorphic to the vector bundle $\bb{E}$   (see \eg \cite[Chapter 3, Proposition 5.8]{husemoller-94}). Let $f:\bb{E}\to \bb{E}'$ be such an isomorphism and consider the automorphism $\Phi'\in{\rm Aut}(\bb{E}')$ defined by $\Phi':=f\circ \Phi\circ f^{-1}$. By construction
$$
\bb{E}'|_x\;:=\;\Big\{\big(x,(\Sigma, {\rm v})\big)\in X\times \bb{T}_m^n\ \big|\ \psi(x)=\Sigma,\ \ {\rm v}\in\Sigma \Big\}\;\simeq\; \Sigma\;.
$$
This shows
that
$\bb{E}'$ is a vector subbundle of the product bundle $X\times\C^n$ and one has the direct sum decomposition $\bb{E}'\oplus \bb{E}''= X\times\C^n$  for some complement vector bundle 
$\bb{E}''\to X$. Let ${\rm Id}_{\bb{E}''}$ be the identity map on $\bb{E}''$ and consider the 
automorphism of $X\times\C^n$ induced by $\Phi'\oplus {\rm Id}_{\bb{E}''}$. This implies the existence of a map $g : X \to \n{U}(n)$ such that 
$$
\Phi'\oplus {\rm Id}_{\bb{E}''}\;:\;(x,{\rm v})\;\longmapsto\; (x, g(x)\cdot {\rm v})\;,\qquad\quad \forall\ (x,{\rm v})\in X\times\C^n\;.
$$
The image of the map $(\psi, g) : X \to G_m(\C^n) \times \n{U}(n)$ is contained
in $\chi_m(\C^n)$ since $\psi(x)=\Sigma\simeq\bb{E}'|_x$ and
$g(x)\big(\bb{E}'|_x\big)=\bb{E}'|_x$ for all $x\in X$.
Hence,  the following map is well-defined:
$$
\begin{aligned}
\varphi\;:\;&\ X&\;\longrightarrow\;&\ \ \chi_m(\C^n)&\\
&\ x&\;\longmapsto\;&\ \ \big(\psi(x),g(x)\big)\;.&
\end{aligned}
$$
By construction
\begin{equation}\label{eq:teo_classif1}
\varphi^*{~^{\chi}\!\bb{T}_m^n}|_x\;:=\;\Big\{\Big(x,(\Sigma,u, {\rm v})\Big)\in X\times {~^{\chi}\!\bb{T}_m^n}\ \Big|\ \psi(x)=\Sigma,\ \ g(x)=u,\ \ u(\Sigma)=\Sigma,\ \ {\rm v}\in\Sigma \Big\}\;\simeq\; \Sigma\;
\end{equation}
which proves the isomorphism
$\varphi^*{~^{\chi}\!\bb{T}_m^n}\simeq\bb{E}'$ 
of complex vector bundles. Moreover,
$$
\varphi^*\Phi_m^n\;:\;\Big(x,(\Sigma,u, {\rm v})\Big)\;\longmapsto\;\Big(x,\Phi_m^n(\Sigma,u, {\rm v})\Big)\;=\;\Big(x,(\Sigma,u, u\cdot{\rm v})\Big)
$$
by definition, hence $\varphi^*\Phi_m^n$ agrees with $\Phi'$ under the isomorphism
\eqref{eq:teo_classif1}. Therefore, $\varphi^*({~^{\chi}\!\bb{T}_m^n},\Phi_m^n)\approx (\bb{E},\Phi)$. To conclude the proof it is enough to recall that $\n{B}_\chi^m$ is defined as
the direct limit of $\chi_m(\C^n)$.
\qed

\medskip

\begin{remark}{\upshape 
The condition of compactness in the statement of Theorem \ref{teo:iso_pullback}, and consequently also in Corollary \ref{cor:iso_pullback}, can be relaxed by requiring only \emph{paracompactness}. In this case  the proof follows along the same lines
apart for the fact that $n$ has to be replaced by $\infty$ as suggested in the analogous argument of \cite[Chapter 3, Theorem 5.5]{husemoller-94}. However, we will not need this kind of generalization in the following. 
}\hfill $\blacktriangleleft$
\end{remark}

\section{Topology of the classifying space}\label{sec:top_class}
In this section we investigate the topology of the classifying space $\n{B}_\chi^m$ by computing its homotopy and cohomology. These results are strongly based on the fact that $\n{B}_\chi^m$ can be identified with the total space of a fibration
\begin{equation}\label{eq:fib_seq_m_body}
\n{U}(m)\;\longrightarrow\;\n{B}_\chi^m\;\stackrel{\pi}{\longrightarrow}\; G_m\big(\C^\infty\big)
\end{equation}
where the fiber projection $\pi={\rm pr}_1\circ \imath$ is defined in  \eqref{eq:class_space33}.
The proof of this technical but important fact is postponed to Appendix \ref{sec:sub_ident}.

\subsection{Homotopy of the classifying space}\label{sec:homot_B}

We start with the case $m = 1$. In this situation we can use the identification 
\begin{equation}\label{eq:B1_chi_prod}
\n{B}_\chi^1\;\simeq\;\C P^\infty\;\times\;\n{U}(1)\;
\end{equation}
proved in Corollary \ref{corol:m=1} in order
 to compute the complete set of  homotopy groups of $\n{B}_\chi^1$.

\begin{proposition}
$$
\pi_k\big(\n{B}_\chi^1\big)\;\simeq\;
\left\{
\begin{aligned}
&\Z&\ \ \ \ &k=1,2\\
&0&\ \ \ \ &k\in\N\cup\{0\}\;,\quad k\neq 1,2\;.
\end{aligned}
\right.
$$
\end{proposition}
\proof
The above identification leads to
$$
\pi_k\big(\n{B}_\chi^1\big)\;\simeq\;\pi_k\big(\C P^\infty\big)\;\oplus\; \pi_k\big(\n{U}(1)\big)\;.
$$
The proof is completed by  $\pi_k\big(\n{U}(1)\big)\simeq \delta_{k,1}\Z$  and  
$\pi_k\big(\C P^\infty\big)\simeq \delta_{k,2}\Z$ \cite[Chapter 4, Example 4.50]{hatcher-02}.
\qed

\medskip

The  general case $m>1$ can be studied by considering the fiber sequence \eqref{eq:fib_seq_m_body}
which induces a long exact sequence of homotopy groups
\begin{equation}\label{eq:esac_seq_homot_m>1}
\ldots\:\pi_{k}\big(\n{U}(m)\big)\;\longrightarrow\;\pi_{k}\big(\n{B}_\chi^m\big)\;\stackrel{\pi_*}{\longrightarrow}\;\pi_{k}\big(G_m(\C^{\infty})\big)\;\longrightarrow\;\pi_{k-1}\big(\n{U}(m)\big)\;\ldots
\end{equation}
\begin{theorem}\label{theo_homot_B^m} For all $m\in\N$ there are isomorphisms of groups
$$
\pi_k\big(\n{B}_\chi^m\big)\;\simeq\;
\left\{
\begin{aligned}
&\pi_{k}\big(\n{U}(m)\big)\;\oplus\; \pi_{k-1}\big(\n{U}(m)\big)&\ \ \ \ &k\in\N\\
&0&\ \ \ \ &k=0\;.
\end{aligned}
\right.
$$
\end{theorem}
\proof
The proof of the fiber sequence \ref{eq:fib_seq_m_body} is based on the identification of $\n{B}_\chi^m$ with the total space ${\rm Ad}\big(\bb{S}_m^\infty\big)$ of an adjoint bundle (\cf Proposition \ref{prop:identif1}).
Sections of the adjoint bundle
$\pi:{\rm Ad}\big(\bb{S}_m^\infty\big)\to G_m\big(\C^\infty\big)$ are in one-to-one correspondence
with automorphisms of the principal $\n{U}(m)$-bundle $\bb{S}_m^\infty\to G_m\big(\C^\infty\big)$ \cite[Chapter 7, Section 1]{husemoller-94}. Therefore, the identity automorphism ${\rm Id}_{\bb{S}_m^\infty}$ identifies a section $\rr{s}: G_m\big(\C^\infty\big)\to {\rm Ad}\big(\bb{S}_m^\infty\big) \simeq\n{B}_\chi^m$. The existence of this section induces a homomorphism
$\rr{s}_*:\pi_{k}(G_m(\C^{\infty}))\to \pi_{k}(\n{B}_\chi^m)$ such that $\pi_*\circ \rr{s}_*={\rm Id}_{\pi_{k}(G_m(\C^{\infty}))}$. It follows that $\pi_*$ is surjective so that the long exact sequence \eqref{eq:esac_seq_homot_m>1} splits into several short exact sequences
\begin{equation}\label{eq:esac_seq_short_homot_m>1}
0\;\longrightarrow\;\pi_{k}\big(\n{U}(m)\big)\;\longrightarrow\;\pi_{k}\big(\n{B}_\chi^m\big)\;\pile{\stackrel{\pi_*}{\longrightarrow}\\\stackbin[\rr{s}_*]{}{\longleftarrow}}\;\pi_{k}\big(G_m(\C^{\infty})\big)\;\longrightarrow\;0
\end{equation}
which are still splitting. Hence, $\pi_{k}(\n{B}_\chi^m)\simeq \pi_{k}(\n{U}(m))\oplus \pi_{k}(G_m(\C^{\infty}))$ and the result follows by a comparison with \eqref{eq:App05}. Finally $\pi_0(\n{B}_\chi^m)=0$ is  equivalent to the connectedness of $\n{B}_\chi^m$.
\qed

\medskip

\noindent
A comparison with the values  in Table {\rm A.1} provides an explicit determination of the low dimensional homotopy groups of $\n{B}_\chi^m$:
$$
\pi_1\big(\n{B}_\chi^m\big)\;\simeq\;\pi_2\big(\n{B}_\chi^m\big)\;\simeq\;\Z\;,\qquad\quad \forall\ m\in\N
$$
and
$$
\pi_3\big(\n{B}_\chi^m\big)\;\simeq\;
\left\{
\begin{aligned}
&0&\ \ \ \ &m=1\\
&\Z&\ \ \ \ &m\geqslant2\;,
\end{aligned}
\right.
\;\qquad\qquad
\pi_4\big(\n{B}_\chi^m\big)\;\simeq\;
\left\{
\begin{aligned}
&0&\ \ \ \ &m=1\\
&\Z_2\oplus\Z&\ \ \ \ &m=2\\
&\Z&\ \ \ \ &m\geqslant3\;.
\end{aligned}
\right.
$$

\medskip

The fundamental group of a path-connected space $X$ acts on its higher homotopy
groups by automorphisms, \ie $\pi_1(X)\ni[\gamma]\mapsto \beta_\gamma\in{\rm Aut}(\pi_k(X))$
(see   \cite[Section 16]{steenrod-51} or \cite[Warning 17.6]{bott-tu-82} for a precise definition).
The following result will be relevant in the sequel.

\begin{proposition}\label{propos:iso_homot_B^m}
The action of $\pi_1(\n{B}_\chi^m)$ on $\pi_k(\n{B}_\chi^m)$ is trivial for all $k\in \N$. In particular one has the isomorphisms
$$
\pi_k\big(\n{B}_\chi^m\big)\;\simeq\; \big[\n{S}^k,\n{B}_\chi^m\big]\;,\qquad\qquad k\in\N
$$
which allow to neglect the role of base points in the computation of the homotopy groups.
\end{proposition}
\proof
{The split exact sequence \eqref{eq:esac_seq_short_homot_m>1} shows that the action of 
$\pi_{1}(\n{B}_\chi^m)\simeq \pi_{1}(\n{U}(m))\oplus 0$ on $\pi_{k}(\n{B}_\chi^m)\simeq \pi_{k}(\n{U}(m))\oplus \pi_{k}(G_m(\C^{\infty}))$ reduces just to the action of $\pi_{1}(\n{U}(m))$ on $\pi_{k}(\n{U}(m))$ which is trivial \cite[Theorem 16.9]{steenrod-51}. The last part of the claim follows from \cite[Proposition 17.6.1]{bott-tu-82}.}
\qed

\subsection{Cohomology of the classifying space}\label{sec:cohom_B}
By construction $\n{B}_\chi^m$ is a subset of $G_m(\C^{\infty})\times\n{U}(\infty)$  with an inclusion map $\imath$ as in \eqref{eq:class_space33}. We also know 
from \eqref{eq:fib_seq_m_body}
that 
$\n{B}_\chi^m$
is  identifiable with the total space of a fiber bundle over $G_m(\C^{\infty})$ with bundle map $\pi$ and fiber $\n{U}(m)$. Moreover, the inclusion  $\imath$ and the projection $\pi$ are compatible as shown by 
the
 following commutative diagram 
\beql{eq:cohom_diag1}
\begin{diagram}
 \n{B}_\chi^m&                       &           \rTo^{\imath}               &                        & G_m(\C^\infty)\times \n{U}(\infty)&& \rTo^{\rm pr_2}&& \n{U}(\infty)      \\
                                                                                     & \rdTo_{\pi} &                             &  \ldTo_{\rm pr_1} & &&&&\\
                                                                                    &                         &   G_m(\C^\infty)    &                         &  &&&&\\
 \end{diagram}\;
\eeq
where ${\rm pr_1}$ and ${\rm pr_2}$ are the first and second component projection, respectively (\cf Appendix \ref{sec:sub_ident}). These facts are 
basic for the proof of the next result.
\begin{theorem}\label{theo:cohom_calss_spac}
 For all $m\in\N$
the cohomology ring of  the classifying space  $\n{B}_\chi^m$
\begin{equation}\label{eq:univ_wind_class_B}
H^\bullet\big(\n{B}_\chi^m,\Z\big)\;\simeq\;\left(\Z\left[\rr{c}^\chi_1,\ldots,\rr{c}^\chi_m\right]\right)\;\otimes_\Z\;\left({\bigwedge}_\Z\;\left[\rr{w}_1^\chi,\ldots,\rr{w}_{m}^\chi\right]\right)\;,\qquad\quad 
\left\{
\begin{aligned}
&\rr{c}^\chi_k\in H^{2k}\big(\n{B}_\chi^m,\Z\big)\\
&\rr{w}_k^\chi\in H^{2k-1}\big(\n{B}_\chi^m,\Z\big)\\
\end{aligned}
\right.
\end{equation}
 is the tensor product of an integer coefficient  polynomial ring in $m$ even-degree free generators $\rr{c}^\chi_k$ and an
  exterior algebra generated by $m$ odd-degree classes $\rr{w}^\chi_k$. The even-degree generators $\rr{c}^\chi_k:=({\rm pr_1}\circ\imath)^*\rr{c}_k$
are   the pullbacks of the universal Chern classes $\rr{c}_k$  which generate the cohomology ring  $H^\bullet(G_m(\C^\infty),\Z)$. Similarly, the odd-degree generators $\rr{w}^\chi_k:=({\rm pr_2}\circ\imath)^*\rr{w}_k$
are    the pullbacks of the universal odd Chern classes $\rr{w}_k$  which generate   $H^\bullet(\n{U}(\infty),\Z)$ (\cf Appendix \ref{app:topU&G}).
 \end{theorem}

\medskip

\noindent
Our proof of Theorem \ref{theo:cohom_calss_spac} requires the application of the \emph{Leray-Serre spectral sequences} associated to the fibrations $\pi$ and ${\rm pr}_1$. For more details on this technique we refer to \cite[Appendix D]{denittis-gomi-15} and references therein.
Since $\imath$ is a fiber bundle map, we have also an induced map $\imath^*$ between the spectral sequences. For the $E_2$-pages of these spectral sequences this map reads
\begin{equation}\label{eq:E_2term}
\begin{diagram}
 E^{p,q}_2({\rm pr}_1)&\;=\;  H^p\Big(G_m(\C^\infty),H^q\big(\n{U}(\infty),\Z\big)\Big)&\;\simeq\; H^p\big(G_m(\C^\infty),\Z\big)\;\otimes_\Z\;H^q\big(\n{U}(\infty),\Z\big)   \\
     \dTo_{\imath^*}&\\
 E^{p,q}_2(\pi)&\;=\; H^p\Big(G_m(\C^\infty),H^q\big(\n{U}(m),\Z\big)\Big)&\;\simeq\;H^p\big(G_m(\C^\infty),\Z\big)\;\otimes_\Z\;H^q\big(\n{U}(m),\Z\big)\;.\\
 \end{diagram}\;
\end{equation}
The last two isomorphisms are a consequence of $\pi_1(G_m(\C^\infty))=0$ 
along with the fact that the groups  $H^q\big(\n{U}(m),\Z\big)$ are free (also for $m=\infty$). This implies that the effect of any local system of coefficients on the cohomology can be factorized with a tensor product (see \eg \cite[Section 5.2]{davis-kirk-01}). 
For proving Theorem \ref{theo:cohom_calss_spac}  we need to anticipate a technical result.
\begin{lemma}\label{lemma:spec_seq1}
For all $p,q\in\N\cup\{0\}$ the map $\imath^*: E^{p,q}_2({\rm pr}_1)\to  E^{p,q}_2(\pi)$ is surjective.
\end{lemma}
\proof
It is enough to show that at each point $\Sigma\in G_m(\C^\infty)$ the fiberwise restriction 
$$
\imath|_\Sigma\;:\; \n{B}^\infty_\chi|_\Sigma\;\simeq\;\n{U}(m)\;\longrightarrow\;\{\Sigma\}\;\times\;\n{U}(\infty)\;\simeq\;\n{U}(\infty) 
$$
induces a surjection $\imath|_\Sigma^*$ from $H^\bullet\big(\n{U}(\infty),\Z\big)$ onto $H^\bullet\big(\n{U}(m),\Z\big)$ (see \eg \cite[Section 21]{fomenko-fuchs-gutenmacher-86}). From its very definition it is clear that $\imath|_\Sigma$ agrees with the standard inclusion $\n{U}(m)\hookrightarrow \n{U}(\infty)$, hence its pullback induces the surjection in
cohomology. The later  is determined in terms of generators  by the relations $\imath|_\Sigma^*(\rr{w}_k)=\rr{w}_k$ if $k=1,\ldots,m$ and 
$\imath|_\Sigma^*(\rr{w}_k)=0$ for $k>m$ (\cf Appendix \ref{app:topU&G}).
\qed

\medskip
\medskip

\proof[Proof of Theorem \ref{theo:cohom_calss_spac}]
Since ${\rm pr}_1:G_m(\C^\infty)\times \n{U}(\infty)\to G_m(\C^\infty)$ is a product bundle the associated spectral sequence degenerates at the $E_2$-page, \ie all the differentials $\delta_{r}$ are trivial for $r\geqslant2$. This implies that $E^{p,q}_2({\rm pr}_1)=E^{p,q}_3({\rm pr}_1)=\ldots=E^{p,q}_\infty({\rm pr}_1)$. By Lemma \ref{lemma:spec_seq1} the maps $\imath^*:E^{p,q}_2({\rm pr}_1)\to E^{p,q}_2(\pi)$ are surjective. In particular all the non-trivial generators $\rr{c}_k\otimes\rr{w}_j\in E^{2k,2j-1}_2(\pi)$ come from the related generators $\rr{c}_k\otimes\rr{w}_j$ in $E^{2k,2j-1}_2({\rm pr}_1)$. Hence, the triviality of the differentials $\delta_r: E^{p,q}_r({\rm pr}_1)\to E^{p+r,q-r+1}_r({\rm pr}_1)$  implies the triviality of the differentials $\delta_r: E^{p,q}_r(\pi)\to E^{p+r,q-r+1}_r(\pi)$. Because of that, one gets $E^{p,q}_2(\pi)= E^{p,q}_\infty(\pi)$. In order to recover the cohomology of $\n{B}_\chi^m$ from $E^{p,q}_\infty(\pi)$ one has to solve the extension problems. However, in the present case 
 all the abelian
groups $E^{p,q}_\infty(\pi)$ are free, so that all the extensions are split.
As a result, one has the following isomorphism of abelian groups:
$$
H^k\big(\n{B}_\chi^m,\Z\big)\;\simeq\;\bigoplus_{p+q=k}E^{p,q}_\infty(\pi)\;\simeq\;H^k\Big(G_m(\C^\infty)\times\n{U}(m),\Z\Big) \;.
$$
where the last isomorphism is a consequence of the K\"{u}nneth formula for cohomology.
This isomorphism gives rise to a ring  isomorphism. Indeed, as a consequence of Lemma \ref{lemma:spec_seq1}  we can infer that the ring map
$$
\imath^*\;:\;H^\bullet\Big(G_m(\C^\infty)\times\n{U}(\infty),\Z\Big)\;\longrightarrow \;H^\bullet\big(\n{B}_\chi^m,\Z\big)
$$
is surjective and acts on the generators of $H^\bullet\big(G_m(\C^\infty)\times\n{U}(\infty),\Z\big)$ as follows: $\imath^*(\rr{c}_k)=\rr{c}_k$  and $\imath^\ast(\rr{w}_k)=\rr{w}_k$ for all $k=1,\ldots,m$   and 
$\imath^*(\rr{w}_k)=0$ for $k>m$.\qed

\medskip

\begin{remark}[The winding number]\label{rk:wind_num}{\upshape 
Let us give a closer look at the generator $\rr{w}_1^\chi$ of $H^1(\n{B}_\chi^m,\Z)\simeq\Z$.
We  recall that the  first integral cohomology group of a space $X$ has the realization $H^1(X,\Z)\simeq[X,\n{U}(1)]$. This fact is discussed in \cite[Section 3.1, Exercise 13]{hatcher-02} but the reader can also prove it
by applying the technique described in Section \ref{sec:postmikov}. Due to Theorem \ref{theo:cohom_calss_spac} we know that 
the generator $\rr{w}_1^\chi$ is the pullback with respect to the map ${\rm pr_2}\circ\imath$ of the first universal odd Chern classes $\rr{w}_1$
which generates   $H^1(\n{U}(\infty),\Z)$ (\cf Appendix \ref{app:topU&G}). This generator is identifiable with the class $[{\rm det}]\in[\n{U}(\infty),\n{U}(1)]$ associated to the \virg{limit} determinant map ${\rm det}:\n{U}(\infty)\to\n{U}(1)$. Thus the composition
$$
\begin{aligned}
{\rm det}\circ{\rm pr_2}\circ\imath\;:\;&\ \ \n{B}_\chi^m&\;\longrightarrow\;&\ \ \n{U}(1)&\\
&(\Sigma,u)&\;\longmapsto\;&\ \ {\rm det}(u)&
\end{aligned}
$$
provides a representative for the generator $[{\rm det}\circ{\rm pr_2}\circ\imath]\in [\n{B}_\chi^m,\n{U}(1)]\simeq H^1(\n{B}_\chi^m,\Z)$. Finally, the \emph{winding number} of this map provides the identification with $\Z$.	
}\hfill $\blacktriangleleft$
\end{remark}

%---------------------------------------------%

\section{Classification of chiral vector bundles}
In this section we will combine the knowledge of the topology of the classifying space $\n{B}_\chi^m$ developed in Section \ref{sec:top_class} with the criterion of the
homotopy classification  proved in Corollary \ref{cor:iso_pullback} in order to define the appropriate family of \emph{characteristic classes}
that classify  chiral vector bundles. In effect, we will show that these classes provide a suitable classification scheme in the case that the base space $X$ is a CW-complex of low dimension $d\leqslant 4$. The special  cases of $X=\n{S}^d$ and   $X=\n{T}^d$ will be discussed in detail.

\subsection{Characteristic classes}\label{sec:charact_class}
Usually one uses the  cohomology of the classifying space of a given category of vector (or principal) bundles to define the associated characteristic classes by pullback with respect to the classifying maps. We apply the same strategy here to define the
characteristic classes for chiral vector bundles.
\begin{definition}[Chiral characteristic classes]\label{def:chi_class}
Let $(\bb{E},\Phi)$ be a rank $m$ chiral vector bundle over $X$ and $\varphi:X\to \n{B}_\chi^m$ a representative for the  classifying map
which classifies $[(\bb{E},\Phi)]$.
We define the $k$-th  \emph{even chiral} class of $(\bb{E},\Phi)$ to be
$$
c_k(\bb{E},\Phi)\;:=\;\varphi^*\rr{c}^\chi_k\;\in\; H^{2k}\big(X,\Z\big)\;,\qquad\quad k=1,\ldots,m\;.
$$
In much the same way, we refer to
$$
w_k(\bb{E},\Phi)\;:=\;\varphi^*\rr{w}^\chi_k\;\in\; H^{2k-1}\big(X,\Z\big)\;,\qquad\quad k=1,\ldots,m\;.
$$
as the  $k$-th  \emph{odd chiral} class of $(\bb{E},\Phi)$.
\end{definition}

\medskip

\begin{proposition}\label{propos:Equiv_chern}
The \emph{even chiral} classes $c_k(\bb{E},\Phi)$ of the chiral vector bundle $(\bb{E},\Phi)$ coincide with the Chern classes $c_k(\bb{E})$ of the underlying complex vector bundle $\bb{E}$.
\end{proposition}
\proof
As is shown in the proof of Theorem \ref{teo:iso_pullback}, if the  chiral vector bundle $(\bb{E}, \Phi)$ is classified by a map $\varphi : X \to \n{B}_\chi^m$, then the map $\psi : X\to G_m(\C^n)$ (for some $n$ big enough), given by $\psi:=({\rm pr}_1 \circ \iota)\circ\varphi$
classifies $\bb{E}$ as complex vector bundle.
Now, by the very definition of the chiral  class, it follows that
$c_k(E, \Phi) = \varphi^*\rr{c}_k^\chi = \psi^*\rr{c}_k = c_k(E)$.
\qed

\medskip

\begin{remark}[Simplified nomenclature]{\upshape 
We can take advantage from the previous proposition to simplify the nomenclature concerning the characteristic classes for chiral vector bundles.
Indeed, the equality $c_k(\bb{E},\Phi)=c_k(\bb{E})$ shows that 
we can properly refer to the {even} chiral classes simply as Chern classes. In this way   we can reserve the name \emph{chiral classes}
only for the {odd}  classes $w_k(\bb{E},\Phi)$.
This choice of names presents the advantage to make clear that the extra topological information induced on the vector bundle $\bb{E}$ by a chiral structure $\Phi$ is totally encoded in the odd classes $w_k(\bb{E},\Phi)$.
}\hfill $\blacktriangleleft$
\end{remark}

\medskip

{
\begin{remark}[Differential geometric representation]\label{rk_diff_aspect}{\upshape 
In this work we are interested in a topological classification of chiral vector bundles valid for topological base spaces with a CW-complex
structure. However, when the base space is a differential manifold (with no torsion) one can argue from Definition \ref{def:chi_class} (and Proposition \ref{propos:Equiv_chern})
that the Chiral characteristic classes can be represented by differential forms by an application of the  Chern-Weil theory. More precisely   
the even classes  $c_k(\bb{E},\Phi)$ can be represented by the standard Chern-Weil  forms $c_k^{\rm CW}(\Omega)$ of order $2k$ builded from any curvature $\Omega$ by means of the standard formula
$$
{\rm det}\left(\frac{\ii t}{2\pi}\Omega\;+\;\n{1}\right)\;=\;\sum_{k}c_k^{\rm CW}(\Omega)\; t^k\;.
$$
In much of the same way, the odd classes $w_k(\bb{E},\Phi)$ can be represented 
 by the differential forms 
 $$
 \omega_{2k-1}^\Phi\;:=\;
 \frac{1}{(\ii 2\pi)^k}\frac{((k-1)!)^2}{(2k-1)!}{\rm Tr}\left[\left(\Phi^{-1}\nabla\Phi\right)^{\wedge 2k-1}\right]
 $$
where $\nabla\Phi$  is the
covariant derivative of the automorphism $\Phi$ with respect to
a connection on $\bb{E}$. This last point deserves a deeper analysis.
}\hfill $\blacktriangleleft$
\end{remark}
}

\medskip

The chiral classes posses a kind of additive behavior induced by the composition of automorphisms.
\begin{proposition}\label{prop:group_wind}
Let $(\bb{E},\Phi_1)$, $(\bb{E},\Phi_2)$ and $(\bb{E},\Phi_1\circ \Phi_2)$ be three chiral vector bundles which share the same  underlying
complex vector bundle $\bb{E}\to X$. Then, the related chiral classes obey
\begin{equation}\label{ea:group_w_1}
w_k(\bb{E},\Phi_1\circ \Phi_2)\;=\;w_k(\bb{E},\Phi_1)\;+\;w_k(\bb{E},\Phi_2)\;.
\end{equation}
The above relations are completed by
\begin{equation}\label{ea:group_w_2}
w_k(\bb{E},{\rm Id}_{\bb{E}})\;=\;0\;.
\end{equation}
\end{proposition}
\proof
Let $\varphi_1,\varphi_2,\varphi:X\to\n{B}_\chi^m$ be the classifying maps for $(\bb{E},\Phi_1)$, $(\bb{E},\Phi_2)$ and $(\bb{E},\Phi_1\circ \Phi_2)$ respectively. From the very definition of the classifying maps one can verify that $({\rm pr}_2 \circ \iota)\circ\varphi:X\to\n{U}(\infty)$ agrees with the pointwise product of the maps $({\rm pr}_2 \circ \iota)\circ\varphi_i:X\to\n{U}(\infty)$, $i=1,2$.
Therefore, one has
$$
\begin{aligned}
w_k(\bb{E},\Phi_1\circ \Phi_2)\;&=\;\varphi^*\circ({\rm pr}_2 \circ \iota)^*\rr{w}_k\;=\; (\varphi_1\varphi_2)^*\circ({\rm pr}_2 \circ \iota)^*\rr{w}_k\\
&=\;\varphi_1^*\circ({\rm pr}_2 \circ \iota)^*\rr{w}_k\;+\; \varphi_2^*\circ({\rm pr}_2 \circ \iota)^*\rr{w}_k\;=\;w_k(\bb{E},\Phi_1)\;+\;w_k(\bb{E},\Phi_2)\;.
\end{aligned}
$$
Finally, we can classify $(\bb{E},{\rm Id}_{\bb{E}})$ by a map $\varphi:X\to\n{B}_\chi^m$ such that $({\rm pr}_2 \circ \iota)\circ\varphi=\n{1}$. This immediately leads to \eqref{ea:group_w_2}. 
\qed

\subsection{The Picard group of chiral line bundles}
Given two rank $m$ chiral vector bundles $(\bb{E}_1,\Phi_1)$, $(\bb{E}_2,\Phi_2)$ over $X$ we can define their tensor product $(\bb{E}_1,\Phi_1)\otimes(\bb{E}_2,\Phi_2):=(\bb{E}_1\otimes \bb{E}_2,\Phi_1\otimes\Phi_2)$ as the rank $m^2$ chiral vector bundle with 
underlying
complex vector bundle $\pi:\bb{E}_1\otimes \bb{E}_2\to X$ given by the tensor product of $\pi_1:\bb{E}_1\to X$ and $\pi_2:\bb{E}_2\to X$ (in the sense of \cite[Chapter 6, Section 6]{husemoller-94}) and automorphism $\Phi_1\otimes\Phi_2\in{\rm Aut}(\bb{E}_1\otimes \bb{E}_2)$ defined by
$\Phi_1\otimes\Phi_2(p_1\otimes p_2):=\Phi_1(p_1)\otimes\Phi_2(p_2)$ for all $p_1\in \bb{E}_1$ and $p_2\in\bb{E}_2$ such that $\pi_1(p_1)=\pi_2(p_2)$. Just by considering tensor product homotopies of chiral vector bundles one can easily show  that the operation $\otimes$ is compatible with the notion of equivalence in Definition \ref{def:eq_rel_chi_VB}. In particular, one has that
$$
\otimes\;:\;{{\rm Vec}}_\chi^1(X)\;\times\; {{\rm Vec}}_\chi^1(X)\;\longrightarrow\;{{\rm Vec}}_\chi^1(X) 
$$
endows ${{\rm Vec}}_\chi^1(X)$  with an abelian group structure.
According to a standard terminology, we refer to  ${{\rm Vec}}_\chi^1(X)$ as the \emph{chiral Picard group}.
\begin{proposition}[Classification of chiral line bundles]\label{propos:chi_lin_bun}
There is a group isomorphism
$$
(w_1,c_1)\;:\;{{\rm Vec}}_\chi^1(X)\;\longrightarrow\;H^1(X,\Z)\;\oplus\;H^2(X,\Z)
$$
induced by the first chiral class $w_1$ and the first Chern class $c_1$.
\end{proposition}
\proof
As a consequence of the identification \eqref{eq:B1_chi_prod}  one has
$$
{{\rm Vec}}_\chi^1(X)\;\simeq\;[X, \n{B}_\chi^1]\;\simeq\;[X,\n{U}(1)]\;\times\;[X,\C P^\infty]\;.
$$
By combining Remark \ref{rk:wind_num} and Proposition \ref{prop:group_wind} one can show that $w_1$ sets the group isomorphism $[X,\n{U}(1)]\simeq H^1(X,\Z)$.  The proof is completed by recalling the well-known
group isomorphism
$[X,\C P^\infty]\simeq {{\rm Vec}}_\C^1(X)\simeq H^2(X,\Z)$ induced by 
$c_1$ (\cf  \cite[Section 3.2]{denittis-gomi-14} and references therein).
\qed

\medskip

When we apply the above result to spheres of dimension $d$ we get the following isomorphisms
\begin{equation}\label{eq:class_spher_m=1}
\begin{aligned}
w_1\;:\;&{{\rm Vec}}_\chi^1\big(\n{S}^1\big)\;\longrightarrow\;H^1\big(\n{S}^1,\Z\big)\;\simeq\;\Z\\
c_1\;:\;&{{\rm Vec}}_\chi^1\big(\n{S}^2\big)\;\longrightarrow\;H^2\big(\n{S}^2,\Z\big)\;\simeq\;\Z\\
&{{\rm Vec}}_\chi^1\big(\n{S}^d\big)\;\simeq\;0\qquad\text{if}\quad d\geqslant 3\;.
\end{aligned}
\end{equation}
Instead, for tori of dimension $d$ one has the isomorphisms
$$
(w_1,c_1)\;:\;{{\rm Vec}}_\chi^1\big(\n{T}^d\big)\;\longrightarrow\;\Z^d\;\oplus\;\Z^{\frac{1}{2}d(d-1)}\;.
$$
where we used $H^k(\T^d,\Z)\simeq\Z^{{d \choose k}}$.

%-----%
\subsection{Chiral vector bundles over spheres}\label{sect:class_spher}
The classification of chiral vector bundles over spheres (\cf Theorem \ref{theo:class_spher}) is a direct consequence of the homotopy classification described in Corollary \ref{cor:iso_pullback} combined with Proposition \ref{propos:iso_homot_B^m} and Theorem \ref{theo_homot_B^m}. 

\medskip

In the case of  low-dimensional spheres the classification can be described in terms of characteristic classes by the following bijections
\begin{equation}\label{eq:class_spher_m>1}
\begin{aligned}
w_1\;:\;&{{\rm Vec}}_\chi^m\big(\n{S}^1\big)\;\longrightarrow\;H^1\big(\n{S}^1,\Z\big)\;\simeq\;\Z\\
c_1\;:\;&{{\rm Vec}}_\chi^m\big(\n{S}^2\big)\;\longrightarrow\;H^2\big(\n{S}^2,\Z\big)\;\simeq\;\Z\\
w_2\;:\;&{{\rm Vec}}_\chi^m\big(\n{S}^3\big)\;\longrightarrow\;H^3\big(\n{S}^3,\Z\big)\;\simeq\;\Z\qquad\quad (m>1)\;.\\
\end{aligned}
\end{equation}
When $d=1$ one has $\pi_1(\n{B}_\chi^m)\simeq \pi_1(\n{U}(m))\simeq\pi_1(\n{U}(1))\simeq H^1\big(\n{S}^1,\Z\big)$ where the last isomorphism has been discussed in Remark \ref{rk:wind_num}. Similarly, for $d=3$ one has  $\pi_3(\n{B}_\chi^m)\simeq \pi_3(\n{U}(m))\simeq\pi_3({\rm S}\n{U}(2))\simeq\pi_3(\n{S}^3)\simeq H^3\big(\n{S}^3,\Z\big)$  where we used the standard identification between ${\rm S}\n{U}(2)$ and $\n{S}^3$
and the last isomorphism can be understood in terms of the Brouwer's degree \cite[Remark 5.8]{denittis-gomi-14}. In the even case $d=2$ one has 
$\pi_2(\n{B}_\chi^m)\simeq \pi_2(G_m(\C^\infty))\simeq[\n{S}^2,G_m(\C^\infty)]\simeq {{\rm Vec}}_\C^m(\n{S}^2)\simeq H^2(\n{S}^2,\Z)$ where the last isomorphism is given, as usual, by the first Chern class $c_1$.

\medskip

The case $d=4$ is more involved and hence interesting. A direct computation shows that
\begin{equation}\label{eq:class_spher_m>1_d=4}
{{\rm Vec}}_\chi^m\big(\n{S}^4\big)\;\simeq\; \pi_{4}\big(\n{U}(m)\big)\;\oplus\; \pi_{4}\big(G_m(\C^{\infty})\big)\;\simeq\;
\left\{
\begin{aligned}
\Z_2\oplus\Z&\qquad&\text{if}\ \ m=2\\
\Z&\qquad&\text{if}\ \ m\geqslant 3
\end{aligned}
\right.
\end{equation}
where the $\Z$ summand is given by $\pi_{4}(G_m(\C^{\infty}))\simeq[\n{S}^4,G_m(\C^{\infty}]\simeq{{\rm Vec}}_\C^m(\n{S}^4)\simeq H^4(\n{S}^4,\Z\big)\simeq\Z$ (if $m>1$) and is described by the second Chern class $c_2$.
The $\Z_2$ summand is given by the unstable group $\pi_{4}(\n{U}(2))\simeq \pi_{4}({\rm S}\n{U}(2))\simeq\Z_2$ (\cf Table A.1) which is non-trivial only when $m=2$.
This
torsion summand  is not accessible by the \virg{primary} characteristic classes. One way to understand this invariant is to look at the map $f:\n{S}^4\to \n{S}^3$ 
which generates $\pi_{4}(\n{S}^3)\simeq \pi_{4}\big({\rm S}\n{U}(2)\big)$. This map agrees with the (reduced or unreduced) suspension of the Hopf's map $h:\n{S}^3\to \n{S}^2$ which generates $\pi_3(\n{S}^2)$.
An explicit realization of $f$ is given by
$$
f(k_0,k_1,k_2,k_3,k_4)\;:=\; \frac{2}{1+k_0^2}\left(k_0,k_1k_3-k_2k_4,k_1k_4+k_2k_3,\frac{k_1^2+k_2^2-k_3^2-k_4^2}{2}\right)
$$
for all $(k_0,k_1,k_2,k_3,k_4)\in\R^5$ which fulfill  the constraint $\sum_{i=0}^4k^2_i=1$. One can check that when $k_0=0$ the map $f$  restricts  exactly to the Hopf's map. A differential geometric approach to the study of $\pi_{4}\big({\rm S}\n{U}(2)\big)\simeq\Z_2$ is discussed in the final part of \cite{witten-83} (see also \cite[Section 1.4]{kori-14}). Given a map $f:\n{S}^4\to {\rm S}\n{U}(2)$ and the standard inclusion $\jmath:{\rm S}\n{U}(2)\hookrightarrow
{\rm S}\n{U}(3)$ 
one can consider the composition $\jmath\circ f:\n{S}^4\to {\rm S}\n{U}(3)$. Since $\pi_4({\rm S}\n{U}(3))=0$ there exists a map $F:\n{D}^5\to {\rm S}\n{U}(3)$ defined on the unit ball $\n{D}^5\subset\R^5$ such that $F|_{\partial \n{D}^5}= \jmath\circ f$. In \cite{witten-83} E. Witten showed that the integral (a pure gauge Chern-Simons form)
$$
CS_5(f)\;:=\;\frac{\ii}{240\pi^3}\int_{\n{D}^5}
{\rm Tr}_{\C^3}\left[\left(F^{-1}\dd F\right)^{\wedge 5}\right]
$$
depends only on the homotopy class $[f]\in \pi_4({\rm S}\n{U}(2))$ modulo $\Z$. Moreover,  the homotopy invariant $\epsilon(f):=\expo{\ii 2\pi CS_5(f)}\in\{\pm1\}$ takes the value $-1$ when $f$ is a representative for the non-trivial element of $\pi_4({\rm S}\n{U}(2))$. See also  
\cite[Proposition 1.11]{kori-14} for more details.

%-----%
\subsection{Chiral vector bundles in low dimension}
\label{sec:postmikov}
We proved in Section \ref{sec:homot_B} that  $\pi_0(\n{B}^m_\chi)=0$ and $\pi_1(\n{B}^m_\chi)\simeq\Z$ acts trivially on each $\pi_k(\n{B}^m_\chi)$, $k\geqslant1$. Moreover, $\n{B}^m_\chi$ has a  CW-complex structure. All these facts imply that 
$\n{B}^m_\chi$ is a \emph{simple} space, a fact which assures that $\n{B}^m_\chi$ admits a \emph{Postnikov tower} of principal fibrations \cite[Theorem 4.69]{hatcher-02} (see also \cite[Section 7]{arlettaz-00} or  \cite[Chapter 7]{arkowitz-11}). More precisely this means that there exists a sequence of spaces $\bb{B}_j^m$ along with maps $\alpha_{j}:\n{B}^m_\chi\to \bb{B}_j^m$ and $p_{j+1}:\bb{B}_{j+1}^m\to \bb{B}_j^m$ such that the diagram
\beql{eq:cohom_diag_postnikov}
\begin{diagram}
           &      &K(\pi_{j+2},j+3)  &        &      K(\pi_{j+1},j+2)    &       &  K(\pi_{j},j+1)&      &         &      &  K(\pi_{2},3)     \\
           &      &\uTo^{\kappa^{j+3}}                    &        &      \uTo^{\kappa^{j+2}}                       &       &  \uTo^{\kappa^{j+1}}                &     &         &      & \uTo^{\kappa^{3}}      \\
  %----
  \ldots&\rTo^{p_{j+2}}&\bb{B}^m_{j+1}      & \rTo^{p_{j+1}} &       \bb{B}^m_j               &\rTo^{p_{j}} &  \bb{B}^m_{j-1}&\rTo^{p_{j-1}}& \ldots&\rTo^{p_{2}}& \bb{B}_{1}^m&\hspace{-2mm}=K(\pi_{1},1)     \\
          &      &                       & \luTo_{\alpha_{j+1}} &               \uTo ^{\alpha_{j}}             &  \ruTo^{\alpha_{j-1}} & &&& \ruTo^{\alpha_{1}}(6,2)&\\
 &&                                                                                   &                         &   \n{B}^m_\chi    &                         &  &&&&\\
 \end{diagram}\;
\eeq
 is commutative, \ie  $p_{j+1}\circ \alpha_{j+1}=\alpha_j$ for all $j\geqslant 1$. The maps $p_{j}$ and $\kappa^{j+1}$ define (principal)  fibration sequences
\begin{equation}\label{eq:Postnikov1}
K(\pi_{j},j)\;\longrightarrow\; \bb{B}_{j}^m\;\stackrel{p_{j}}{\longrightarrow}\; \bb{B}_{j-1}^m\;\stackrel{\kappa^{j+1}}{\longrightarrow}\; K(\pi_{j},j+1)
\end{equation}
where the symbols $K(\pi_{j},n)$ denote
  the   Eilenberg-MacLane spaces  associated to $\n{B}^m_\chi$. 
In particular $\bb{B}_{j}^m$, usually called $j$-th  \emph{Postnikov section}, turns out to be (up to weak homotopy equivalence) the homotopy fiber of the map 
$\kappa^{j+1}$.
 We recall that  $K(\pi_{j},n)$
 is a connected  space  (with a uniquely specified homotopy type) defined by the following property \cite[Section 4.2]{hatcher-02}
$$
\pi_{k}\big(K(\pi_{j},n)\big)\;\simeq\;\left\{
\begin{aligned}
&\pi_{j}(\n{B}^m_\chi)&\qquad&\text{if}\ k= n\\
&0&\qquad&\text{if}\ k\neq n\;.
\end{aligned}
\right.
$$
Moreover, one has  the isomorphisms \cite[Theorem 4.57]{hatcher-02}
\begin{equation}\label{eq:cohom_EML_space}
[X, K(\pi_{j},n)]\;\simeq\;H^n\big(X, \pi_j(\n{B}^m_\chi)\big)
\end{equation}
given by $[f]\mapsto f^*(\xi)$ with $\xi\in H^n(K(\pi_{j},n),\pi_j(\n{B}^m_\chi))$ being the \emph{basic} or \emph{fundamental}  class \cite[Definition 5.3.1]{arkowitz-11}.
In our particular case we can choose $K(\pi_{1},1)=\n{S}^1$.
The map $\alpha_{j}:\n{B}^m_\chi\to \bb{B}_j$ is a \emph{$(j+1)$-equivalence} in the sense that it  induces isomorphisms $\pi_k(\n{B}^m_\chi)\simeq\pi_k(\bb{B}_j^m)$ for all $k\leqslant j$ and one has that 
$\pi_k(\bb{B}_j^m)=0$ for all $k>j$.  Said differently, the \virg{auxiliary} spaces $\bb{B}_j^m$ (which still have the homotopy type of  CW-complexes) approximate the system of homotopy groups of $\n{B}^m_\chi$ up to the level $j$.
The collection of maps $\alpha_j$ induces a weak homotopy equivalence between $\n{B}^m_\chi$ and the inverse limit generated by the {auxiliary} spaces $\bb{B}^m_j$ (the inverse limit generally will not have the homotopy type of a CW-complex). The \virg{$\kappa$-invariants} $\kappa^{j+1}\in H^{j+1}(\bb{B}^m_{j-1}, \pi_{j}(\n{B}^m_\chi))$
 are to be regarded as cohomology classes.
These classes together with the homotopy groups $\pi_k(\n{B}^m_\chi)$ specify the weak homotopy
type of $\n{B}^m_\chi$. In  particular $\kappa^{j+1}=0$ as cohomological class means that $\kappa^{j+1}$ is homotopic to the constant map in the  fibration sequences
\eqref{eq:Postnikov1} and, in this case, one has that the induced fibration $p_j$ is trivial, \ie $\bb{B}_{j}^m\simeq\bb{B}_{j-1}^m\times K(\pi_{j},j)$ (for more details see \cite[Lemma 7.3]{arlettaz-00}).

\medskip

The importance of the Postnikov tower for the classification of chiral vector bundles  lies in the  following general result.
\begin{proposition}\label{prop:postnikov}
Let $X$ be as in Assumption \ref{ass:1} and assume that the maximal dimension of its cells is $d$. Then
$$
{{\rm Vec}}_\chi^m\big(X\big)\;\simeq\; \big[X,\bb{B}^m_d\big]\;.
$$ 
\end{proposition}
\proof
The proof of the claim is just a combination of the homotopy classification for chiral vector bundles provided by Corollary \ref{cor:iso_pullback} with the classical result \cite[Lemma 4.1]{james-thomas-65}.
\qed

\medskip

The concrete utility of Proposition \ref{prop:postnikov} is related to the ability to compute the {auxiliary} spaces $\bb{B}_j^m$, a problem which usually is of difficult solution.  However, if one restricts the interest to low dimensions a rigorous computation becomes reasonably doable.
\begin{proposition}[Classification in low dimension]\label{propos:class_low_dim}
Let $X$ be as in Assumption \ref{ass:1} and assume that the maximal dimension of its cells is $d$. 
\begin{enumerate}
\item[(i)] If $1\leqslant d\leqslant 3$, there are bijections of sets 
\begin{equation}\label{eq:class_low_dim}
{{\rm Vec}}_\chi^m\big(X\big)\;\simeq\;\bigoplus_{j=1}^d H^j(X,\Z)\;,\qquad\quad m\geqslant2
\end{equation}
 induced by the characteristic classes $w_1,c_1,w_2$ up to the suitable dimension.
\vspace{1.3mm}
\item[(ii)] If $d=4$, there are bijections of sets  
\begin{equation}\label{eq:class_low_dim4}
{{\rm Vec}}_\chi^m\big(X\big)\;\simeq\;\bigoplus_{j=1}^4 H^j(X,\Z)\;,\qquad\quad m\geqslant3
\end{equation}
 induced by the characteristic classes $w_1,c_1,w_2,c_2$.
\end{enumerate}
\end{proposition}
\proof
We know that $\bb{B}^m_1=K(\pi_{1},1)=\n{S}^1$ and from Proposition \ref{prop:postnikov} and the isomorphism $[X,\n{S}^1]\simeq H^1(X,\Z)$
(see eq. \eqref{eq:cohom_EML_space} or Remark \ref{rk:wind_num})
we can conclude the proof of \eqref{eq:class_low_dim}  for the case $d=1$. In particular, as a consequence of  Lemma \ref{lemma:cohomU_inf} and Remark \ref{rk:postnikov_sect_grass}, we know that the basic class of 
 $H^1(\n{S}^1,\Z)$, which determines the isomorphism  \eqref{eq:cohom_EML_space}, can be identified  with the generator $\rr{w}_1$ of $H^\bullet(\n{U}(\infty),\Z)$
 and consequently with the first chiral generator
 $\rr{w}_1^\chi$ of $H^\bullet(\n{B}_\chi^m,\Z)$ according to Theorem \ref{theo:cohom_calss_spac}.

\medskip

The study of the  case $d=2$ in Proposition \ref{prop:postnikov} needs the computation of the Postnikov  section $\bb{B}_{2}^m$. Equation \eqref{eq:Postnikov1} says that 
$\bb{B}_{2}^m$ is the total space of a principal fibration over  $\bb{B}^m_1=\n{S}^1$ with fiber $K(\pi_{2},2)$. Since $\pi_2(\n{B}_\chi^m)\simeq\Z$
we can use the identification $K(\Z,2)=\C P^\infty$. Moreover, 
$H^3(\bb{B}_{1}^m, \pi_3(\n{B}_\chi^m))=H^3(\n{S}^1,\Z)=0$ implies the vanishing of the $\kappa$-invariant $\kappa^{3}=0$.
This assures the triviality of the fibration \eqref{eq:Postnikov1}, namely $\bb{B}_{2}^m=\n{S}^1\times \C P^\infty$. At this point Proposition \ref{prop:postnikov} provides
$$
{{\rm Vec}}_\chi^m\big(X\big)\;\simeq\;[X,\n{S}^1]\;\times\;[X,\C P^\infty]\;\simeq\; H^1(X,\Z)\;\oplus\;H^2(X,\Z)\;.
$$
and the basic 
class in $ H^2(K(\Z,2),\Z)$ which induces the
isomorphism $[X,K(\Z,2)]\simeq H^2(X,\Z)$ as in \eqref{eq:cohom_EML_space} can be identified with the first universal Chern class
$\rr{c}_1\in H^2(G_m(\C^{\infty}),\Z)$ as discussed in
Remark \ref{rk:postnikov_sect_grassA}.

\medskip

The case $d=3$ is more involved. The Postnikov  section  $\bb{B}_{3}^m$ is the total space of the fibration 
\begin{equation}\label{eq:PostnikovB-3}
K(\Z,3)\;\longrightarrow\; \bb{B}_{3}^m\;\stackrel{p_{3}}{\longrightarrow}\; \bb{B}_{2}^m\;=\;\n{S}^1\;\times\; \C P^\infty
\end{equation}
where we used $\pi_3(\n{B}_\chi^m)\simeq\Z$ when $m\geqslant 2$. This section is determined by the 
invariant $\kappa^4\in H^4(\n{S}^1\times \C P^\infty,\Z)\simeq H^4( \C P^\infty,\Z)\simeq \Z$ and we want to prove that $\kappa^4=0$. 
From one hand we know that
the map $\alpha_3:\n{B}_\chi^m\to  \bb{B}_{3}^m$ is a $4$-equivalence so that $H^j(\n{B}_\chi^m,\Z)\simeq H^j(\bb{B}_{3}^m,\Z)$ for all $j\leqslant 3$ 
and $\alpha_3^*:H^4(\bb{B}_{3}^m,\Z)\to H^4(\n{B}_\chi^m,\Z)$ is injective (\cf Note \ref{note_A1} in Remark \ref{rk:postnikov_sect_grass}).
Our knowledge of the cohomology of $\n{B}_\chi^m$ implies that 
$$
H^1\big(\bb{B}_{3}^m,\Z\big)\;\simeq\; H^2\big(\bb{B}_{3}^m,\Z\big)\;\simeq\;\Z\;,\qquad\qquad
H^3\big(\bb{B}_{3}^m,\Z\big)\;\simeq\;\Z^2
$$
and $H^4\big(\bb{B}_{3}^m,\Z\big)$ is a subgroup of $\Z^3$ with no torsion.
On the other hand we can compute this cohomology by means of the Leray-Serre spectral
sequence associated with the fibration \eqref{eq:PostnikovB-3} (we refer to \cite[Appendix D]{denittis-gomi-15} and references therein for a summary about spectral sequences). The 2-page of this spectral sequence   is given by
\begin{equation}\label{eq:spec_seq_a1*}
E_2^{p,q}\;=\;H^p\Big(\bb{B}_{2}^m, H^q\big(K(\Z,3),\Z\big )\Big)\;.
\end{equation}
As a consequence of \cite[Theorem 16.9]{steenrod-51} one has that the  group $\pi_1(\bb{B}_{2}^m)\simeq\pi_1(\n{S}^1)\simeq\Z$ acts trivially on 
higher homotopy groups 
$$
\pi_k\big(\bb{B}_{2}^m\big)\;\simeq\; \pi_k\big(\n{S}^1\big)\;\times\; \pi_k\big(K(\Z,2)\big)\;,\qquad\quad k\geqslant 1\;.
$$ 
This fact implies  that the system of coefficients in \eqref{eq:spec_seq_a1*} is constant and not local.
In particular, by using that $\bb{B}_{2}^m$ is path-connected one obtains from \eqref{eq:spec_seq_a1*} the following isomorphisms
\begin{equation}\label{eq:spec_seq_a1}
E_2^{0,q}\;\simeq\; H^q\big(K(\Z,3),\Z\big )\;.
\end{equation}
The explicit knowledge of the cohomology of 
$K(\Z,3)$ (see \eg \cite[Section 18]{bott-tu-82}) 
 \begin{table}[h]\label{tab:K=3}
 \begin{tabular}{c||c|c|c|c|c|c|c|c|c|c|}
 \rule[-3mm]{0mm}{9mm}
 & $k=0$ & $k=1$ & $k=2$ & $k=3$ & $k=4$& $k=5$& $k=6$& $k=7$& $k=8$\\
\hline
 \hline
 \rule[-3mm]{0mm}{9mm}
$H^k\big(K(\Z,3),\Z\big)$& $\Z$ &$0$  &$0$ &$\Z$ &$0$ &$0$&$\Z_2$&$0$&$\Z_3$\\ 
\end{tabular}
 \end{table}
\noindent
leads to the computation of $E_2^{p,q}$ showed in the following table:
 \begin{table}[h]\label{tab:E_2_d=3}
 \begin{tabular}{c||c|c|c|c|c|c}
 \cline{1-6}
 \rule[-3mm]{0mm}{9mm}
$q=4$ & $0$ & $0$ & $0$ &$0$&$0$\\
\cline{1-6}
 \rule[-3mm]{0mm}{9mm}
$q=3$ & $\Z$ & $\Z$ & $\Z$ &$\Z$&$\Z$\\
 \hline
 \rule[-3mm]{0mm}{9mm}
 $q=2$& $0$ & $0$ & $0$ & $0$ & $0$      \\
\cline{1-6}
 \rule[-3mm]{0mm}{9mm}
$q=1$ & $0$ & $0$ & $0$ &$0$&$0$\\
\cline{1-6}
 \rule[-3mm]{0mm}{9mm}
$q=0$ & $\Z$ & $\Z$ & $\Z$ &$\Z$&$\Z$\\
\hline
 \hline
 \rule[-3mm]{0mm}{9mm}
$E^{p,q}_2$& $p=0$ & $p=1$ &$p=2$&$p=3$&$p=4$\\ 
\end{tabular}
 \end{table}

\noindent
Since the sequence is concentrate in the first quadrant (\ie $E_r^{p,q}=0$ if $p<0$ or $q<0$)
one obtains by a recursive application of the formula
\begin{equation}\label{eq;spec_seq_gen_for}
E^{p,q}_{r+1}\;:=\;\frac{{\rm Ker}\Big(\delta_r:E^{p,q}_r\to E_r^{p+r,q-r+1}\Big)}{{\rm Im}\Big(\delta_r:E^{p-r,q+r-1}_r\to E_r^{p,q}\Big)}
\end{equation}
 that $E^{0,3}_4\simeq E^{0,3}_3\simeq E^{0,3}_2$ and $E^{4,0}_4\simeq E^{4,0}_3\simeq E^{4,0}_2$. Moreover, one has that
$$
E^{0,3}_\infty\;\simeq\;\ldots\;\simeq E^{0,3}_6\;\simeq\; E^{0,3}_5\;:=\; {\rm Ker}\Big(\delta_4:E^{0,3}_4\to E_4^{4,0}\Big)\;
$$
and the isomorphisms $E^{0,3}_4\simeq E^{0,3}_2\simeq H^3(K(\Z,3),\Z)$ and $E^{4,0}_4\simeq E^{4,0}_2\simeq H^4(\C P^\infty,\Z)$ allow to write
$$
E^{0,3}_\infty\;\simeq\;{\rm Ker}\Big(\delta_4:H^3\big(K(\Z,3),\Z\big)\to H^4\big(\C P^\infty,\Z\big)\Big)\;.
$$
The map $\delta_4$ relates the basic class
 $\rr{w}_2\in H^3(K(\Z,3),\Z)$ with the invariant $\kappa^4\in H^4(\C P^\infty,\Z)$ according to the formula  
$\delta_4(\rr{w}_2)=-\kappa^4$ (\cf \cite[Remark 7.2.6 (4)]{arkowitz-11} or \cite[Lemma 3.4.2]{may-ponto-12}).
The convergence of the spectral sequence provides the following short exact sequences\footnote{For the precise  derivation of the short exact sequences the reader can refer to  \cite[eq. D8]{denittis-gomi-15} (and reference therein). Here we used the shorter notational convention
$$
F^jH^k\big(X,\Z\big)\;:=\;{\rm Ker}\Big(H^k\big(X,\Z\big)\to H^k\big(X_{j-1},\Z\big)\Big)\;\subset\; H^k\big(X,\Z\big)
$$
where $\empty=X_1\subset X_0\subset X_1\subset\ldots\subset X_p\subset\ldots\subset X$ is a  filtration for the space $X$.} 
$$
\begin{aligned}
& 0\;\longrightarrow\; F^1H^3\big(\bb{B}_{3}^m,\Z\big)\;\longrightarrow\;H^3\big(\bb{B}_{3}^m,\Z\big)\;\longrightarrow\;E^{0,3}_\infty\;\longrightarrow\;0\;\\
& 0\;\longrightarrow\; F^2H^3\big(\bb{B}_{3}^m,\Z\big)\;\longrightarrow\;F^1H^3\big(\bb{B}_{3}^m,\Z\big)\;\longrightarrow\;E^{1,2}_\infty\;\longrightarrow\;0\;\\
& 0\;\longrightarrow\; F^3H^3\big(\bb{B}_{3}^m,\Z\big)\;\longrightarrow\;F^2H^3\big(\bb{B}_{3}^m,\Z\big)\;\longrightarrow\;E^{2,1}_\infty\;\longrightarrow\;0\;.\\
\end{aligned}
$$
Since $E^{2,1}_\infty\simeq\ldots \simeq E^{2,1}_2=0$ 
and similarly $E^{1,2}_\infty\simeq\ldots \simeq E^{1,2}_2=0$ one gets that 
$$
F^1H^3\big(\bb{B}_{3}^m,\Z\big)\;\simeq\; F^2H^3\big(\bb{B}_{3}^m,\Z\big)
\;\simeq\; F^3H^3\big(\bb{B}_{3}^m,\Z\big)\;.
$$
By observing that $F^3H^3\big(\bb{B}_{3}^m,\Z\big)\simeq E^{3,0}_\infty$,
one ends up with the following short
exact sequence
\begin{equation}
0\;\longrightarrow\;E^{3,0}_\infty\;\longrightarrow\;H^3\big(\bb{B}_{3}^m,\Z\big)\;\longrightarrow\;E^{0,3}_\infty\;\longrightarrow\;0\;.
\end{equation}
Since we know that $H^3\big(\bb{B}_{3}^m,\Z\big)\simeq \Z^2$ and we  can compute that $E^{3,0}_\infty\simeq\ldots\simeq E^{3,0}_3\simeq E^{3,0}_2\simeq\Z$ we  can immediately conclude from the above exact sequence and $E^{0,3}_\infty\subseteq H^3(K(\Z,3),\Z)\simeq\Z$ that $E^{0,3}_\infty\simeq\Z$ which in turn implies that $\delta_4=0$ acts as the trivial map. Finally the vanishing of the Postnikov invariant $\kappa^4=0$  assures that 
$
\bb{B}_{3}^m=\n{S}^1\times \C P^\infty\times K(\Z,3)
$.
 Proposition \ref{prop:postnikov} provides
$$
{{\rm Vec}}_\chi^m\big(X\big)\;\simeq\;[X,\n{S}^1]\;\times\;[X,\C P^\infty]\;\times\;[X,K(\Z,3)]\;\simeq\; H^1(X,\Z)\;\oplus\;H^2(X,\Z)\;\oplus\;H^3(X,\Z)\;
$$
and the 
basic class in $H^3(K(\Z,3),\Z)$  which induces the
isomorphism $[X,K(\Z,3)]\simeq H^3(X,\Z)$ described in \eqref{eq:cohom_EML_space}  can be identified with the second generator $\rr{w}_2\in H^3(\n{U}(\infty),\Z)$, as discussed in Lemma \ref{lemma:cohomU_inf} and Remark \ref{rk:postnikov_sect_grass}, and consequently, with the second chiral generator
 $\rr{w}_2^\chi$ of $H^\bullet(\n{B}_\chi^m,\Z)$ according to Theorem \ref{theo:cohom_calss_spac}.

\medskip

The  next case $d=4$, $m\geqslant 3$ can be discussed along the same lines  as those of the previous case. The Postnikov  section  $\bb{B}_{4}^m$ is the total space of the fibration 
\begin{equation}\label{eq:PostnikovB-4}
K(\Z,4)\;\longrightarrow\; \bb{B}_{4}^m\;\stackrel{p_{4}}{\longrightarrow}\; \bb{B}_{3}^m\;=\;\n{S}^1\;\times\; \C P^\infty\;\times\;  K(\Z,3)
\end{equation}
where we used $\pi_4(\n{B}_\chi^m)\simeq\Z$ when $m\geqslant 3$. We point out that  for $m=2$ the group 
$\pi_4(\n{B}_\chi^2)$ has a torsion part and \eqref{eq:PostnikovB-4} needs to be modified (see Section \ref{sect_tori_4}).
 This section is determined by the 
invariant 
$$
\begin{aligned}
\kappa^5\in H^5\big(\n{S}^1\times \C P^\infty\times K(\Z,3),\Z\big)\;&\simeq\;\Big(\underbrace{H^2( \C P^\infty,\Z)\otimes_\Z H^3( K(\Z,3),\Z)}_{\simeq\Z}\Big)\;\oplus\; \Big(\underbrace{H^1( \n{S}^1,\Z) \otimes_\Z H^4( \C P^\infty,\Z)}_{\simeq\Z} \Big)\\
\end{aligned}
$$
and we want to prove that $\kappa^5=0$. 
Since
the map $\alpha_4:\n{B}_\chi^m\to  \bb{B}_{4}^m$ is a $5$-equivalence we can deduce from our knowledge of the cohomology of $\n{B}_\chi^m$  that 
$$
H^1\big(\bb{B}_{4}^m,\Z\big)\;\simeq\; H^2\big(\bb{B}_{4}^m,\Z\big)\;\simeq\;\Z\;,\qquad
H^3\big(\bb{B}_{4}^m,\Z\big)\;\simeq\;\Z^2\; \qquad
H^4\big(\bb{B}_{4}^m,\Z\big)\;\simeq\;\Z^3\;.
$$
On the other hand we can compute this cohomology by means of the Leray-Serre spectral
sequence associated with the fibration \eqref{eq:PostnikovB-4}. The 2-page of this spectral sequence   is given by
\begin{equation}\label{eq:spec_seq_a1**}
E_2^{p,q}\;=\;H^p\Big(\bb{B}_{3}^m, H^q\big(K(\Z,4),\Z\big )\Big)
\end{equation}
with constant system of coefficients since
$\pi_1(\bb{B}_{3}^m)\simeq\pi_1(\n{S}^1)\simeq\Z$
  acts trivially on 
higher homotopy groups 
$$
\pi_k\big(\bb{B}_{3}^m\big)\;\simeq\; \pi_k\big(\n{S}^1\big)\;\times\; \pi_k\big(K(\Z,2)\big)\;\times\; \pi_k\big(K(\Z,3)\big)\;,\qquad\quad k\geqslant 1\;.
$$ 
In particular, since $\bb{B}_{3}^m$ is path-connected one obtains from \eqref{eq:spec_seq_a1**} the following isomorphisms
\begin{equation}\label{eq:spec_seq_a1_d4}
E_2^{0,q}\;\simeq\; H^q\big(K(\Z,4),\Z\big)\;.
\end{equation}
The cohomology of 
$K(\Z,4)$ can be computed as in \cite[Section 18]{bott-tu-82} (see also \cite[Table 1]{percy-10}) and one obtains:
 \begin{table}[h]\label{tab:K=4}
 \begin{tabular}{c||c|c|c|c|c|c|c|c|c|c|}
 \rule[-3mm]{0mm}{9mm}
 & $k=0$ & $k=1$ & $k=2$ & $k=3$ & $k=4$& $k=5$& $k=6$& $k=7$& $k=8$\\
\hline
 \hline
 \rule[-3mm]{0mm}{9mm}
$H^k\big(K(\Z,4),\Z\big)$& $\Z$ & $0$  &$0$  &$0$ &$\Z$ &$0$&$0$&$\Z_2$&$\Z$\\ 
\end{tabular}
 \end{table}

\noindent
This allows us to compute the values of the 2-page  $E_2^{p,q}$ as showed in the following table:
 \begin{table}[h]\label{tab:E_2_d=4}
 \begin{tabular}{c||c|c|c|c|c|c|c|c}
\cline{1-8}
 \rule[-3mm]{0mm}{9mm}
$q=6$ & $0$ & $0$ & $0$ &$0$&$0$&$0$&$0$\\ 
 \cline{1-8}
 \rule[-3mm]{0mm}{9mm}
$q=5$ & $0$ & $0$ & $0$ &$0$&$0$&$0$&$0$\\
\cline{1-8}
 \rule[-3mm]{0mm}{9mm}
$q=4$ & $\Z$ & $\Z$ & $\Z$ &$\Z^2$&$\Z^2$&$\Z^2$&$\Z^2\oplus\Z_2$ \\
\cline{1-8}
 \rule[-3mm]{0mm}{9mm}
$q=3$ & $0$ & $0$ & $0$ &$0$&$0$&$0$&$0$\\
 \hline
 \rule[-3mm]{0mm}{9mm}
 $q=2$& $0$ & $0$ & $0$ & $0$ & $0$& $0$& $0$      \\
\cline{1-8}
 \rule[-3mm]{0mm}{9mm}
$q=1$ & $0$ & $0$ & $0$ &$0$&$0$&$0$&$0$\\
\cline{1-8}
 \rule[-3mm]{0mm}{9mm}
$q=0$ & $\Z$ & $\Z$ & $\Z$ &$\Z^2$&$\Z^2$&$\Z^2$&$\Z^2\oplus\Z_2$\\
\hline
 \hline
 \rule[-3mm]{0mm}{9mm}
$E^{p,q}_2$& $p=0$ & $p=1$ &$p=2$&$p=3$&$p=4$&$p=5$&$p=6$\\ 
\end{tabular}
 \end{table}

\noindent
A recursive application of the formula
 \eqref{eq;spec_seq_gen_for} provides
$$
E^{0,4}_\infty\;\simeq\;\ldots\;\simeq E^{0,4}_7\;\simeq\; E^{0,4}_6\;:=\; {\rm Ker}\Big(\delta_5:E^{0,4}_5\to E_5^{5,0}\Big)\;
$$ 
and the isomorphisms 
$$
\begin{aligned}
&E^{0,4}_5\;\simeq\; E^{0,4}_4\;\simeq\; E^{0,4}_3\;\simeq\; E^{0,4}_2\;\simeq\; H^4\big(K(\Z,4),\Z\big)\;\simeq\;\Z\\
&E^{5,0}_5\;\simeq\; E^{5,0}_4\;\simeq\; E^{5,0}_3\;\simeq\; E^{5,0}_2\;\simeq\; H^5\big(\bb{B}_{3}^m, \Z\big)\;\simeq\;\Z^2
\end{aligned}
$$ 
 allow us to write
$$
E^{0,4}_\infty\;\simeq\;{\rm Ker}\Big(\delta_5:H^4\big(K(\Z,4),\Z\big)\to H^5\big(\bb{B}_{3}^m, \Z\big)\Big)\;.
$$
The map $\delta_5$ relates the basic class
 $\rr{c}_2\in H^4(K(\Z,4),\Z)$  with the invariant $\kappa^5\in H^5\big(\bb{B}_{3}^m, \Z\big)$ according to the formula  
$\delta_5(\rr{c}_2)=-\kappa^5$ (\cf \cite[Remark 7.2.6 (4)]{arkowitz-11} or \cite[Lemma 3.4.2]{may-ponto-12}). The convergence of the spectral sequence provides in exactly the same way as above the 
short
exact sequence
\begin{equation}\label{eq:XVVAS}
0\;\longrightarrow\;E^{4,0}_\infty\;\longrightarrow\;H^4\big(\bb{B}_{4}^m,\Z\big)\;\longrightarrow\;E^{0,4}_\infty\;\longrightarrow\;0\;.
\end{equation}
Since we know that $H^4(\bb{B}_{4}^m,\Z)\simeq \Z^3$ and we  can compute that $E^{4,0}_\infty\simeq\ldots\simeq E^{4,0}_3\simeq E^{4,0}_2\simeq\Z^2$ we  can immediately conclude from \eqref{eq:XVVAS} and $E^{0,4}_\infty\subseteq H^4(K(\Z,4),\Z)\simeq\Z$ that $E^{0,4}_\infty\simeq\Z$ which in turn implies that $\delta_5=0$ acts as the trivial map. Finally the vanishing of the Postnikov invariant $\kappa^5=0$  assures that 
$
\bb{B}_{4}^m=\n{S}^1\times \C P^\infty\times K(\Z,3) \times K(\Z,4)
$ if $m\geqslant 3$.
 Proposition \ref{prop:postnikov} provides
$$
{{\rm Vec}}_\chi^m\big(X\big)\;\simeq\;[X,\n{S}^1]\;\times\;[X,\C P^\infty]\;\times\;[X,K(\Z,3)]\;\times\;[X,K(\Z,4)]\;\simeq\; \bigoplus_{j=1}^4 H^j(X,\Z)\;,\qquad\quad m\geqslant3
$$
and the 
basic class in $H^4(K(\Z,4),\Z)$  which induces the
isomorphism $[X,K(\Z,4)]\simeq H^4(X,\Z)$ described in \eqref{eq:cohom_EML_space}  can be identified with the second universal Chern class
$\rr{c}_2\in H^4(G_m(\C^{\infty}),\Z)$ as discussed in
Remark \ref{rk:postnikov_sect_grassA}.\qed

\medskip

\begin{remark}{\upshape
In the statement of Proposition \ref{propos:class_low_dim} we pointed out that
 the equations \eqref{eq:class_low_dim} and \eqref{eq:class_low_dim4} have to be understood as bijections of sets rather than group isomorphisms.
This is because the standard group structure of direct sums of cohomology groups does not coincide \emph{a priori} 
with a possible group structure on classes of chiral vector bundles
induced by some geometric operations such as the Whitney sum.
For this reason it should be better to use expressions like $H^1(X,\Z)\times H^2(X,\Z)\times \ldots$ instead $H^1(X,\Z)\oplus H^2(X,\Z)\oplus \ldots$ in \eqref{eq:class_low_dim} and \eqref{eq:class_low_dim4}.
However, in accordance with a  modern (maybe inaccurate) custom in the classification of topological phases we chose to keep with the use of the symbol
$\oplus$ instead of $\times$,
always keeping in mind that the identification is at level of sets and not of groups. The main reason and the first advantage of this choice of notation  is that Table 1.1 and Table 1.2 
 result directly comparable with the existing literature.
In any case this \virg{conflict} in the notation opens the necessity for a deeper investigation of the group structure induced on the classes of chiral vector bundles by  Whitney sum. We plan to investigate this aspect in an upcoming work.}\hfill $\blacktriangleleft$
\end{remark}

\medskip

\begin{remark}{\upshape 
The next step in the construction of the 
Postinikov tower of $\n{B}^m_\chi$   is the determination
of the 
invariant $\kappa^6$. However, this invariant turns out to be non-trivial even
in the stable range $m\geqslant 3$ 
for a reason that we will sketch below. First of all from $\bb{B}_{4}^m=\bb{B}_{3}^m\times K(\Z,4)$ and an application of the K\"{u}nneth formula for cohomology one obtains 
$$
H^6\big(\bb{B}_{4}^m, \Z\big)\;\simeq\; \Big(\underbrace{H^2\big(\bb{B}_{3}^m, \Z\big)\;\otimes_\Z\;H^4\big(K(\Z,4), \Z\big)}_{\simeq\Z}\Big)\;\oplus\;\Big(\underbrace{H^6\big(\bb{B}_{3}^m, \Z\big)\;\otimes_\Z\;H^0\big(K(\Z,4), \Z\big)}_{\simeq H^6\big(\bb{B}_{3}^m, \Z\big)}\Big)\;.
$$
Hence $H^6(\bb{B}_{4}^m, \Z)$ has a torsion subgroup $\Z_2$ which comes from
$
 H^6(\bb{B}_{3}^m, \Z)\simeq\Z^2\oplus\Z_2
$. 
By following the same lines of reasoning of the proof of Proposition \ref{propos:class_low_dim} one concludes that the Postinikov invariant $\kappa^6$ is, up to a sign, the image of the basic class in $K(\Z,5)$ under the differential map 
$$
H^5\big(K(\Z,5),\Z\big )\;\simeq\;E_2^{0,5}\;\simeq\;\ldots\simeq\;E_6^{0,5}\;\stackrel{\delta_6}{\longrightarrow}\;E_6^{6,0}\;\simeq\;\ldots\simeq\;E_2^{6,0}\;\simeq\;H^6\big(\bb{B}_{4}^m, \Z\big)
$$
associated to the Leray-Serre spectral
sequence of the fibration 
$K(\Z,5)\to \bb{B}_{5}^m\to \bb{B}_{4}^m$. Thus, the map $\delta_6$  fits in the exact sequence
$$
0\;\longrightarrow\;E_\infty^{0,5}\;\longrightarrow\;\underbrace{\Z}_{\simeq E_6^{0,5}}\;\stackrel{\delta_6}{\longrightarrow}\;\underbrace{\Z^3\oplus\Z_2}_{\simeq E_6^{6,0}}\;\longrightarrow\;\underbrace{E_\infty^{6,0}}_{\text{free group}}\;\longrightarrow\;0
$$
where the absence of torsion in $E_\infty^{6,0}$ 
is justified by the fact that $E_\infty^{6,0}$ 
is a subgroup of $H^6(\bb{B}_{5}^m,\Z)$ which in turn is injected in $H^6(\n{B}_{\chi}^m,\Z)\simeq\Z^5$ since $\alpha_5:\n{B}_\chi^m\to  \bb{B}_{5}^m$ is a $6$-equivalence. Therefore, 
the absence of torsion in $E_\infty^{6,0}$ assures that
$\delta_6\neq0$ and so $\kappa^6$ is non trivial since it has to generate at least the torsion part $\Z_2$. 
}\hfill $\blacktriangleleft$
\end{remark}

%-------------------------------------%

\subsection{The unstable case $d=4$, $m=2$}\label{sect_tori_4}
This case, in principle, can be handled with the same technique of the proof of Preposition \ref{propos:class_low_dim}. However, the presence of a torsion term in  $\pi_4(\n{B}_\chi^2)\simeq\Z_2\oplus\Z$ (\cf Section \ref{sec:homot_B}) makes the argument much more involved. 
\medskip

The computation of  the
 Postnikov  section  $\bb{B}_{4}^2$  requires the investigation of the fibration
\begin{equation}\label{eq:PostnikovB-d4m2}
K(\Z_2\oplus\Z,4)\;\longrightarrow\; \bb{B}_{4}^2\;\stackrel{p_{4}}{\longrightarrow}\; \underbrace{\n{S}^1\times \C P^\infty\times K(\Z,3)}_{\bb{B}_{3}^2}\;\stackrel{\kappa^{5}}{\longrightarrow}\; K(\Z_2\oplus\Z,5)
\end{equation}
and in particular we need
to determine
$
\kappa^5\in H^5(\bb{B}_{3}^2,\Z_2\oplus\Z)
$. From the very definition of the singular cohomology one obtains the natural direct sum decomposition
\begin{equation}\label{eq:cohom_unst_mag}
H^k\big(\bb{B}_{3}^2,\Z_2\oplus\Z\big)\;\simeq\; H^k\big(\bb{B}_{3}^2,\Z_2\big)\;\oplus\; H^k\big(\bb{B}_{3}^2,\Z\big)\;.
\end{equation}
These cohomology groups can be computed with the help of the following table:
 \begin{table}[h]
 \begin{tabular}{c||c|c|c|c|c|c|c|c|c|c|}
 \rule[-3mm]{0mm}{9mm}
 & $k=0$ & $k=1$ & $k=2$ & $k=3$ & $k=4$& $k=5$\\
\hline
 \hline
 \rule[-3mm]{0mm}{9mm}
$H^k(\bb{B}_{3}^2,\Z)$& $\Z$ & $\Z$  &$\Z$  &$\Z^2$ &$\Z^2$ &$\Z^2$\\
 \hline
 \rule[-3mm]{0mm}{9mm}
$H^k(\bb{B}_{3}^2,\Z_2)$& $\Z_2$ & $\Z_2$  &$\Z_2$  &${\Z_2}^2$ &${\Z_2}^2$ &${\Z_2}^3$\\
 \hline
\end{tabular}
 \end{table}
%

%\newpage
\noindent
In particular one has that
$$
\kappa^5\;\in\; H^5\big(\bb{B}_{3}^2,\Z_2\oplus\Z\big)\;\simeq\;{\Z_2}^3\;\oplus\;\Z^2\;.
$$

\medskip

We can obtain information about the cohomology of $\bb{B}_{4}^2$ by recalling
that
the map $\alpha_4:\n{B}_\chi^2\to  \bb{B}_{4}^2$ is a $5$-equivalence. Our knowledge of the cohomology of $\n{B}_\chi^2$  provides 
$$
H^1\big(\bb{B}_{4}^2,\Z\big)\;\simeq\; H^2\big(\bb{B}_{4}^2,\Z\big)\;\simeq\;\Z\;,\qquad
H^3\big(\bb{B}_{4}^2,\Z\big)\;\simeq\;\Z^2\; \qquad
H^4\big(\bb{B}_{4}^2,\Z\big)\;\simeq\;\Z^3\;.
$$
Moreover $H^5(\bb{B}_{4}^2,\Z)$ is injected in $H^5(\n{B}_\chi^2,\Z)\simeq\Z^3$ and so it has no torsion. This also implies information about homology
(\cf \cite[Corollary 3.3]{hatcher-02}) and with the help of   the universal coefficient theorem one obtains
$$
H^k\big(\bb{B}_{4}^2,\Z_2\oplus\Z\big)\;\simeq\;
{\rm Hom}_\Z\big(H^k\big(\bb{B}_{4}^2,\Z\big),\Z_2\big) \;\oplus\; H^k\big(\bb{B}_{4}^2,\Z\big)\;,\qquad\quad  k=1,2,3,4\;.
$$
In particular for $k=4$ one has
$$
H^4\big(\bb{B}_{4}^2,\Z_2\oplus\Z\big)\;\simeq\;{\Z_2}^3\;\oplus\;\Z^3\;.
$$

\medskip

We can study the cohomology of $\bb{B}_{4}^2$ {with coefficients $\Z_2\oplus\Z$} also with the help of the spectral sequence associated to the principal fibration
\begin{equation}
K(\Z_2\oplus\Z,4)\;\longrightarrow\; \bb{B}_{4}^2\;\stackrel{p_{4}}{\longrightarrow}\; \bb{B}_{3}^2\;.
\end{equation}
Since we know that $\pi_1(\bb{B}_{3}^2)$ acts trivially on higher homotopy groups the 
 2-page of this spectral sequence   is given by
\begin{equation}\label{eq:spec_seq_a1_unstable}
E_2^{p,q}\;=\;H^p\Big(\bb{B}_{3}^2, H^q\big(K(\Z_2\oplus\Z,4),{\Z_2\oplus\Z}\big )\Big)
\end{equation}
with constant system of coefficients. 
In particular the path-connectedness of  $\bb{B}_{3}^m$ implies
\begin{equation}\label{eq:spec_seq_a1_d4_unstable}
E_2^{0,q}\;\simeq\; H^q\big(K(\Z_2\oplus\Z,4),{\Z_2\oplus\Z}\big)\;.
\end{equation}
The computation of the spectral sequence requires the cohomology of $K(\Z_2\oplus\Z,4)$.
One can use the relation
$
K(\n{G}_1\oplus\n{G}_2,j) \simeq K(\n{G}_1,j) \times  K(\n{G}_2,j)
$  and the K\"{u}nneth formula for cohomology to reduce the problem to the computation of the cohomology of $K(\Z,4)$ and $K(\Z_2,4)$. Since we already used the cohomology of $K(\Z,4)$ in the proof of Proposition \ref{propos:class_low_dim} we need to compute only the cohomology of $K(\Z_2,4)$. this can be done along the lines sketched in \cite[Section 18]{bott-tu-82} or with the \virg{Eilenberg-MacLane machine} described in \cite{clement-02}.
The results are showed in the following table:
 \begin{table}[h]
 \begin{tabular}{c||c|c|c|c|c|c|c|c|c|c|}
 \rule[-3mm]{0mm}{9mm}
 & $k=0$ & $k=1$ & $k=2$ & $k=3$ & $k=4$& $k=5$& $k=6$& $k=7$\\
\hline
 \hline
 \rule[-3mm]{0mm}{9mm}
$H^k\big(K(\Z,4),\Z\big)$& $\Z$ & $0$  &$0$  &$0$ &$\Z$ &$0$&$0$&$\Z_2$\\ 
\hline
\rule[-3mm]{0mm}{9mm}
$H^k\big(K(\Z_2,4),\Z\big)$& $\Z$ & ${0}$  & ${0}$    & ${0}$   &  ${0}$  &${\Z_2}$   & ${0}$  & ${\Z_2}$     \\ 
\hline
 \hline
\rule[-3mm]{0mm}{9mm}
$H^k\big(K(\Z_2\oplus\Z,4),\Z\big)$& $\Z$ & $0$  & $0$  & $0$ & $\Z$  &$\Z_2$ &  $0$& ${\Z_2}^2$\\ 
\hline
\rule[-3mm]{0mm}{9mm}
$H_k\big(K(\Z_2\oplus\Z,4),\Z\big)$& $\Z$ &  $0$ & $0$  & $0$ & $\Z_2\oplus\Z$  & $0$  &  ${\Z_2}^2$ & $\Z_2$  \\ 
\hline
\rule[-3mm]{0mm}{9mm}
$H^k\big(K(\Z_2\oplus\Z,4),\Z_2\oplus\Z\big)$& $\Z_2\oplus\Z$ & $0$  & $0$  & $0$ & ${\Z_2}^2\oplus\Z$  &  ${\Z_2}^2$ & ${\Z_2}^2$  & ${\Z_2}^3$   \\ 
\hline
\end{tabular}
 \end{table}

\noindent
With these results we can compute the values of $E^{p,q}_2$ showed in the following table:
 \begin{table}[h]\label{tab:E_2_d=4_unstable}
 \begin{tabular}{c||c|c|c|c|c|c|c|}
\hline
 \rule[-3mm]{0mm}{9mm}
$q=5$ & ${\Z_2}^2$ &  $\dots$  &   & & &  \\
\hline
 \rule[-3mm]{0mm}{9mm}
$q=4$ & ${\Z_2}^2\oplus\Z$ & ${\Z_2}^2\oplus\Z$ &  $\dots$ & & &   \\
\hline
 \rule[-3mm]{0mm}{9mm}
$q=3$ & $0$ & $0$ & $0$ &$0$&$0$&$0$\\
 \hline
 \rule[-3mm]{0mm}{9mm}
 $q=2$& $0$ & $0$ & $0$ & $0$ & $0$& $0$      \\
\hline
 \rule[-3mm]{0mm}{9mm}
$q=1$ & $0$ & $0$ & $0$ &$0$&$0$&$0$\\
\hline
 \rule[-3mm]{0mm}{9mm}
$q=0$ & $\Z_2\oplus\Z$ & $\Z_2\oplus\Z$ &  $\Z_2\oplus\Z$&${\Z_2}^2\oplus\Z^2$&${\Z_2}^2\oplus\Z^2$&${\Z_2}^3\oplus\Z^2$\\
\hline
 \hline
 \rule[-3mm]{0mm}{9mm}
$E^{p,q}_2$& $p=0$ & $p=1$ &$p=2$&$p=3$&$p=4$&$p=5$
\end{tabular}
 \end{table}

\noindent
A recursive application of the formula
 \eqref{eq;spec_seq_gen_for} provides
$$
E^{0,4}_\infty\;\simeq\;\ldots\;\simeq E^{0,4}_7\;\simeq\; E^{0,4}_6\;:=\; {\rm Ker}\Big(\delta_5:E^{0,4}_5\to E_5^{5,0}\Big)\;
$$ 
and the isomorphisms 
$$
\begin{aligned}
&E^{0,4}_5\;\simeq\; E^{0,4}_4\;\simeq\; E^{0,4}_3\;\simeq \; E^{0,4}_2\;\simeq\; H^4\big(K(\Z_2\oplus\Z,4),{\Z_2\oplus\Z}\big)\;\simeq\;{\Z_2}^2\;\oplus\;\Z\\
&E^{5,0}_5\;\simeq\; E^{5,0}_4\;\simeq\; E^{5,0}_3\;\simeq\; E^{5,0}_2\;\simeq\; H^5\big(\bb{B}_{3}^2, {\Z_2\oplus\Z}\big)\;\simeq\;{\Z_2}^3\;\oplus\;\Z^2
\end{aligned}
$$ 
 allow to write
$$
E^{0,4}_\infty\;\simeq\;{\rm Ker}\Big(\delta_5:H^4\big(K(\Z_2\oplus\Z,4),{\Z_2\oplus\Z}\big)\to H^5\big(\bb{B}_{3}^2, {\Z_2\oplus\Z}\big)\Big)\;.
$$
The map $\delta_5$ relates the basic class
 $\rr{d}\in H^4(K(\Z_2\oplus\Z,4),{\Z_2\oplus\Z})$ with the invariant $\kappa^5\in H^5\big(\bb{B}_{3}^2, {\Z_2\oplus\Z}\big)$ according to the formula  
$\delta_5(\rr{d})=-\kappa^5$ (\cf \cite[Remark 7.2.6 (4)]{arkowitz-11} or \cite[Lemma 3.4.2]{may-ponto-12}). 
The convergence of the spectral sequence provides in exactly the same way as in the proof of Proposition \ref{propos:class_low_dim} the 
short
exact sequence 
\begin{equation}\label{eq:exact_unst_emine*}
0\;\longrightarrow\;E^{4,0}_\infty\;\longrightarrow\;H^4\big(\bb{B}_{4}^2,\Z_2\oplus\Z\big)\;\longrightarrow\;E^{0,4}_\infty\;\longrightarrow\;0\;.
\end{equation}
Since we know that $H^4(\bb{B}_{4}^2,\Z_2\oplus\Z)\simeq {\Z_2}^3\oplus\Z^3$ and we  can compute that $E^{4,0}_\infty\simeq\ldots\simeq E^{4,0}_3\simeq E^{4,0}_2\simeq {\Z_2}^2\oplus\Z^2$ we  can  conclude that:
\begin{lemma}\label{lemma:E04_inf_unstab}
$E^{0,4}_\infty\simeq {\Z_2}\oplus\Z$ .
\end{lemma}
\proof
The key observation is that the spectral sequence $E^{p,q}_r$ induced by the 2-page \eqref{eq:spec_seq_a1_unstable} is the direct sum of two 
spectral sequences $E^{p,q}_r(\Z_2)$ and $E^{p,q}_r(\Z)$ induced by the natural splitting $E^{p,q}_2\simeq E^{p,q}_2(\Z_2)\oplus E^{p,q}_2(\Z)$ where
$$
\begin{aligned}
E^{p,q}_2(\Z_2)\;&:=\;H^p\Big(\bb{B}_{3}^2, H^q\big(K(\Z_2\oplus\Z,4),{\Z_2}\big )\Big)&& \Rightarrow\ H^\bullet\big(\bb{B}_{4}^2,\Z_2\big)\\
E^{p,q}_2(\Z)\;&:=\;H^p\Big(\bb{B}_{3}^2, H^q\big(K(\Z_2\oplus\Z,4),{\Z}\big )\Big)&& \Rightarrow\ H^\bullet\big(\bb{B}_{4}^2,\Z\big)\;.
\end{aligned}
$$
Consequently, also the  short
exact sequence \eqref{eq:exact_unst_emine*} splits as
\begin{equation}\label{eq:exact_unst_emine}
\begin{aligned}
0\;&\longrightarrow\;\ E^{4,0}_\infty(\Z_2)&&\longrightarrow\;H^4\big(\bb{B}_{4}^2,\Z\big)&&\longrightarrow\;E^{0,4}_\infty(\Z_2)&&\longrightarrow\;0\;\\
0\;&\longrightarrow\;\ E^{4,0}_\infty(\Z)&&\longrightarrow\;H^4\big(\bb{B}_{4}^2,\Z_2\big)&&\longrightarrow\;E^{0,4}_\infty(\Z)&&\longrightarrow\;0\;\;.
\end{aligned}
\end{equation}
By observing that $E^{0,4}_\infty\simeq E^{0,4}_\infty(\Z_2)\oplus E^{0,4}_\infty(\Z)$
with $E^{0,4}_\infty(\Z_2)\subseteq H^4(K(\Z_2\oplus\Z,4),{\Z_2})\simeq{\Z_2}^2$ and $E^{0,4}_\infty(\Z)\subseteq H^4(K(\Z_2\oplus\Z,4),{\Z})\simeq\Z$
and from $E^{4,0}_\infty(\Z_2)\simeq{\Z_2}^2$, $E^{4,0}_\infty(\Z)\simeq{\Z}^2$ and $H^4\big(\bb{B}_{4}^2,\Z_2\big)\simeq{\Z_2}^3$,
$H^4\big(\bb{B}_{4}^2,\Z\big)\simeq{\Z}^3$ one immediately gets $E^{0,4}_\infty(\Z_2)\simeq\Z_2$ and $E^{0,4}_\infty(\Z)\simeq\Z$. 
\qed

\medskip

\noindent
The fact that  $E^{0,4}_\infty={\rm Ker}(\delta_5)\neq H^4\big(K(\Z_2\oplus\Z,4),{\Z_2\oplus\Z}\big)$ implies that $\delta_5$ cannot be 
the trivial map. In particular this implies that $\kappa^5=-\delta_5(\rr{d})\neq 0$. 
Even though this argument  does not suffice to completely specify the invariant $\kappa^5$, the fact that  the $\Z$ summand is in the kernel of the differential $\delta_5$  gives us  an extra piece of information about the structure of $\kappa^5$.
\begin{corollary}\label{corol_k5_order2}
The non-trivial invariant $\kappa^5$ is of order $2$. More precisely $\kappa^5$ takes value in the torsion subgroup ${\Z_2}^3$  
of $H^k(\bb{B}_{3}^2,\Z_2\oplus\Z)$ with respect to the splitting \eqref{eq:cohom_unst_mag}.
\end{corollary}

\medskip

\noindent
Let us recall that $\kappa^5$ can be identified as an element of 
$$[\bb{B}_{3}^2,K(\Z_2\oplus\Z,5)]\;\simeq\;[\bb{B}_{3}^2,K(\Z_2,5)]\;\times \;[\bb{B}_{3}^2,K(\Z,5)]\;\simeq\;H^5(\bb{B}_{3}^2,\Z_2)\;\oplus\; H^5(\bb{B}_{3}^2,\Z)\;.
$$
Thus, as a consequence of Corollary \ref{corol_k5_order2} one has that the invariant  $\kappa^5$ behaves as a map
$$
\kappa^5:=(0,\zeta^5)\;:\;\bb{B}_{3}^2\;\stackrel{}{\longrightarrow}\; K(\Z,5)\;\times\; K(\Z_2,5)
$$
where we used $0$ for the constant map.
\begin{proposition}\label{conj:key_unstab}
\begin{equation}
\bb{B}_{4}^2\;=\;\Big(\bb{B}_{3}^2\;\times_{\zeta^5}\; K(\Z_2,4)\Big)\;\times\; K(\Z,4)\;.  
\end{equation}
{where   the \virg{twisted} product $\bb{B}_{3}^2\;\times_{\zeta^5}\; K(\Z_2,4)$ is defined in the proof and $\zeta^5$
can be identified with a non trivial class in $H^5(\bb{B}_{3}^2,\Z_2)\simeq{\Z_2}^3$.}
\end{proposition}

\proof
The Postnikov section $\bb{B}_{4}^2$ is a principal fibration which is built from the following diagram  
$$
\begin{diagram}
K(\Z,4)\times K(\Z_2,4)& \rTo^{\simeq}&K(\Z,4)\times K(\Z_2,4)\\
 \dTo &  & \dTo\\
\bb{B}_{4}^2&\rTo&PK(\Z,5)\times PK(\Z_2,5)      \\
 \dTo &  & \dTo_{\wp=(\wp_1,\wp_2)}&\\
 \bb{B}_{3}^2&\rTo^{(0,\zeta^5)\ \ \ \ \ }&K(\Z,5)\times K(\Z_2,5)
 \end{diagram}\;
$$
where $\wp=(\wp_1,\wp_2)$ is the path fibration and the bottom square is the homotopy
pull-back. Therefore, one has that
$$
\begin{aligned}
\bb{B}_{4}^2\;&\simeq\;\Big\{(x,y,z)\;\in\;\  \bb{B}_{3}^2\times PK(\Z,5)\times PK(\Z_2,5) \;\big|\; \wp_1(y)=0(x)=\ast,\   \wp_2(z)=\zeta^5(x)\Big\}\\
\;&\simeq\;\wp_1^{-1}(\ast)\;\times\; \Big\{(x,z)\;\in\;\  \bb{B}_{3}^2\times  PK(\Z_2,5) \;\big|\;  \wp_2(z)=\zeta^5(x)\Big\}\;.
\end{aligned}
$$
The claim follows by setting $\bb{B}_{3}^2 \times_{\zeta^5}  K(\Z_2,4):=\{(x,z)\in \bb{B}_{3}^2\times  PK(\Z_2,5)\;|\;\wp_2(z)=\zeta^5(x)\}$ and by observing that $\wp_1^{-1}(\ast)\simeq K(\Z,4)$.
\qed

\medskip

\noindent
As a consequence of Proposition \ref{conj:key_unstab} (and Proposition \ref{prop:postnikov}) one has that rank 2 chiral vector bundles over a CW-complex $X$ of dimension $d=4$ are classified by  
\begin{equation}\label{eq:unst_part_blog}
{{\rm Vec}}_\chi^2\big(X\big)\;\simeq\; \big[X,\big(\bb{B}_{3}^2 \times_{\zeta^5}  K(\Z_2,4)\big)\big]\;\oplus\; H^4\big(X,\Z\big)
\end{equation}
where the contribution of $H^4\big(X,\Z\big)$ is given by the second Chern class. Unfortunately, the determination of the space of homotopic equivalent  maps $[X,(\bb{B}_{3}^2 \times_{\zeta^5}  K(\Z_2,4))]$ is difficult in  general and at the moment it seems to be beyond our capabilities.
However, under some extra assumptions on $X$, a description of the above set seems possible.

\begin{proposition}\label{propos:biject_unst}
Let $X$ be such that $H^5(X,\Z_2)=0$. Then, there is a bijection of sets
$$
\big[X,\big(\bb{B}_{3}^2 \times_{\zeta^5}  K(\Z_2,4)\big)\big]\;\simeq\; \big[X,\bb{B}_{3}^2 \big]\;\times\; \big[X,  K(\Z_2,4)\big]\;.
$$
\end{proposition}
 \proof
 The fibration projection $\pi: (\bb{B}_{3}^2 \times_{\zeta^5}  K(\Z_2,4))\to \bb{B}_{3}^2$ induces a map
 $$
 \pi_*\;:\;\big[X,\big(\bb{B}_{3}^2 \times_{\zeta^5}  K(\Z_2,4)\big)\big]\;\longrightarrow\; \big[X,\bb{B}_{3}^2 \big]\;.
 $$
  To prove the claim we need to show  that:
 (i) the inverse image $\pi_*^{-1}([f])$ of every element $[f]\in [X,\bb{B}_{3}^2 \big]$ can be identified with the set $[X,K(\Z_2; 4)]$, (ii) $\pi_*$ is surjective. First of all we notice that $\pi_*^{-1}([f])$ is
homotopy equivalent to the homotopy classes of lifts of $f:X\to \bb{B}_{3}^2$, \ie
$$
\pi_*^{-1}([f])\;\simeq\; \Big\{\tilde{f}: X\to 
\big(\bb{B}_{3}^2 \times_{\zeta^5}  K(\Z_2,4)\big)\ \big|\ \pi\circ \tilde{f}= f\Big\}\;\Big/\;\text{homotopy}\;.
$$
Let us consider the pullback fibration $\tilde{\pi}: {f}^*(\bb{B}_{3}^2 \times_{\zeta^5}  K(\Z_2,4))\to X$. Sections $s$ of this fibration are in one-to-one correspondence with the lifts $\tilde{f}$ by means of the formula
$s(x):=(x, \tilde{f}(x))$ for all $x\in\ X$. Hence 
\begin{equation}\label{eq:biject_XXY}
\pi_*^{-1}([f])\;\simeq\; \text{Sections}\Big({f}^*\big(\bb{B}_{3}^2 \times_{\zeta^5}  K(\Z_2,4)\big)\Big)\;\Big/\;\text{homotopy}\;.
\end{equation}
As a principal fibration ${f}^*(\bb{B}_{3}^2 \times_{\zeta^5}  K(\Z_2,4))$ is characterized by $f^*\zeta^5\in H^5(X,\Z_2)=0$. Therefore, there is a homotopy
equivalence $f^*(\bb{B}_{3}^2 \times_{\zeta^5}  K(\Z_2,4))\simeq X\times K(\Z_2,4)$. This implies that the sections of $f^*(\bb{B}_{3}^2 \times_{\zeta^5}  K(\Z_2,4))$ can be identified with maps on $X$ into $K(\Z_2,4)$ and so from \eqref{eq:biject_XXY} one gets
\begin{equation}\label{eq:biject_XXZ}
\pi_*^{-1}([f])\;\simeq\;\big[X,K(\Z_2,4)\big]
\end{equation}
as desired for point (i). The point (ii) which concerns the surjectivity of $\pi_*$ follows from the exact sequence of pointed sets
$$
\big[X,\big(\bb{B}_{3}^2 \times_{\zeta^5}  K(\Z_2,4)\big)\big]_* \;\longrightarrow\; \big[X,\bb{B}_{3}^2 \big]_*  \;\stackrel{\zeta_*^5}{\longrightarrow}\; \big[X,  K(\Z_2,5)\big]_*\;\simeq\; H^5(X,\Z_2) \;=\;0
$$
induced by the fibration $\bb{B}_{3}^2 \times_{\zeta^5}  K(\Z_2,4)\to \bb{B}_{3}^2\to K(\Z_2,5)$. We used $[\;,\;]_*$ for the homotopy classes of base point preserving maps. The fact that    $\bb{B}_{3}^2$ has  an H-space structure 
implies the bijection $[X,\bb{B}_{3}^2 ]_*\simeq[X,\bb{B}_{3}^2 ]$ (\cf  \cite[Example 4A.3]{hatcher-02}).
Finally, the surjectivity of $\pi_*$ comes from the commutativity of the following diagram
$$
\begin{diagram}
\big[X,(\bb{B}_{3}^2 \times_{\zeta^5}  K(\Z_2,4))\big]_*&\rTo^{\rm surj.}&\big[X,\bb{B}_{3}^2\big ]_*      \\
 \dTo^{\imath} &  & \dCorresponds_{\simeq}&\\
\big[X,(\bb{B}_{3}^2 \times_{\zeta^5}  K(\Z_2,4))\big]&\rTo^{\pi_*\ }&\big[X,\bb{B}_{3}^2 \big] \end{diagram}\;
$$
where $\imath$ is the obvious inclusion which forgets the base point.
\qed

\medskip

\noindent
Let us point out that the bijection \eqref{eq:biject_XXZ}
 depends on a choice of a
homotopy equivalence between 
$f^*(\bb{B}_{3}^2 \times_{\zeta^5}  K(\Z_2,4))$ and $X\times K(\Z_2,4)$. 
 This suggests that, a priori,
different equivalences will lead to
different  bijections. Said differently
the bijection claimed in Proposition \eqref{propos:biject_unst}  could not be \emph{natural}.

%-----%
\subsection{Chiral vector bundles over tori}\label{sect:class_tori}
Proposition \ref{propos:class_low_dim} applies, in particular, to the torus $\n{T}^d$ for $1\leqslant d\leqslant 4$. However, the unstable case of rank $2$ chiral vector bundles over $\n{T}^4$ necessitates  an extra argument.
\begin{corollary}\label{corol:torus4unst}
Let $X$ be a closed oriented $4$-dimensional manifold. The following bijection of sets
$$
{{\rm Vec}}_\chi^2\big(X\big)\;\simeq\; {{\rm Vec}}_\chi^{m>2}\big(X\big)\;\oplus\; \Z_2
$$
holds true.
\end{corollary}
\proof
The key argument is contained in Proposition \ref{propos:biject_unst} that can be applied here since $H^5(X,\Z_2)=0$. 
The result then follows just by combining equation \eqref{eq:unst_part_blog} with the classification of ${{\rm Vec}}_\chi^{m>2}\big(X\big)$ in Proposition \ref{propos:class_low_dim} and the fact that $[X,K(\Z_2,4)]\simeq H^4(X,\Z_2)\simeq\Z_2$.
\qed

\medskip

\noindent
Since the torus $\n{T}^4$  fulfills all the conditions of Corollary \ref{corol:torus4unst} one immediately gets
$$
{{\rm Vec}}_\chi^2\big(\T^4\big)\;\simeq\; \Z^{15}\;\oplus\;\Z_2\;.
$$

%--------------------%
\section{From topological quantum systems of class {\bf AIII} to chiral Bloch-bundles}\label{sect:chi_bloch_bun}

%-------%
\subsection{Chiral vector bundles vs. Clifford vector bundles}\label{sect:chi_VB_clifford}
Vector bundles endowed with a Clifford action provide a geometric model for the $K$-theoretical version of the \virg{Bott-clock} 
\cite[Chapter III]{karoubi-97}. For this reason, these objects are usually taken as base for the justification of the   Kitaev's \virg{Periodic Table} for topological insulators  \cite{freed-moore-13, thiang-14}. In this section we investigate how
vector bundles endowed with a  Clifford action can be related to $\chi$-bundles.

\medskip

{
Let us  introduce the fundamental 
notation for Clifford algebras. For a more complete introduction we refer to \cite[Chapter III, Section 3]{karoubi-97} or \cite{lee-48}.
Let $C\ell^{p,n-p}$ be the Clifford algebra
generated over $\R$ by a collection of symbols $e_1,\ldots,e_n$ subjected to the  relations $(e_i)^2=-1$ if $1\leqslant i\leqslant p$;  $(e_i)^2=+1$ if $p+1\leqslant i\leqslant n$ and $e_ie_j=-e_je_i$ if $i\neq j$. These Clifford algebras have been completely classified and for our purpose, we need only the isomorphisms  $C\ell^{1,0}\simeq\C$ and $C\ell^{0,1}\simeq\R\oplus\R$.}

\medskip

Clifford algebras are compatible with the structure of (complex) vector bundles in the following sense:
\begin{definition}[Clifford vector bundle]
Let $\bb{E}\to X$ be a rank $m$ complex vector bundle over $X$ and denote with ${\rm End}(\bb{E})$ the set of the (vector bundle) endomorphisms of $\bb{E}$. We can endow $\bb{E}$ with a {Clifford action} by means of  an $\R$-algebra homomorphism $\rho: C\ell^{p,q}\to {\rm End}(\bb{E})$. We call the pair $(\bb{E},\rho)$ a \emph{Clifford  vector bundle} of type $(p,q)$. A morphism between two Clifford  vector bundles of the same type $(\bb{E},\rho)$ and $(\bb{E}',\rho')$ is a vector bundle morphism $f:\bb{E}\to \bb{E}'$ such that $f\circ \rho(a)=\rho'(a) \circ f$ for all $a\in C\ell^{p,q}$.
\end{definition}

\medskip

We want to investigate the  connection between chiral vector bundles and Clifford vector bundles of type $(0,1)$. Let us remark  that $C\ell^{0,1}$ is generated over $\R$ by a  unique symbol $e$ such that $e^2=+1$. This 
implies that given an object $\bb{A}$ in a  Banach category (\eg a vector bundle) and an endomorphism $g\in{\rm End}(\bb{A})$
such that $g^2={\rm Id}_{\bb{A}}$ there is a unique  $\R$-algebra homomorphism $\rho: C\ell^{0,1}\to {\rm End}(\bb{A})$
completely specified by $\rho(e)=g$ and $\rho(+1)={\rm Id}_{\bb{A}}$.

\begin{lemma}\label{lem:iso_cliff1}
For a given   rank $m$ chiral vector bundle $(\bb{E}, \Phi)$ over $X$, the pair  $(\hat{\bb{E}},\rho)$ given by
\begin{equation}\label{eq:iso_cliff1}
\hat{\bb{E}}\;:=\;\bb{E}\;\oplus\; \bb{E}\;,\qquad\qquad \rho(e)\;:=\;\left(\begin{array}{cc} 0 & \Phi \\  \Phi^{-1} & 0 \\ \end{array}\right)
\end{equation}
defines a rank $2m$ Clifford vector bundle of type $(0,1)$ over $X$. Moreover, this assignment is compatible with isomorphisms in the sense that isomorphic chiral vector bundles define isomorphic Clifford vector bundles.
 \end{lemma}
\proof 
Let $g=\rho(e)\in  {\rm End}(\hat{\bb{E}})$ as specified by \eqref{eq:iso_cliff1}. For the proof that $(\hat{\bb{E}},\rho)$ is a  Clifford vector bundle  with a $C\ell^{0,1}$-action
it is enough to observe that $g^2={\rm Id}_{\hat{\bb{E}}}$. Now, let us consider 
 two isomorphic $\chi$-bundles $(\bb{E}_1, \Phi_1)\approx (\bb{E}_2, \Phi_2)$ with isomorphism given by a map $f$ in the sense of diagram \eqref{eq:diag2}. Let $(\hat{\bb{E}}_1,\rho_1)$ and $(\hat{\bb{E}}_2,\rho_2)$ be the  Clifford vector bundles
 associated to $(\bb{E}_1, \Phi_1)$ and $(\bb{E}_2, \Phi_2)$, respectively. The map 
 $$
 \hat{f}\;:=\;{ \left(\begin{array}{cc} f & 0 \\  0 & f \\ \end{array}\right)}
 $$ 
 provides an isomorphism between $(\bb{E}_1, \Phi_1)$ and $(\bb{E}_2, \Phi_2)$ in the category of Clifford vector bundles since $\hat{f}\circ \rho_1(e)=\rho_2(e)\circ \hat{f}$.
\qed

\noindent 
The Clifford vector bundle $(\hat{\bb{E}},\rho)$ built from a $\chi$-bundle $(\bb{E}, \Phi)$ according to the prescription in Lemma \ref{lem:iso_cliff1} has more structure. Indeed,  it is endowed with an \emph{odd symmetric gradation} $\Gamma\in {\rm End}(\hat{\bb{E}})$ defined by
\begin{equation}\label{eq:iso_cliff020}
\Gamma\;:=\;{\left(\begin{array}{cc} +{\rm Id}_{\bb{E}} & 0 \\  0 & -{\rm Id}_{\bb{E}} \\ \end{array}\right)}\;.
\end{equation}
A direct computation shows that 
\begin{equation}\label{eq:iso_cliff2}
\Gamma^2\;=\;{\rm Id}_{\hat{\bb{E}}}\;,\qquad\qquad \Gamma\circ \rho(e)\;=\;-\rho(e)\circ\Gamma\;
\end{equation}
and this two relations agree with the following general definition. 
\begin{definition}[Odd symmetric gradation]
Let $\bb{E}\to X$ be a rank $m$ complex vector bundle over $X$.   A \emph{gradation} of ${\bb{E}}$ is a $\Gamma\in {\rm End}({\bb{E}})$
such that $\Gamma^2={\rm Id}_{{\bb{E}}}$. The two endomorphisms $\Pi_\pm\in {\rm End}({\bb{E}})$ defined by 
\begin{equation}\label{eq:iso_cliff3}
\Pi_\pm\;:=\;\frac{1}{2}\Big({\rm Id}_{ {\bb{E}}}\;\pm\;\Gamma\Big)
\end{equation}
are idempotent, \ie $\Pi_\pm^2=\Pi_\pm$. 
Let $X$ be as in Assumption \ref{ass:1}.
This assures that the numbers $n_{\pm}:={\rm dim}\;{\rm Ker}(\Pi_\pm|_x)$ do not depend of the choice of $x\in X$ (see \eg the argument in
\cite[Theorem 2.10]{gracia-varilly-figueroa-01}) and $n_++n_-=m$. 
The numbers $n_\pm$ are called the \emph{indeces} of $\Gamma$.
The two vector subbundles $\bb{E}_\pm:={\rm Ker}(\Pi_\pm)$ \cite[Chapter 3, Theorem 8.2]{husemoller-94} provide a splitting 
$$
\bb{E}\;=\;\bb{E}_+\;\oplus\;\bb{E}_-\;.
$$
The gradation $\Gamma$ is called \emph{symmetric} when $n_+=n_-$ (which implies that $m$ has to be even). If $(\bb{E},\rho)$ is a 
Clifford vector bundle of type $(p,q)$
 we say that $\Gamma$ is an \emph{odd}  gradation for $(\bb{E},\rho)$ if 
in addition $\Gamma\circ \rho(e_i)=-\rho(e_i)\circ\Gamma$ for all $1\leqslant i\leqslant p+q$.
\end{definition}

\begin{lemma}\label{lemma:chi_cliff1}
Let $(\hat{\bb{E}},\rho)$ be a rank $2m$ Clifford vector bundle of type $(0,1)$ over a space $X$ that verifies Assumption \ref{ass:1}.
Let $\Gamma\in {\rm End}(\hat{\bb{E}})$ be an odd symmetric gradation and $\bb{E}_\pm:={\rm Ker}(\Pi_\pm)$  the two rank $m$ vector bundles defined by the idempotents \eqref{eq:iso_cliff3}. Then $\bb{E}_+$ and $\bb{E}_-$ are isomorphic as complex vector bundles and there is an isomorphisms $\Theta:\bb{E}_-\to \bb{E}_+$ such that
\begin{equation}\label{eq:iso_cliff1_biss}
\hat{\bb{E}}\;=\;\bb{E}_+\;\oplus\; \bb{E}_-\;,\qquad\qquad \rho(e)\;=\;\left(\begin{array}{cc} 0 & \Theta \\  \Theta^{-1} & 0 \\ \end{array}\right)\;.
\end{equation}
Moreover, each pair of isomorphisms $h_\pm:\bb{E}_\pm\to \bb{E}$ defines a rank $m$ chiral vector bundle $(\bb{E}, \Phi_h)$ with  $\Phi_h:=h_+\circ\Theta\circ h^{-1}_-$.
\end{lemma}
\proof
The first part  follows  by 
observing that $\rho(e)\circ\Pi_\pm\circ\rho(e)=\Pi_\mp$
which implies that
$\Pi_+\circ\rho(e)\circ\Pi_-$ is an endomorphism of $\hat{\bb{E}}$ which restricts to a
linear
 isomorphism  between corresponding fibers of $\bb{E}_-$ and $\bb{E}_+$. This assures by \cite[Chapter 3, Theorem 2.5]{husemoller-94} that the restriction of $\Pi_+\circ\rho(e)\circ\Pi_-$ defines an isomorphism $\Theta:\bb{E}_-\to \bb{E}_+$ with inverse given by $\Pi_-\circ\rho(e)\circ\Pi_+$. The second part of the claim is just a consequence of the fact that $\Phi_h$ is  an endomorphisms of $\bb{E}$ by construction.
\qed

\medskip

\begin{remark}\label{rk:cliff_chiral}{\upshape 
Lemma  \ref{lemma:chi_cliff1} deserves some comments.
\begin{itemize}
\item[(a)] There is a natural notion of isomorphism 
between two Clifford  vector bundles of type $(0,1)$
endowed with symmetric gradations
$(\hat{\bb{E}},\rho, \Gamma)$ and
$(\hat{\bb{E}}',\rho',\Gamma')$. This is given by an  isomorphism of complex vector bundles $\hat{f}:\hat{\bb{E}}\to \hat{\bb{E}}'$ such that $\hat{f}\circ\Gamma=\Gamma'\circ\hat{f}$ and $\hat{f}\circ\rho(e)=\rho'(e)\circ\hat{f}$. In terms of the splitting \eqref{eq:iso_cliff1_biss} this is equivalent to the existence of two isomorphisms $f_\pm:\bb{E}_\pm\to \bb{E}_\pm'$ such that
\begin{equation}\label{eq:iso_cliff0001}
\hat{f}\;=\;\left(\begin{array}{cc} f_+ & 0 \\  0 & f_- \\ \end{array}\right)\;, \qquad\qquad 
f_+\circ\Theta\;=\; \Theta'\circ f_-\;.
\end{equation}

\vspace{1.3mm}

\item[(b)] There is an asymmetry between Lemma \ref{lem:iso_cliff1} and Lemma \ref{lemma:chi_cliff1}. The first proves that a chiral vector bundle  uniquely specifies (up to isomorphisms) a Clifford vector bundle of type $(0,1)$ endowed with an odd symmetric gradation \eqref{eq:iso_cliff020}. Conversely, Lemma  \ref{lemma:chi_cliff1} shows that in order to reconstruct a chiral vector bundle $(\bb{E}_h,\Phi_h)$
 from a Clifford  vector bundle of type $(0,1)$ with odd symmetric gradation $(\hat{\bb{E}},\rho, \Gamma)$
 one needs to make a choice of a pair of isomorphisms $h_\pm:\bb{E}_\pm\to\bb{E}$ to a \emph{target} vector bundle $\bb{E}$. However, the new Clifford  vector bundle of type $(0,1)$ with odd symmetric gradation $(\hat{\bb{E}}_h,\rho_h, \Gamma_h)$
obtained from $(\bb{E}_h,\Phi_h)$ by following the construction in Lemma \ref{lem:iso_cliff1} turns out to be isomorphic (in the sense of point (a) above) to the original 
triplet $(\hat{\bb{E}},\rho, \Gamma)$ through the isomorphism 
$$
\hat{h}: (\hat{\bb{E}},\rho, \Gamma)\;\to\; (\hat{\bb{E}}_h,\rho_h, \Gamma_h)\;,\qquad\quad \hat{h}\;=\;\left(\begin{array}{cc} h_+ & 0 \\  0 & h_- \\ \end{array}\right)\;.
$$ 
\vspace{1.3mm}

\item[(c)]
The construction of the $\chi$-bundle $(\bb{E}_h, \Phi_h)$ is based  on the choice of a pair of isomorphisms $h_\pm:\bb{E}_\pm\to\bb{E}$. However,  only
the \emph{difference} between $h_+$ and $h_-$ is really important.
More precisely, let us consider the  pair of isomorphisms $h_\pm^{(+)}:\bb{E}_\pm\to\bb{E}_+$ 
defined by $h_+^{(+)}:={\rm Id}_{\bb{E}_+}$ and 
$h_-^{(+)}:=h_+^{-1}\circ h_-$ and the related chiral vector bundle $(\bb{E}_+, \Phi_{h_+})$
with $\Phi_{h_+}:=\Theta\circ h_-^{-1}\circ h_+$. Then, $(\bb{E}_+, \Phi_{h_+}) \approx (\bb{E}, \Phi_h)$ as chiral vector bundles with isomorphism given by the map $h_+$. Similarly, one can consider the pair of maps 
$h_\pm^{(-)}:\bb{E}_\pm\to\bb{E}_-$ 
defined by $h_+^{(-)}:= h_-^{-1}\circ h_+$ and 
$h_-^{(-)}:={\rm Id}_{\bb{E}_-}$ and the associated chiral vector bundle $(\bb{E}_-, \Phi_{h_-})$. Then the map $h_-$ provides a $\chi$-bundle isomorphism $(\bb{E}_-, \Phi_{h_-}) \approx (\bb{E}, \Phi_h)$. Finally, by transitivity one gets
$$
(\bb{E}_+, \Phi_{h_+}) \;\approx \; (\bb{E}, \Phi_h)\;\approx \; (\bb{E}_-, \Phi_{h_-})\;.  
$$
In conclusion, in order to realize a $\chi$-bundle
from a triplet $(\hat{\bb{E}},\rho, \Gamma)$ one only needs to specify a \emph{reference} map $h_{\rm ref}:\bb{E}_-\to \bb{E}_+$ and to consider the associated \emph{standard} $\chi$-bundle $(\bb{E}_-,\Phi)$ with $\Phi:= h_{\rm ref}^{-1}\circ \Theta$. Let $h_{\rm ref}':\bb{E}_-\to \bb{E}_+$ be a second reference map. Then $(\bb{E}_-,\Phi)\approx (\bb{E}_-,\Phi')$ as chiral vector bundles if and only if there is a $f\in{\rm Aut}(\bb{E}_-)$ such that $\Phi'=f\circ \Phi\circ f^{-1}$, namely if and only if
$$
h_{\rm ref}'\;=\;\Theta\circ f \circ\Theta^{-1}\circ h_{\rm ref}\circ f^{-1}
$$
This equation shows that the freedom in the choice of $h_{\rm ref}$ is measured by ${\rm Aut}(\bb{E}_-)$.

\vspace{1.3mm}

\item[(d)]
Consider now two of isomorphic triplets 
$(\hat{\bb{E}},\rho, \Gamma)$ and $(\hat{\bb{E}}',\rho', \Gamma')$ with isomorphism $\hat{f}$ as in \eqref{eq:iso_cliff0001}. Let $h_{\rm ref}:\bb{E}_-\to \bb{E}_+$ and 
$h_{\rm ref}':\bb{E}'_-\to \bb{E}'_+$ be two reference maps and $(\bb{E}_-,\Phi)$ and $(\bb{E}'_-,\Phi')$
the two associated {standard} $\chi$-bundles
described in (c). 
If it happens that
\begin{equation}\label{eq:iso_cliff0002}
f_+\circ h_{\rm ref}\;=\; h_{\rm ref}'\circ f_-
\end{equation}
then 
$f_-:(\bb{E}_-,\Phi)\to(\bb{E}'_-,\Phi')$
is an isomorphism  of chiral vector bundles. Indeed
the condition $\Phi=f_-^{-1}\circ\Phi'\circ f_-$ follows by exploiting the relations \eqref{eq:iso_cliff0001} and
\eqref{eq:iso_cliff0002}. This observation is, to some extent, an inverse of the content of Lemma
\ref{lem:iso_cliff1}.

\vspace{1.3mm}

\item[(e)]
The effect of the {reference} map $h_{\rm ref}:\bb{E}_-\to \bb{E}_+$ on the triplet $(\hat{\bb{E}},\rho, \Gamma)$ can be also  understood as the introduction of a second  $C\ell^{0,1}$ Clifford-action
\begin{equation}\label{eq_rho_ref}
\rho_{\rm ref}(e)\;=\;\left(\begin{array}{cc} 0 & h_{\rm ref} \\  h_{\rm ref}^{-1} & 0 \\ \end{array}\right)
\end{equation}
which evidently anti-commute with the gradation $\Gamma$.
\end{itemize}
The sequence (a)-(e) of the above  remarks suggests a strong relation between $\chi$-bundles and 
Clifford vector bundle with a \emph{double} $C\ell^{0,1}$-action endowed with an odd symmetric gradation.
}\hfill $\blacktriangleleft$
\end{remark}

\medskip

Let us consider quadruplets $(\hat{\bb{E}},\rho_1,\rho_2, \Gamma)$ given by a rank $2m$ complex vector bundle
$\hat{\bb{E}}$,  an odd symmetric gradation  $\Gamma$ and two Clifford actions
$\rho_i:C\ell^{0,1}\to {\rm End}(\hat{\bb{E}})$, $i=1,2$, which anti-commute with the gradation, $\Gamma\circ \rho_i(e)=-\rho_i(e)\circ\Gamma$. An isomorphism between two quadruplets $(\hat{\bb{E}},\rho_1,\rho_2, \Gamma)$ and $(\hat{\bb{E}}',\rho'_1,\rho'_2, \Gamma')$ is an isomorphism $\hat{f}:\hat{\bb{E}}\to\hat{\bb{E}}'$ which intertwines  the gradations $\hat{f}\circ\Gamma=\Gamma'\circ\hat{f}$ and the $C\ell^{0,1}$-actions $\hat{f}\circ\rho_i(e)=\rho'_i(e)\circ\hat{f}$.
Lemma \ref{lem:iso_cliff1}, Lemma
\ref{lemma:chi_cliff1} and the various items (a)-(e) in Remark \ref{rk:cliff_chiral} can be summarized in the following claim.
\begin{proposition}\label{propo:iso_cliff_chiral}
There is a one-to-one correspondence between isomorphism classes  of rank $m$ chiral vector bundles  and isomorphism  classes of rank $2m$ Clifford vector bundles with a {double} $(0,1)$ structure and odd symmetric gradation.
\end{proposition}

\medskip

Finally, few words about the homotopy.
The notion of homotopy equivalence 
between quadruplets $(\hat{\bb{E}},\rho_1,\rho_2, \Gamma)$ and $(\hat{\bb{E}}',\rho'_1,\rho'_2, \Gamma')$ over the space $X$ can be introduced by  mimicking Definition \ref{def:eq_rel_chi_VB}. We say that $(\hat{\bb{E}},\rho_1,\rho_2, \Gamma)\sim (\hat{\bb{E}}',\rho'_1,\rho'_2, \Gamma')$ if and only if there is an intertwining quadruplet $(\tilde{\bb{E}},\tilde{\rho}_1,\tilde{\rho}_2, \tilde{\Gamma})$
 over $X\times[0,1]$ such that
$$
\begin{aligned}
\left(\hat{\bb{E}},\rho_1,\rho_2, \Gamma\right)\;&\simeq\; \left(\tilde{\bb{E}}|_{X\times \{0\}},\tilde{\rho}_1|_{X\times \{0\}},\tilde{\rho}_2|_{X\times \{0\}}, \tilde{\Gamma}|_{X\times \{0\}}\right)\\
\left(\hat{\bb{E}}',\rho_1',\rho_2', \Gamma'\right)\;&\simeq\; \left(\tilde{\bb{E}}|_{X\times \{1\}},\tilde{\rho}_1|_{X\times \{1\}},\tilde{\rho}_2|_{X\times \{1\}}, \tilde{\Gamma}|_{X\times \{1\}}\right)\\
\end{aligned}
$$
where $\simeq$ denotes the isomorphism between
Clifford vector bundles with a {double} $(0,1)$ structure and odd symmetric gradation.
The following result is an immediate consequence of Proposition \ref{propo:iso_cliff_chiral} and the notion of homotopy.

\begin{theorem}
There is a one-to-one correspondence between homotopic classes $[(\bb{E},\Phi)]$ of rank $m$ chiral vector bundles  and homotopic  classes $[(\hat{\bb{E}},\rho_1,\rho_2, \Gamma)]$ of rank $2m$ Clifford vector bundles with a {double} $C\ell^{0,1}$-action and odd symmetric gradation.
\end{theorem}

\subsection{Construction of the chiral Bloch-bundle}
\label{sec:bloc_bund_top_phas}

A standard construction associates to each  topological quantum system
\eqref{eq:tqsA1} with an {isolated family} of $m$ energy bands \eqref{eq:tqsA2} a complex vector bundle  of rank $m$ over $X$ which, 
to some extent, can be named \emph{Bloch-bundle}\footnote{This name is justified by the fact that 
the \emph{Bloch theory} for electrons in a crystal  (see \eg \cite{ashcroft-mermin-76}) provides 
some of the most interesting
examples of topological quantum systems.}. 
The first step of this construction is the realization of a continuous map of rank $m$ spectral projections $X\ni x\mapsto P(x)$
by means of the \emph{Riesz-Dunford integral}:
 \begin{equation}\label{eq:proj}
P_\Omega(x)\;:=\;\frac{\ii}{2\pi}\int_{\s{C}}\dd z\;\Big(H(x)-z\;\n{1}_{\s{H}}\Big)^{-1}
\end{equation}
associated with any isolated family of energy bands
$\Omega$ which verifies
\eqref{eq:tqsA2}. Here $\s{C}\subset\C$ is any regular closed path which encloses, without touching, the spectral range $\Omega$. 
  The second step
 turns out to be a concrete application of the \emph{Serre-Swan Theorem} \cite{swan-62} (see also \cite[Theorem 2.10]{gracia-varilly-figueroa-01}) which relates vector bundles and continuous family of projections by
\begin{equation}\label{eq:tot_spac}
\hat{\bb{E}}_\Omega\;:=\; \bigsqcup_{x\in X}\; {\rm Ran}\; {P}_\Omega(x)\;.
\end{equation}
Sometimes, $x\mapsto P_\Omega(x)$ is equivariant with respect to a \emph{free} action of a topological group $\n{G}$ over $X$. In this case a quotient procedure of the type \cite[Proposition 1.6.1]{atiyah-67} is required. For instance,  this kind of construction is necessary for the Harper operator \cite{denittis-landi-12} which describes the strong magnetic filed limit of the Landau operator.

\medskip

 By borrowing  the accepted terminology for topological insulators we can refer to systems 
which verify only \eqref{eq:tqsA1} and \eqref{eq:tqsA2}
 as  topological quantum systems
 in class \emph{{\bf A}} (see \eg \cite{schnyder-ryu-furusaki-ludwig-08}). In this case the (Cartan) label {\bf A} expresses  the absence of any kind of (pseudo-)symmetry or other extra structures.
Since topological quantum systems
 in class {\bf A}
  lead to complex vector bundles (without any other extra structure) they are topologically classified by the set ${\rm Vec}_\C^m(X)$ of  the  isomorphism classes of rank $m$ complex vector bundles over the base space $X$.  The classification of complex vector bundles is a classical, well-studied, problem in topology. Under rather general assumptions on the base space $X$, the set ${\rm Vec}_\C^m\big(X\big)$ can be classified by using homotopy theory techniques and, for dimension $d\leqslant 4$ (and for all $m$) a complete description can be done in terms of cohomology groups and Chern classes \cite{peterson-59}  (see \eg \cite[Section 3]{denittis-gomi-14} for a review of these standard results).

\medskip

The presence of a chiral symmetry \eqref{eq:tqsA3} introduces more structure. The first consequence is that the spectrum of the operator $H(x)$ is symmetric with respect to the zero energy, \ie    
\begin{equation}
+\lambda(x)\;\in\;\sigma\Big(H(x)\Big)\qquad\Leftrightarrow\qquad-\lambda(x)\;\in\;\sigma\Big(H(x)\Big)\qquad\quad \forall\ x\in X\;.
\end{equation}
Indeed, if $\psi(x)\in\s{H}$ is an eigenvector of
$H(x)$ with eigenvalue $\lambda(x)$ then $\psi'(x):=\chi(x)\psi(x)$ is still an eigenvector but with eigenvalue $-\lambda(x)$. This symmetry justifies  the following assumption which is usually verified in the most common physical situations.

 \begin{assumption}[Zero energy gap\footnote{{This is the standard work assumption used by physicists working on topological phases for
BdG Hamiltonians. In any case Assumption 6.2 can be
weakened by requiring the existence of a family of strictly positive (eq. negative) bands.}}]\label{ass:2}
Let $X\ni x\mapsto H(x)$ be a topological quantum systems with chiral symmetry $X\ni x\mapsto \chi(x)$ as in Definition \ref{def:tsq_Chi}. We will assume that $H(x)$ has an energy gap in zero, \ie
$0\notin\sigma(H(x))$ for all $x\in X$.
\end{assumption}

\noindent
Due to this assumption we can select an isolated family of energy bands of type $\Omega=\Omega_-\cup\Omega_+$ where the sets $\Omega_\pm:=\{\lambda_{\pm j_1}(\cdot),\ldots,\lambda_{\pm j_m}(\cdot)\}$ contain energy bands of fixed sign
and pair of bands $\lambda_{+ j_i},\lambda_{- j_i}$ are related by the chiral symmetry $\chi$. We refer to 
$\Omega_+$ and $\Omega_-$ as the \emph{positive} and \emph{negative energy sector}, respectively. The spectral projection associated to $\Omega$ by \eqref{eq:proj} splits accordingly
in a positive and a negative part
$$
P_\Omega(x)\;=\;P_{\Omega_+}(x)\;+\;P_{\Omega_-}(x)\;,\qquad\quad P_{\Omega_\pm}(x)\; P_{\Omega_\mp}(x)\;=\;0
$$
since the zero energy gap assumption. 
Moreover, the chiral symmetry intertwines the two energy sectors,
\begin{equation}\label{eq:split0}
\chi(x)\; P_{\Omega_\pm}(x)\; \chi(x)\;=\; P_{\Omega_\mp}(x)\;.
\end{equation}
The splitting of $P_\Omega$ induces a splitting of the Bloch-bundle
\begin{equation}\label{eq:split1}
\hat{\bb{E}}_\Omega\;=\;{\bb{E}}_{\Omega_+}\;\oplus\; {\bb{E}}_{\Omega_-} \;,\qquad\quad {\bb{E}}_{\Omega_\pm}\;:=\; \bigsqcup_{x\in X}\; {\rm Ran}\; {P}_{\Omega_\pm}(x)\;
\end{equation}
and the relation \eqref{eq:split0} implies that the two summands  are isomorphic as complex vector bundles, ${\bb{E}}_{\Omega_+}\simeq {\bb{E}}_{\Omega_-}$.
The operator
\begin{equation}\label{eq:bloch_gradu}
\Gamma_\Omega(x)\;:=\;P_\Omega(x)\;\chi(x)\; P_\Omega(x)\;,\qquad\quad \Gamma_\Omega(x)^2 \;=\;P_\Omega(x)\;
\end{equation}
has the structure of an involution on the space 
${\rm Ran} {P_\Omega}(x)$ and so it identifies a gradation 
 $\Gamma_\Omega\in{\rm End}(\hat{\bb{E}}_\Omega)$, $\Gamma_\Omega^2={\rm Id}_{\hat{\bb{E}}_\Omega}$,
of the rank $2m$ Bloch-bundle \eqref{eq:split1}. This gradation 
 turns out to be symmetric.
\begin{lemma}\label{lemma:gradiation_splitting}
Let $\Gamma_\Omega$ be the gradation 
 of the 
Bloch-bundle $\hat{\bb{E}}_\Omega\to X$ defined fiberwise by \eqref{eq:bloch_gradu}. Define the 
 two idempotent  endomorphisms  $\Pi_{\Omega,\pm}\in {\rm End}(\hat{\bb{E}}_\Omega)$ as in \eqref{eq:iso_cliff3}. Then
  ${\rm dim}\;{\rm Ker}(\Pi_{\Omega,\pm})=m$
 showing that $\Gamma_\Omega$  is a symmetric gradation.
\end{lemma}
\proof
Since $\Pi_{\Omega,\pm}$ are idempotent  endomorphisms of $\hat{\bb{E}}_\Omega$, the dimension of their kernels are locally constant on $X$
(see \eg the argument in
\cite[Theorem 2.10]{gracia-varilly-figueroa-01}). 
The connectedness of $X$
 assures that  the dimension of the kernels are globally constant. This means that we can make the computation just by looking at a single fiber over a point  $x\in X$. Here the idempotents $\Pi_{\Omega,\pm}$ have the  expression
\begin{equation}\label{eq:proj_to_proj}
\Pi_{\Omega,\pm}(x)\;=\;\frac{1}{2}\Big(P_\Omega(x)\;\pm\;P_\Omega(x)\;\chi(x)\; P_\Omega(x)\Big)\;.
\end{equation}
Let $\{\psi_{-j}\}$, $j=1,\ldots,m$, be a basis for the negative energy sector eigenspace $P_{\Omega_-}(x)\s{H}$ and $\{\psi_{+j}\}$ the related basis for the positive energy sector eigenspace $P_{\Omega_+}(x)\s{H}$. The relation between the two bases is fixed by $\psi_{\pm j}=\chi(x) \psi_{\mp j}$. Let $\{\phi_{\pm j}\}$, $j=1,\ldots,m$, be the new basis  for the full space $P_{\Omega}(x)\s{H}$ given by
\begin{equation}\label{eq:chi_basis}
\phi_{\pm j}\;:=\;\frac{1}{2}\Big(\psi_{+j}\;\pm\; \psi_{-j}\Big)
\end{equation}
Since $\chi(x)\phi_{\pm j}=\pm\phi_{\pm j}$ and $P_\Omega(x)\phi_{\pm j}=\phi_{\pm j}$ one  checks that $\Pi_{\Omega,\pm}(x)\phi_{\pm j}=\phi_{\pm j}$ and $\Pi_{\Omega,\pm}(x)\phi_{\mp j}=0$. This concludes the proof.
\qed

\medskip

The evident relations $P(x)= \Pi_{\Omega,+}(x)+\Pi_{\Omega,-}(x)$ and $\Pi_{\Omega,\pm}(x)\Pi_{\Omega,\mp}(x)=0$ imply a different splitting of the Bloch-bundle
\begin{equation}\label{eq:split2}
\hat{\bb{E}}_\Omega\;=\;{\bb{E}}_+\;\oplus\; {\bb{E}}_- \;,\qquad\quad {\bb{E}}_\pm\;:=\; \bigsqcup_{x\in X}\; {\rm Ran}\; \Pi_{\Omega,\pm}(x)\;.
\end{equation}
The \emph{flattened} (restricted) Hamiltonian
\begin{equation}\label{eq:split2bis}
\rho_\Omega(x)\;:=\;P_\Omega(x)\;\frac{H(x)}{|H(x)|}\;P_\Omega(x)=\; P_{\Omega_+}(x)\;-\;P_{\Omega_-}(x)
\end{equation}
turns out to be well-defined due to Assumption \ref{ass:2} and the two relations
\begin{equation}\label{eq:split3}
\rho_\Omega(x)^2\;=\;P_\Omega(x)\;,\qquad\quad \Gamma_\Omega(x)\;\rho_\Omega(x)\;=\;- \rho_\Omega(x)\;\Gamma_\Omega(x)
\end{equation}
prove that the operator $\rho_\Omega$ introduces a $C\ell^{0,1}$-action on the total Bloch-bundle $\hat{\bb{E}}_\Omega$ and the gradation 
 $\Gamma_\Omega$ is odd with respect to this action.
The relations
$$
\rho_\Omega(x)\;\Pi_{\Omega,\pm}(x)\;\rho_\Omega(x)\;=\; \Pi_{\Omega,\mp}(x)\;,\qquad\quad \Gamma_\Omega(x)\;=\;\Pi_{\Omega,+}(x)\;-\;\Pi_{\Omega,-}(x)
$$
imply the isomorphism ${\bb{E}}_+\simeq {\bb{E}}_-$ of the complex vector bundles in the splitting \eqref{eq:split2}. Moreover, with respect to such a gradation   the Clifford action $\rho_\Omega(e)$ and the gradation 
 $\Gamma_\Omega$ are represented as
$$
\Gamma_\Omega\;=\;\left(\begin{array}{cc} +{\rm Id}_{{\bb{E}}_+} & 0 \\  0 & -{\rm Id}_{{\bb{E}}_-} \\ \end{array}\right)\;,\qquad\quad\rho_\Omega(e)\;=\;\left(\begin{array}{cc} 0 & \Theta_\chi \\  \Theta_\chi^{-1} & 0 \\ \end{array}\right)\;.
$$
 The isomorphism $\Theta_\chi:{\bb{E}}_-\to {\bb{E}}_+$ and its inverse $\Theta_\chi^{-1}$ are described fiberwise by
$$
\begin{aligned}
\Theta_\chi(x)\;&:=\;\Pi_{\Omega,+}(x)\;\rho_\Omega(x)\;\Pi_{\Omega,-}(x)\\
\Theta_\chi^{-1}(x)\;&:=\;\Pi_{\Omega,-}(x)\;\rho_\Omega(x)\;\Pi_{\Omega,+}(x)\;.
\end{aligned}
$$

\medskip

Summarizing, up to now
we had  shown how from a topological quantum system with chiral symmetry and zero energy gap 
one can associate a rank $2m$ Bloch-bundle $\hat{\bb{E}}_\Omega$
with a $C\ell^{0,1}$-action $\rho_\Omega$ and a symmetric odd gradation 
 $\Gamma_\Omega$, provided that there exists a family of $m$ negative (resp. positive) energy bands $\Omega_-$ (resp. $\Omega_+$) separated from the rest of the spectrum. 
However, as discussed at the end of Section \ref{sect:chi_VB_clifford}, the data contained in the triplet $(\hat{\bb{E}}_\Omega,\rho_\Omega, \Gamma_\Omega)$ are not enough alone to determine a chiral vector bundle.
We need a second \emph{reference} isomorphism $h_{\rm ref}:{\bb{E}}_-\to {\bb{E}}_+$ or equivalently a second  $C\ell^{0,1}$-action $\rho_{\rm ref}$ compatible with the gradation $\Gamma_\Omega$. Once such an $h_{\rm ref}$ has been given we can uniquely specify the $\chi$-bundle $({\bb{E}}_-, \Phi_{\chi})$ with $\Phi_{\chi}:=h_{\rm ref}^{-1}\circ \Theta_\chi$. This construction justifies the Assumption \ref{defi:top_phases} given in the introduction.

\medskip

\begin{remark}\label{rk:cliff_chiral_bis}{\upshape 
In order to  complete the picture of the relation between topological quantum systems with chiral symmetry and $\chi$-bundles
a couple of remarks  are still necessary.

\begin{itemize}
\item[(a)]The choice of the {reference} isomorphism $h_{\rm ref}$ corresponds to the choice of a (local) system of coordinates for the vector bundles $\pi_\pm:{\bb{E}}_{\pm}\to X$. As a matter of fact
 $\bb{E}_-$ and $\bb{E}_+$ are isomorphic hence they can be trivialized on the same open covering $\{\f{U}_\alpha\}$ of $X$. Let $h_{\alpha,\pm}:\pi^{-1}_\pm(\f{U}_\alpha)\to \f{U}_\alpha\times \C^m$ be a choice of a local trivialization.  On the overlappings $\f{U}_\alpha\cap \f{U}_\beta$ both systems of local trivializations transform with the same system of transition functions $g_{\beta \alpha}:\f{U}_\alpha\cap \f{U}_\beta\to\n{U}(m)$
according to $h_{\beta,\pm}=g_{\beta \alpha}\circ h_{\alpha,\pm}$. This implies that the collection of local isomorphisms $h_{\alpha,{\rm ref}}:= h_{\alpha,+}^{-1}\circ h_{\alpha,-}: \pi^{-1}_-(\f{U}_\alpha)\to \pi^{-1}_+(\f{U}_\alpha)$ \virg{glues} together and defines a global isomorphism $h_{{\rm ref}}:{\bb{E}}_{-}\to {\bb{E}}_{+}$. 
  \vspace{1.3mm}
\item[(b)] The Clifford $C\ell^{0,1}$-action $\rho_{\rm ref}$ associated to the reference isomorphisms $h_{\rm ref}$ according to \eqref{eq_rho_ref} has fiberwise the expression
$$
\rho_{\rm ref}(x)\;=\;\Pi_{\Omega,+}(x)\;h_{{\rm ref}}(x)\;\Pi_{\Omega,-}(x)\;+\;\Pi_{\Omega,-}(x)\;h_{{\rm ref}}^{-1}(x)\;\Pi_{\Omega,+}(x)\;.
$$
After expressing the projections $\Pi_{\Omega,\pm}(x)$ in terms of the Fermi projection $P_\Omega(x)$ according to \eqref{eq:proj_to_proj}
one gets after some algebra
\begin{equation*}
\rho_{\rm ref}(x)\;=\;P_\Omega(x)\;\left(\frac{\n{1}+\chi(x)}{2}\;h_{\rm ref}(x)\; \frac{\n{1}-\chi(x)}{2}\;+\;\frac{\n{1}-\chi(x)}{2}\;h_{\rm ref}^{-1}(x)\; \frac{\n{1}+\chi(x)}{2}\right)\;P_\Omega(x)\;.
\end{equation*}
By choosing the map $x\mapsto h_{\rm ref}(x)$ to be  unitary-valued (this implies no loss of generality) one can see that $\rho_{\rm ref}(x)$ turns out to be a self-adjoint operator acting on the relevant spectral subspace $P_\Omega(x)\s{H}$ and endowed with the (evident) chiral symmetry $\chi(x)\rho_{\rm ref}(x)\chi(x)=-\rho_{\rm ref}(x)$. In this sense $\rho_{\rm ref}(x)$ can be considered as a \emph{reference} chiral Hamiltonian
for the initial chiral Hamiltonian $H(x)$.
\end{itemize}
Some of the observations in (a) and (b) above will acquire a certain importance inside the context of Section \ref{sect:com_lit}. 
}\hfill $\blacktriangleleft$
\end{remark}

We conclude this section with a comparison between the  \emph{energy splitting} \eqref{eq:split1}  and  the  \emph{chiral splitting} \eqref{eq:split2} 
of the Bloch-bundle  $\hat{\bb{E}}_\Omega$. 

\begin{lemma}\label{lemm:iso_ener_bund}
Let ${\bb{E}}_{\Omega_\pm}$ be the two rank $m$ vector bundles which provide the
energy splitting  \eqref{eq:split1} of the Bloch-bundle  $\hat{\bb{E}}_\Omega$. Similarly, let  
${\bb{E}}_{\pm}$ be the two rank $m$ vector bundles which provide the
chiral splitting  \eqref{eq:split1} of $\hat{\bb{E}}_\Omega$. Then one has
$$
{\bb{E}}_{\Omega_-}\;\simeq\;{\bb{E}}_{\Omega_+}\;\simeq\;{\bb{E}}_{+}\;\simeq\;{\bb{E}}_{-}
$$
where all the isomorphisms are meant in the category of complex vector bundles.
\end{lemma}
\proof
The isomorphisms ${\bb{E}}_{\Omega_-} \simeq {\bb{E}}_{\Omega_+}$ and ${\bb{E}}_{+} \simeq {\bb{E}}_{-}$ have already been discussed in the text. At any rate all the isomorphisms can be checked directly by showing that all the vector bundles have a same system of transition functions $g_{\beta \alpha}$ subordinate to an open covering $\{\f{U}_\alpha\}$ of the base space $X$  \cite[Chapter 5, Theorem 2.7]{husemoller-94}. Let $\f{U}_\alpha$ and $\f{U}_\beta$ be
two elements of the covering with non trivial intersection $\f{U}_\alpha\cap\f{U}_\beta\neq\emptyset$. Let  $\{\psi^{(\alpha)}_{+j}\}$ and $\{\psi^{(\beta)}_{+j}\}$, $j=1,\ldots,m$ be two local frames for ${\bb{E}}_{\Omega_+}$ supported on $\f{U}_\alpha$ and $\f{U}_\beta$, respectively, and related by $\psi^{(\beta)}_{+j}=\sum_{k=1}^mg_{\beta \alpha}^{j,k}\; \psi^{(\alpha)}_{+k}$
on the intersection $\f{U}_\alpha\cap\f{U}_\beta$.
The local systems $\{\psi^{(\alpha)}_{-j}:=\chi\psi^{(\alpha)}_{+j}\}$ and $\{\psi^{(\beta)}_{-j}:=\chi\psi^{(\beta)}_{+j}\}$ provide a local trivialization for ${\bb{E}}_{\Omega_-}$. On the  
intersection $\f{U}_\alpha\cap\f{U}_\beta$ we have that
$$
\psi^{(\beta)}_{-j}\;=\; \chi\;\psi^{(\beta)}_{+j}\;=\;\chi\left( \sum_{k=1}^mg_{\beta \alpha}^{j,k}\; \psi^{(\alpha)}_{+k}\right)\;=\;
\sum_{k=1}^mg_{\beta \alpha}^{j,k}\; \left(\chi\psi^{(\alpha)}_{+k}\right)\;=\; \sum_{k=1}^mg_{\beta \alpha}^{j,k}\; \psi^{(\alpha)}_{-k}
$$
and this shows that $g_{\beta \alpha}$ provides a system of transition functions also for ${\bb{E}}_{\Omega_-}$. Local frames for ${\bb{E}}_{+}$ and ${\bb{E}}_{-}$ can be realized as linear combinations of $\{\psi^{(\beta)}_{+j}\}$ and $\{\psi^{(\beta)}_{-j}\}$ according to equation \eqref{eq:chi_basis}. Since $\{\psi^{(\beta)}_{+j}\}$ and $\{\psi^{(\beta)}_{-j}\}$ transform according to the same system of 
transition functions also their linear combinations will transform according to the same system just by  linearity. 
\qed

\subsection{Comparison with the literature}\label{sect:com_lit}
 The aim of this section is to  compare  our 
 point of view about the topological phases of chiral topological quantum systems (as given in Definition \ref{defi:top_phases}) with various aspects typically  accepted in the literature
concerning the topology of the chiral class {\bf AIII}.
\medskip

\noindent
{\bf Topological insulators of class AIII.} We will briefly sketch the usual derivation of the topological invariants for  topological insulators of class {\bf AIII } as presented in the classical literature on the subject \cite{schnyder-ryu-furusaki-ludwig-08,ryu-schnyder-furusaki-ludwig-10}
as well as  in more recent reviews \cite{qi-zhang-11,budich-trauzettel-13} or new research works \cite{prodan-schulz-baldes-14}.

\medskip

One starts with a  matrix-valued map of the type
\begin{equation}\label{eq:lit1}
X\;\ni\; x\;\longmapsto H(x)\;=\;H(x)^* \;\in\;{\rm Mat}_{\C}(2m)
\end{equation}
where $X$ is usually a $d$-dimensional sphere $\n{S}^d$ (free fermions system) or a  $d$-dimensional torus $\n{T}^d$. Then one assumes the existence of a \emph{constant} involutive unitary matrix $\Gamma\in {\rm Mat}_{\C}(2m)$ such that 
\begin{equation}\label{eq:lit2}
\Gamma\; H(x)\;+\; H(x)\; \Gamma\;=\;0\;,\qquad\quad \Gamma\;=\;\Gamma^*\;=\;\Gamma^{-1}\;.
\end{equation}
The existence of a gap at zero energy (in the same spirit of Assumption \ref{ass:2})
assures that the spectrum of $H(x)$ is symmetric with respect to the zero energy  with equal number of positive and negative eigenvalues $n_+ = n_- =m$. The gradation 
 $\Gamma$ turns out to be symmetric and, up to a choice of a splitting $\C^{2m}=\s{H}_+\oplus\s{H}_-$ (which does \emph{not} fix uniquely a basis!),  it is represented by
\begin{equation}\label{eq:lit3}
\Gamma\; =\; \left(\begin{array}{cc}+\n{1}_ m& 0 \\0 & -\n{1}_m\end{array}\right)
\;
\end{equation}
where $\n{1}_m$ is the $m\times m$ identity matrix. At this point one considers  the flattened Hamiltonian
\begin{equation}\label{eq:lit4}
\rho(x)\; =\; \frac{H(x)}{|H(x)|}\;=\;P_{\Omega_+}(x)\;-\;P_{\Omega_-}(x)
\end{equation}
which is well defined due to  the zero gap condition. The spectral projections $P_{\Omega_\pm}(x)$ provide pointwise a splitting of the ambient space $\C^{2m}$ into the positive and negative energy sector. One of the consequences of \eqref{eq:lit4} is that 
\begin{equation}\label{eq:lit5}
\rho(x)^2\; =\; \n{1}_{2m}\;.
\end{equation}
With respect to the basis which provides the grading 
of $\Gamma$ in \eqref{eq:lit3} one has that 
\begin{equation}\label{eq:lit6}
\rho(x)\; =\; \left(\begin{array}{cc}0& U(x) \\U(x)^* & 0
\end{array}\right)\end{equation}
where the anti-diagonal structure is given by the anti-commutation relation with $\Gamma$ and the presence of the adjoint operator $U(x)^*$ in the lower left corner is due to the self-adjointness of  $\rho(x)$. Finally the constraint \eqref{eq:lit5} implies that $U(x)^*=U(x)^{-1}$ which means that 
$x\mapsto U(x)\in \n{U}(m)$ is a unitary-valued map. At this point one defines the topological phase of the system as
\begin{equation}\label{eq:standard_phase}
[U]\;\in\;[X,\n{U}(m)]\;
\end{equation}
namely as the homotopy class of the map $U:X\to ,\n{U}(m)$. When $X=\n{S}^d$ the phase is classified by $\pi_d(\n{U}(m))$ and $[U]$ coincides with the degree (or winding number) of the map.

\medskip

This derivation seems to be quite close to our arguments in Section \eqref{sec:bloc_bund_top_phas}. However there are  two important differences which make the above approach \emph{less} general than our point of view.

\medskip

\noindent
(1) \emph{- Non-trivial Fermi projection -} 
Model of the type
 \eqref{eq:lit1} with a fiber of finite dimension are usually introduced as effective operators 
which describe the physics related to a relevant energy regime (finite number of filled bands)
of a realistic, generally unbounded, operator. 
 Said differently $H(x)$ has to be interpreted as the restriction $P_\Omega(x) \hat{H}(x) P_\Omega(x)$ where $\hat{H}(x)$ is an operator living in a bigger, possibly infinite dimensional, Hilbert space and $P_\Omega(x)$ is the \emph{Fermi} projection on the relevant set $\Omega=\Omega_++\Omega_-$ of the $m$ positive and $m$ negative energy bands. 
In effect, this is the precise mechanisms which allows to justify the use of   \emph{tight-binding} models
as the representation on the basis of the Wannier functions \cite{wannier-37} of a full  Hamiltonian
 restricted on a finite range of energies (see also \cite{teufel-03,stiepan-teufel-12} for a more modern and general point of view).
The hypothesis  that $H(x)$ is an element of an $x$-independent space  ${\rm Mat}_{\C}(2m)$ is equivalent of assuming that the Fermi projection $P_\Omega(x)$ is constant or that it can be at least continuously deformed to a constant projection. This implies that topology of the Bloch-bundles $\hat{\bb{E}}_\Omega$ associated to $P_\Omega(x)$ is trivial. This fact can be also derived from  the pointwise splitting of the Fermi projection in its positive and negative energy parts $P_\Omega(x)=P_{\Omega_+}(x)+P_{\Omega_-}(x)$. Equations \eqref{eq:lit5} and \eqref{eq:lit4} implies that $\n{1}_{2m}=[P_{\Omega_+}(x)-P_{\Omega_-}(x)]^2=P_\Omega(x)$. Moreover,  
$P_{\Omega_+}(x)$ and $P_{\Omega_-}(x)$ are related by a unitary transformation induced by the gradation 
 $\Gamma$ hence they are topologically equivalent. Since their sum gives a trivial bundle   
they are forced to have order two Chern classes. In many physical situations of interest the base space $X$ has torsion-free integer cohomology groups and this  implies that $P_{\Omega_+}(x)$ and $P_{\Omega_-}(x)$  need to have vanishing Chern classes. 
This is a strong requirement which, for instance, implies  the triviality of $P_{\Omega_+}(x)$ and $P_{\Omega_-}(x)$  in low dimensions $d\leqslant 4$. In conclusion, the definition \eqref{eq:lit1} introduces  from the beginning the \virg{hidden} hypothesis that the system exhibits a collection of $m$ negative (or positive) energy bands with a total trivial Chern charge. 
This is a  hypothesis which is in general \emph{false} {(see \eg the two examples in Section \ref{sec:toy_mos})} and absolutely \emph{non} necessary as showed in Section \ref{sec:bloc_bund_top_phas}.

\medskip

\noindent
(2) \emph{- Arbitrary choice of the eigenbasis of the gradation -} Even if  one accepts the restrictions included in  definition \eqref{eq:lit1}, the construction of the topological invariant for the chiral system based on the decomposition \eqref{eq:lit6} turns out to be ambiguously defined. Indeed, the diagonal form of $\Gamma$ in \eqref{eq:lit3} is preserved by  every (pointwise) reshuffling of the basis of the two eigenspaces $\C^{2m}=\s{H}_+(x)\oplus\s{H}_-(x)$ of  $\Gamma$, namely by the action of any unitary-valued map of the form
$$
x\;\longmapsto\;W(x)\;=\;
\left(\begin{array}{cc} W_1(x)&0 \\
0 & W_2(x)\end{array}\right)
\;,\qquad\quad W_1(x),W_2(x)\in\n{U}(m)\;.
$$ 
Since
$$
\rho(x)\;\longrightarrow W(x)\;\rho(x)\;W(x)^*\;=\;\left(\begin{array}{cc} 0&W_1(x)U(x)W_2(x)^* \\
W_2(x)U(x)^* W_1(x)^*& 0\end{array}\right)
$$
and $[U]\neq[W_1UW_2^*]$ it is evident that the definition of the topological phase is affected by the ambiguity in the choice of the pointwise basis of the 
eigenspaces of  $\Gamma$. Of course, one can notice that the ambiguity can be removed in the case the map $W(x)=W$ is constant-valued, a fact which is equivalent to affirm that the positive and negative eigenspaces of $\Gamma$ can be fixed independently of the base points. However, this is possible only if $\Gamma$ is a constant gradation 
 acting on the trivial bundle $X\times\C^{2m}$, and, as discussed in  (1) above, this is a non-innocent restrictive assumption which excludes many interesting situations.

\medskip

\noindent
{\bf The Karoubi-Thiang approach.}
In a series of two recent brilliant papers \cite{thiang-14,thiang-15} G. Thiang commented about some problematic aspects of the usually accepted definition of topological phases for topological insulators in class {\bf AIII}. He recognized an incompatibility between the notions of isomorphism and homotopic equivalence
by showing  how certain diagonal unitary transforms can change the value of the  winding number. He also noticed that the notion of phase can be well defined only in a relative sense. These kind of criticisms are in effect also contained in questions $\s{Q}$.1) - $\s{Q}$.3) that we posed (and we answered) in the introduction.
In the approach by Thiang the topological phases of topological insulators in class {\bf AIII} are defined
as the $K$-theoretical classes of a $K$-theory introduced by M. Karoubi to classify gradations acting on Banach category endowed with a Clifford action \cite[Chapter III, Section 4]{karoubi-97}. Elements of this $K$-theory are equivalence classes $[\Gamma_1,\Gamma_2]$ of gradations acting on an element $\bb{E}$ of an abelian category (\eg a complex vector bundle) endowed with the action  $\rho$ of a Clifford algebra and the equivalence is meant in the sense of homotopy deformations inside the space of gradations. The meaning reserved by Thiang to $[\Gamma_1,\Gamma_2]$ is that of a \emph{relative} topological phase between the flattened Hamiltonians $\Gamma_1$ and $\Gamma_2$. Modulo a change of roles between gradations  and Clifford actions the point of view by Thiang is very closed with our Definition \ref{defi:top_phases}.
However, we notice that there is an evident \virg{loss of information} in the classification scheme constructed by  Thiang on the basis of  
Karoubi's $K$-theory. In fact, according to Karoubi-Thiang the triviality of the class $[\Gamma_1,\Gamma_2]=0$ is  equivalent to the homotopy $\Gamma_1\sim\Gamma_2$. In this way one completely erases from the theory possible non-trivial topological aspects related to underlying vector bundle  $\bb{E}$. Our point of view, summarized in Definition \ref{defi:top_phases}, can be understood as a generalization of the Karoubi-Thiang theory in which one requires that $[\Gamma_1,\Gamma_2]\in K_0(\bb{E})$ whenever $\Gamma_1\sim\Gamma_2$.

\medskip

\noindent
{\bf The Atiyah-Hopkins-Witten $K_\pm$-theory.}
In 1998, in relation to $K$-theoretic classification of \emph{D-brane charges},
 E. Witten introduced a variant of the $K$-theory, denoted by $K_\pm$ \cite{witten-98}. The geometric situation concerns a manifold $X$ with an involution $\tau$ having a fixed point submanifold $X^\tau$. On $X$ one wants to study a pair of complex vector bundles $(\bb{E}_+, \bb{E}_-)$ with the property that $\tau$ is covered by a map $\varrho$ which interchanges
them. In terms of the \emph{virtual} vector bundle $\bb{E}_+- \bb{E}_-$, then $\varrho$ takes this into its negative, and $K_\pm(X,\tau)$ is meant to be the appropriate $K$-theory of this situation. This \emph{twisted} $K$-theory has been precisely defined and studied by M. Atiyah and M. Hopkins in  a subsequent work
 \cite{atiyah-hopkins-04}. More recently some new aspects have been investigated by K. Gomi in
 \cite{gomi-13}. In the case of a trivial involution $\tau={\rm Id}_X$ the map $\varrho$ can be described in terms of an isomorphism $\Theta:\bb{E}_-\to\bb{E}_+$ as in \eqref{eq:iso_cliff1_biss}.
 This makes evident the link between the $K_\pm$-theory and the classification of chiral vector bundles. Following  \cite[Section 2]{atiyah-hopkins-04}, the $K_\pm$-theory of the space $X$ with trivial action verifies the exact sequence 
\beql{eq:diag:atiyah}
\begin{diagram}
      R(\Z_2)\otimes  K^1(X)  &                        & \rTo^{\phi}             &                       &  K^1(X) &                        & \rTo^{\delta=0}             &                       &  K^1_\pm(X)       \\
                              \uTo                &   &                    &   &     &&&& \dTo \\
                                              K^0_\pm(X)  &                        & \lTo^{\delta=0}             &                       &  K^0(X) &                        & \lTo^{\phi}             &                       & R(\Z_2) \otimes  K^0(X)      \\
 \end{diagram}
\eeq
where $R(\Z_2)\simeq\Z\oplus\Z$ is the \emph{representation ring} (over complex) of $\Z_2$ and it is generated by the  \emph{trivial} representation $1:\Z_2\to\{1\}\subset\C$ and the \emph{sign} representation $e:\Z_2\to\{-1,+1\}\subset\C$. The two homomorphisms $\phi$, which forget the $R(\Z_2)$ component, are surjective with kernels $(1-e)\otimes K^\bullet(X)$ and so it turns out that $\delta=0$.   This implies that $K^j_\pm(X)\simeq{\rm Ker}\phi\simeq K^{j+1}(X)$ where $j$ is meant 
modulo $2$. If from one side the isomorphism $K^0_\pm(X)\simeq  K^{1}(X)$ indicates that $K^0_\pm$ contains the topological information about the automorphism group of vector bundles over $X$, from the other side the  $\delta:K^0(X)\to\{0\}\subset K^0_\pm(X)$ shows that $K^0_\pm(X)$ contains no kind of information about the topology of the underlying vector bundles themselves. Then,  the description provided by the $K_\pm$-theory covers only partially the phenomenology of chiral systems as described in Section \ref{sec:bloc_bund_top_phas}.

%------%
\section{Applications to some toy-models}
\label{sec:toy_mos}
In this section we discuss some toy-models with chiral symmetry and  underlying non-trivial Chern classes. Models of this type appear  naturally 
in problems of condensed matter involving a magnetic field.

\medskip

\begin{example}\label{ex:}{\upshape
 Let us introduce in the Hilbert space $\ell(\Z^2)$   the position operator $X:=(X_1,X_2)$
$$
(X_1\psi)_{n_1,n_2}\;=\;n_1\psi_{n_1,n_2}\;,\qquad\quad (X_2\psi)_{n_1,n_2}\;=\;n_2\psi_{n_1,n_2}
$$
the shift operator $S:=(S_1,S_2)$
$$
(S_1\psi)_{n_1,n_2}\;=\;\psi_{n_1-1,n_2}\;,\qquad\quad (S_2\psi)_{n_1,n_2}\;=\;\psi_{n_1,n_2-1}\;.
$$
Here $\psi:=\{\psi_{n_1,n_2}\}$ is any element of $\ell(\Z^2)$. The $X$'s are sel-adjoint and unbounded while the $S$'s are unitary. They realize the commutation 
\begin{equation}\label{eq:Mod_001}
S_j\;X_k\;S_j^*\;=\;X_k\;-\;\delta_{j,k}\;\n{1}\;,\qquad\quad j,k=1,2\;.
\end{equation}
Given a real parameter $\theta\in\R$ one can define the magnetic operators
$$
S_1^\theta\;:=\;\expo{\ii\pi\theta X_2}S_1\;,\qquad\quad S_2^\theta\;:=\;\expo{-\ii\pi\theta X_1}S_2
$$
From \eqref{eq:Mod_001} it follows
$$
S_1^\theta\;S_2^\theta\;=\;\expo{\ii2\pi\theta}\;S_2^\theta\;S_1^\theta\;.
$$
which are the well-known commutation relations of the non-commutative torus.
The \emph{Hofstadter operator} is defined by 
$$
H_\theta\;:=\;S_1^\theta\;+\;S_2^\theta\;+\;\big(S_1^\theta\big)^*\;+\;\big(S_2^\theta\big)^*\;.
$$
Among the discrete (pseudo-)symmetries of $H_\theta$ one can consider the  \emph{charge conjugation}
$$
Q\;:=\;\expo{\ii\pi(X_1+ X_2)}\;
$$ 
which is unitary, linear and verifies  $Q^2=\n{1}$. A short computation shows that 
\begin{equation}\label{eq:Mod_002}
\left\{
\begin{aligned}
Q\;S_j\; Q\;&=\;-S_j\\
Q\;X_j\; Q\;&=\;X_j
\end{aligned}
\right.
\;,\qquad\quad j=1,2
\end{equation}
which also implies that 
$$
{
Q\;H_\theta\; Q\;=\;-H_\theta\;.
}
$$
This shows that $H_\theta$ possesses a natural chiral symmetry. We point out that the 
(pseudo-)symmetry $Q$ depends only on the position variables and does not need extra degrees of freedom.

\medskip

The spectral structure of $H_\theta$  is well-known. It has been extensively studied in the literature. Moreover, for $\theta=\frac{p}{q}\in\Q$ the operator $H_\theta$ 
turns out to be invariant under (several representations of) a $\Z^2$-action implemented by two commuting unitaries
$$
T_{1,a}\;:=\;\expo{\ii\pi a\theta X_2}S_1^{-a}\;,\qquad\quad T_{2,b}\;:=\;\expo{\ii\pi b\theta X_1}S_2^{b}\;,\qquad\quad a+b=Nq\;,\ \ N\in\N
$$
and
can be fibered over (any multiple of) the \emph{magnetic} Brillouin zone (which is not canonically defined even if one set $N=1$) giving rise to a topological quantum system of the type described in Definition \ref{def:tsq_Chi}.  As case study, let us consider $\theta=\frac{1}{3}$. In this circumstance the spectrum of $H_\theta$ is made by three energy bands separated by gaps: There is one strictly positive energy band with spectral projection $P_{\Omega_+}$, one central band symmetric around the zero energy with  spectral projection $P_0$ and  
one strictly negative energy band with spectral projection $P_{\Omega_-}$. The Chern numbers associated to these projections can be computed with the help of the TKNN-diophantine equation and it is known that $c(P_{\Omega_+})=c(P_{\Omega_-})=1$ and $c(P_0)=-2$ (the computation can be made at an algebraic level without the need of any vector bundle representation).
The operator $Q$ intertwines between $P_{\Omega_+}$ and $P_{\Omega_-}$ and  when restricted on the range of $P_{\Omega_+}+P_{\Omega_-}$ it defines a gradiation $\Gamma_Q$ (still dependents on the position variables) for the flattened Hofstadter operator $\rho=P_{\Omega_+}-P_{\Omega_-}$. Clearly $\rho^2=P_{\Omega_+}+P_{\Omega_-}\neq\n{1}$ (and $c(\rho^2)=2$), in contrast to \eqref{eq:lit5} and consequently $\rho$ cannot admit a block off-diagonal decomposition of type \eqref{eq:lit6} by unitary operators, but only by partial isometries. In this case the \virg{standard} approach for the definition of the topological phases
based on \eqref{eq:standard_phase} fails to be applicable. Conversely the construction of the chiral vector bundle described in Section \ref{sec:bloc_bund_top_phas} can be still performed (up to the election of a reference isomorphism) and one can define the topological chiral phase as the equivalence class of the underlying $\chi$-bundle. 
}\hfill $\blacktriangleleft$
\end{example}

\begin{example}\label{Ex:hofs_C2}{\upshape
One can realize more elaborate toy models by considering Hofstadter-like operators with an extra spinorial degree of freedom (\eg operators of this type has been studied in \cite{denittis-landi-12}). Let us use the  notation introduced in Example \ref{ex:} and 
 in the Hilbert space $\ell(\Z^2)\otimes\C^2$   consider the operators
 $$
 \hat{H}^W_\theta\;:=\;
 \left(\begin{array}{cc}
 H_\theta & 0 \\
 0 & -WH_\theta W^*\end{array}\right)\;,\qquad\quad  \hat{\chi}_W\;:=\;
 \left(\begin{array}{cc}
0 & W^* \\
W & 0\end{array}\right)\;.
 $$
 where $W$ is a unitary operator on $\ell(\Z^2)$. A direct verification shows that 
 $ \hat{\chi}_W$ acts as a chiral symmetry over  $\hat{H}^W_\theta$, namely
$$
\hat{\chi}_W\;  \hat{H}^W_\theta\; \hat{\chi}_W\;=\;- \hat{H}^W_\theta\;.
$$ 
On can specialize this model by choosing $W=S_1^\theta S_2^\theta$ which implies
 $$
 WH_\theta W^*\;=\;\expo{-\ii2\pi\theta}\;\left[S_1^\theta\;+\;\big(S_2^\theta\big)^*\right]\;+\;\expo{\ii2\pi\theta}\;\left[S_2^\theta\;+\;\big(S_1^\theta\big)^*\right]\;.
 $$
 Notice that for $\theta\in\Q$ both $\hat{H}^W_\theta$ and $\hat{\chi}_W$ commute with the magnetic translations $\hat{T}_{1,a}:={T}_{1,a}\otimes\n{1}_2$ and
 $\hat{T}_{2,b}:={T}_{2,b}\otimes\n{1}_2$
 and so they fiber over the magnetic Brillouin zone producing a topological quantum system of the type described in Definition \ref{def:tsq_Chi}. Observe that in this particular case the fibered (pseudo-)symmetry $\hat{\chi}_W(x)$ gets
an explicit dependence  on the point of the base space.
The spectral structure of $\hat{H}^W_\theta$ coincides with that of $H_\theta$ up to a doubling of the degeneracy. Therefore all the arguments developed in Example \ref{ex:} also apply to this case. Observe that $\hat{H}^W_\theta$ admits also a second chiral symmetry given by the diagonal operator $\hat{Q}:=Q\otimes\n{1}_2$
(observe that $[Q,W]=0$). In this case the $\chi$-bundle construction can be used in order to express the \emph{topological difference} between the pairs $(\hat{H}^W_\theta,\hat{\chi}_W)$ and $(\hat{H}^W_\theta,\hat{Q})$ independently of the election of a reference isomorphism.
}\hfill $\blacktriangleleft$
\end{example}

\appendix

\section{Topology of unitary groups and Grassmann manifolds}\label{app:topU&G}

\medskip 

\noindent
{\bf Homotopy.\ }
{
The complete determination of the 
 homotopy groups of the unitary groups $\n{U}(m)$ is still an open problem. The first homotopy groups are showed in the following table.}
\begin{table}[h]
 \begin{tabular}{|c||c|c|c|c|c|c|c|c|c|c|}
\hline
$ \pi_{k}\big(\n{U}(m)\big)$&  $k=1$&$k=2$&$k=3$&$k=4$&$k=5$&$k=6$&$k=7$&$k=8$&$k=9$&$k=10$\\
\hline
 \hline
 \rule[-2mm]{0mm}{7mm}
$m=1$&  $\boxed{\Z}$ & {$0$} &   {$0$}  &{$0$}&{$0$}&{$0$}&{$0$}&{$0$}&{$0$}&{$0$}\\
\hline
 \rule[-2mm]{0mm}{7mm}
$m=2$&  $\boxed{\Z}$ & $\boxed{0}$ &   $\boxed{\Z}$  &{$\Z_2$}&{$\Z_2$}&{$\Z_{12}$}&{$\Z_{2}$}&{$\Z_{2}$}&{$\Z_{3}$}&{$\Z_{15}$}\\
\hline
 \rule[-2mm]{0mm}{7mm}
$m=3$&  $\boxed{\Z}$ & $\boxed{0}$ &   $\boxed{\Z}$ &$\boxed{0}$&$\boxed{\Z}$&$\Z_6$&$0$&$\Z_{12}$&$\Z_{3}$&$\Z_{30}$\\
\hline
 \rule[-2mm]{0mm}{7mm}
$m=4$&  $\boxed{\Z}$ & $\boxed{0}$ &   $\boxed{\Z}$  &$\boxed{0}$&$\boxed{\Z}$&$\boxed{0}$&$\boxed{\Z}$&$\Z_{24}$&$\Z_{2}$&$\Z_2\oplus\Z_{120}$\\
\hline
 \rule[-2mm]{0mm}{7mm}
$m=5$&  $\boxed{\Z}$ & $\boxed{0}$ &   $\boxed{\Z}$  &$\boxed{0}$&$\boxed{\Z}$&$\boxed{0}$&$\boxed{\Z}$&$\boxed{0}$&$\boxed{\Z}$&$\Z_{120}$\\
\hline
 \rule[-2mm]{0mm}{7mm}
$m=6$&  $\boxed{\Z}$ & $\boxed{0}$ &   $\boxed{\Z}$  &$\boxed{0}$&$\boxed{\Z}$&$\boxed{0}$&$\boxed{\Z}$&$\boxed{0}$&$\boxed{\Z}$&$\boxed{0}$\\
\hline
\end{tabular}
\vspace{1mm}
\caption{{\footnotesize First homotopy groups for the unitary groups. The entries enclosed in a box represent the values of the homotopy groups in the stable regime. These groups are subjected to the \emph{Bott periodicity} \cite{bott-58,bott-59}.
 Groups of the type  $\pi_{2m+r}(\n{U}(m))$ with $r=1,2$ have been computed in \cite{kervaire-60}. A larger table with the related references can be found in
\cite{Lundell-92}.
}}
 \end{table}

 \noindent{
The homotopy of the  \emph{Palais unitary group}  $\n{U}(\infty)$ is described by the \emph{Bott periodicity} (see Section \ref{sect:model_classifying_space}).}

{From the fiber sequence ${\rm S}\n{U}(m)\to\n{U}(m)\stackrel{\rm det}{\to}\n{U}(1)$ 
one obtains the isomorphisms
\begin{equation}\label{eq:App03}
\pi_k\big(\n{U}(m)\big)\;\simeq\;\pi_k\big(\n{S}^1\big)\;\oplus\; \pi_k\big({\rm S}\n{U}(m)\big)\;
\qquad\quad k\in\N\cup\{0\}\;.
\end{equation}
Equation \eqref{eq:App03}, along with the computation of the homotopy groups  $\pi_{1}(\n{S}^1)\simeq\Z$ and $\pi_{k}(\n{S}^1)=0$ for all $k\neq 1$ \cite[Proposition 4.1]{hatcher-02}, provides the isomorphisms $\pi_{k}({\rm S}\n{U}(m))\simeq  \pi_{k}(\n{U}(m))$ for all $k\geqslant 2$ with the only difference $\pi_{1}({\rm S}\n{U}(m))=0$ opposed to $\pi_{1}(\n{U}(m))\simeq\Z$. Table {\rm A.1} can be completed with $\pi_{0}(\n{U}(m))= 0=\pi_{0}({\rm S}\n{U}(m))$ which simply means that the groups $\n{U}(m)$ and ${\rm S}\n{U}(m)$ are path-connected.}

\medskip

The computation of the homotopy groups of the Grassmann manifold $G_m(\C^\infty)$
can be reduced to the problem of the computation of the homotopy groups of $\n{U}(m)$ by means of  the fiber sequence associated to the universal classifying principal $\n{U}(m)$-bundle
\begin{equation}\label{eq:App04}
\n{U}(m)\;\longrightarrow\;\bb{S}_m^\infty\;\longrightarrow\;G_m(\C^\infty)
\end{equation}
where $
\bb{S}_m^n:= \n{U}(n)/ \n{U}(n-m)$ is the \emph{Stiefel variety} and $\bb{S}_m^\infty:=\bigcup_{n=1}^{\infty}\;\bb{S}_m^n$ is, as usual, the inductive limit.  
The
universality of the space $\bb{S}_m^\infty$ implies its  contractibility, \ie $\pi_k(\bb{S}_m^\infty)=0$ for all $k$ \cite[Chapter 8, Theorem 5.1]{husemoller-94}. This fact applied to the 
homotopy exact sequence induced by \eqref{eq:App04} gives
\begin{equation}\label{eq:App05}
\pi_k\big( G_m(\C^\infty)\big)\;\simeq\;\pi_{k-1}\big(\n{U}(m)\big)\;\qquad\quad\ k\in\N\;.
\end{equation}
The connectedness of the Grassmann manifold  also implies $\pi_0( G_m(\C^\infty))=0$.

\medskip 

\noindent
{\bf Cohomology.\ }{
The  cohomology ring of the {Grassmann manifold} 
\begin{equation}\label{eq:univ_chern_class}
H^\bullet\big(G_m(\C^\infty),\Z\big)\;\simeq\;\Z[\rr{c}_1,\ldots,\rr{c}_m]\;,\qquad\quad \rr{c}_k\in H^{2k}\big(G_m(\C^\infty),\Z\big)
\end{equation}
is the ring of polynomials with integer coefficients and $m$ generators $\rr{c}_k$ of even degree \cite[Theorem 14.5]{milnor-stasheff-74}. 
These generators $\rr{c}_k$ are called \emph{universal} Chern classes and there are no polynomial relationships between them.
Since isomorphism classes of rank $m$ complex vector bundles over $X$ are classified by maps $[\varphi]\in[X,G_m(\C^\infty)]$, one defines 
the (topological)  Chern classes of a given vector bundle $\bb{E}\to X$  by 
$$
{c}_k(\bb{E})\;:=\;\varphi^\ast(\rr{c}_k)\;\in\;H^{2k}(X,\Z)\qquad\quad k\in\N\;
$$
where $\varphi^\ast:H^k(G_m(\C^\infty),\Z)\to H^k(X,\Z)$ is the  pullback induced by the classifying map $\varphi$.
Isomorphic  vector bundles possess
 the same family of Chern classes. }

\medskip

\begin{remark}[Postnikov sections of the Grassmann manifold]\label{rk:postnikov_sect_grassA}{\upshape 
 The problem of the construction of the Postnikov tower for the spaces 
 $G_m(\C^\infty)$ has been firstly studied in \cite{peterson-59}. 
 With reference to the
content of  Section \ref{sec:postmikov} let $\alpha_j:G_m(\C^\infty)\to  \bb{G}_{j}^m$
be the $j$-th Postnikov section of the Grassmann manifold $G_m(\C^\infty)$.
This section is obtained from the previous one $\bb{G}_{j-1}^m$
according to the 
 (principal)  fibration sequences
\begin{equation}\label{eq:Postnikov_grasmann1}
K(\pi_{j},j)\;\longrightarrow\; \bb{G}_{j}^m\;\stackrel{p_{j}}{\longrightarrow}\; \bb{G}_{j-1}^m\;\stackrel{\kappa^{j+1}}{\longrightarrow}\; K(\pi_{j},j+1)
\end{equation}
where $\pi_{j}:=\pi_{j}(\bb{G}_{j}^m)$ and $\kappa^{j+1}$ define the related Postnikov invariant.
Since 
$\pi_{j}(\bb{G}_{j}^m)=0$ for $j$ odd and $j\leqslant 2m$ one has that $\bb{G}_{2j-1}^m=\bb{G}_{2j-2}^m$ for $j\leqslant m$. Thus, one has to compute $\kappa^{2j+1}\in H^{2j+1}(\bb{G}_{2j-1}^m,\pi_{2j})$ for $j\leqslant m$. Now \cite[Lemma 4.4]{peterson-59} states that for $j\leqslant m$ the invariant $\kappa^{2j+1}$ has order $(j-1)!$. With this information we can immediately conclude that $\bb{G}_{1}^m\simeq\{\ast\}$ which implies $\kappa^3=0$ and so  $\bb{G}_{3}^m\simeq\bb{G}_{2}^m\simeq K(\Z,2)\simeq \C P^\infty$ for all $m\geqslant 1$.  
  For the determination of the next section one needs  $\kappa^5\in H^5(\C P^\infty, \pi_4)\simeq0$ for $m\geqslant 2$.  In conclusion one has
\begin{equation}\label{post_dec_G_4-3}
\bb{G}_{4}^m\;\simeq\;K(\Z,2)\;\times\; K(\Z,4)\;,\qquad\quad \forall\ m\geqslant 2\;.
\end{equation}
Lemma 4.1 in \cite{peterson-59} assures that $H^k(\bb{G}_{4}^m,\Z)\simeq H^k(G_m(\C^\infty),\Z)$ for all $k\leqslant 4$
and with the help of the K\"{u}nneth formula for cohomology and the explicit knowledge of the cohomology groups of $K(\Z,2)$ and $K(\Z,4)$ one obtains that for all $m\geqslant 2$  
$$
\begin{aligned}
H^0\big(G_m(\C^\infty),\Z\big)\;&\simeq\;H^{0}\big( K(\Z,2) , \Z\big)\;\otimes_\Z\; H^{0}\big( K(\Z,4) , \Z\big)\; \simeq\;\Z\\
H^1\big(G_m(\C^\infty),\Z\big)\;&\simeq\;0\\
H^2\big(G_m(\C^\infty),\Z\big)\;&\simeq\;H^{2}\big( K(\Z,2) , \Z\big)\;\simeq\; \Z\\
H^3\big(G_m(\C^\infty),\Z\big)\;&\simeq\;0\\
H^4\big(G_m(\C^\infty),\Z\big)\;&\simeq\;H^{4}\big( K(\Z,2) , \Z\big)\;\oplus\; H^{4}\big( K(\Z,4) , \Z\big)\; \simeq\;\Z^2\;.
\end{aligned}
$$
The above computations show that 
the first two universal Chern classes
$\rr{c}_1$ and  $\rr{c}_2$ can be identified with the \emph{basic class} of $H^{2}( K(\Z,2) , \Z)$ and $H^{4}( K(\Z,4) , \Z)$, respectively (for a definition of basic class see \eg \cite[Definition 5.3.1]{arkowitz-11}).}\hfill $\blacktriangleleft$.
\end{remark}

\medskip

Also  the cohomology ring of the {unitary} group $\n{U}(m)$ is well-known. It is a classical result that
\begin{equation}\label{eq:univ_wind_class}
H^\bullet\big(\n{U}(m),\Z\big)\;\simeq\;{\bigwedge}_\Z\;[\rr{w}_1,\ldots,\rr{w}_{m}]\;,\qquad\quad \rr{w}_k\in H^{2k-1}\big(\n{U}(m),\Z\big)
\end{equation}
 is the exterior algebra generated by $m$ odd-degree classes $\rr{w}_k$ \cite[Th\'{e}or\`{e}me 19.1]{borel-53} or \cite[Section 10]{borel-55}.
By adhering to a modern terminology  (see \eg \cite{prodan-schulz-baldes-14,prodan-schulz-baldes-book-16}) we will refer to the $\rr{w}_k$'s as the \emph{universal odd} Chern classes. The description \eqref{eq:univ_wind_class}
can be generalized to the infinite unitary group $\n{U}(\infty)$.
\begin{lemma}\label{lemma:cohomU_inf}
The cohomology ring of the infinite {unitary} group $\n{U}(\infty)$
\begin{equation}\label{eq:univ_wind_class_inf}
H^\bullet\big(\n{U}(\infty),\Z\big)\;\simeq\;{\bigwedge}_\Z\;[\rr{w}_1,\rr{w}_{2},\ldots]\;,\qquad\quad \rr{w}_k\in H^{2k-1}\big(\n{U}(\infty),\Z\big)\;,\qquad k\in\N
\end{equation}
 is the exterior algebra generated by a countable family of  odd-degree classes $\rr{w}_k$.
\end{lemma}
\proof
The group $\n{U}(\infty)$ is by definition the limit of the direct system of inclusions
$$
\n{U}(1)\;\stackrel{\imath_1}{\hookrightarrow}\;\n{U}(2)\;\stackrel{\imath_2}{\hookrightarrow}\;\n{U}(3)\;\stackrel{\imath_3}{\hookrightarrow}\;\ldots
$$
which induces an inverse system in cohomology
\begin{equation}\label{eq:inv_cohom}
H^\bullet\big(\n{U}(1),\Z\big)\;\stackrel{\imath_1^*}{\longleftarrow}\;H^\bullet\big(\n{U}(2),\Z\big)\;\stackrel{\imath_2^*}{\longleftarrow}\;H^\bullet\big(\n{U}(3),\Z\big)\;\stackrel{\imath_3^*}{\longleftarrow}\;\ldots\;.
\end{equation}
The relation between the inverse system \eqref{eq:inv_cohom} and the cohomology $H^\bullet\big(\n{U}(\infty),\Z\big)$ is specified by the \emph{Milnor exact sequence}. Let us define a cochain complex $(\s{C}^\bullet,\delta)$ by
$$
\s{C}^k\;:=\;
\left\{
\begin{aligned}
&{\prod}_j\; H^\bullet\big(\n{U}(j),\Z\big)&\qquad\quad& j=0,1\\
&0&\qquad\quad& j\geqslant 2\\
\end{aligned}
\right.
$$
where $\prod$ is the direct product of abelian groups and the (only non-trivial) differential $\delta:\s{C}^0\to \s{C}^1$ is defined by
$$
\delta\big(\rr{a}_1,\rr{a}_2,\rr{a}_3,\ldots\big)\;:=\;\big(\rr{a}_1-\imath_1^*(\rr{a}_2),\rr{a}_2-\imath_2^*(\rr{a}_3),\rr{a}_3-\imath_3^*(\rr{a}_4),\ldots\big)\;,\qquad\quad \rr{a}_j\in H^\bullet\big(\n{U}(j),\Z\big)\;.
$$
By definition the inverse limit \eqref{eq:inv_cohom} arises as the $0$-th cohomology group of the cochain complex $(\s{C}^\bullet,\delta)$, \ie
$$
\lim_{\leftarrow j}\ H^\bullet\big(\n{U}(j),\Z\big)\;:=\; H^0\big(\s{C}^\bullet,\delta\big)\;=\;{\rm Ker}(\delta)\;.
$$
The cokernel of $\delta$ is usually called \virg{limit 1}:
$$
{\lim_{\leftarrow j}}^1\ H^\bullet\big(\n{U}(j),\Z\big)\;:=\; H^1\big(\s{C}^\bullet,\delta\big)\;=\;\s{C}^1/{\rm Im}(\delta)\;.
$$
The {Milnor exact sequence} \cite[Lemma 2]{milnor-62} states that
\begin{equation}\label{eq:milnor1}
0\;\longrightarrow\;{\lim_{\leftarrow j}}^1\ H^\bullet\big(\n{U}(j),\Z\big)\;\longrightarrow\;H^\bullet\big(\n{U}(\infty),\Z\big)\;\longrightarrow\;{\lim_{\leftarrow j}}\ H^\bullet\big(\n{U}(j),\Z\big)\;\longrightarrow\;0\;.
\end{equation}
Notice that each $\imath^*_j: H^\bullet(\n{U}(j+1),\Z)\to H^\bullet(\n{U}(j),\Z)$ is surjective since $\imath^*_j(\rr{w}_k)=\rr{w}_k$ for all $k=1,\ldots,j$
and $\imath^*_j(\rr{w}_{j+1})=0$. Then, a simple argument shows that also the differential $\delta:\s{C}^0\to \s{C}^1$ is surjective and so the  \virg{limit 1} is trivial. This implies that
$
H^\bullet\big(\n{U}(\infty),\Z\big)\simeq{\rm Ker}(\delta)
$
and the kernel ${\rm Ker}(\delta)$ is generated by elements of the form $\big(\rr{w}_k,\rr{w}_k,\rr{w}_k,\ldots\big)$ which can be identified with the generators $\rr{w}_k$ for all $k\in\N$. This concludes the proof.
\qed

\medskip

\begin{remark}[Postnikov sections of the Palais unitary group]\label{rk:postnikov_sect_grass}{\upshape 
The Postnikov resolution of  $\n{U}(\infty)$ can be used to provide a different description of the universal odd Chern classes $\rr{w}_k$, at least in low dimension.
With reference to the
 technological apparatus described in Section \ref{sec:postmikov} we want to compute the Postnikov sections $\alpha_j:\n{U}(\infty)\to  \bb{U}_{j}^\infty$. From $\pi_1(\n{U}(\infty))=\Z$ we immediately get $\bb{U}_{1}^\infty=K(\Z,1)\simeq\n{S}^1$. Since we know that $\pi_{2j}(\n{U}(\infty))=0$ and by using the fact that $K(0,j)=\{\ast\}$
  (along with \cite[Corollary 4.63]{hatcher-02}) we obtain from \eqref{eq:Postnikov1} that $\bb{U}_{2j}^\infty\simeq \bb{U}_{2j-1}^\infty$. Hence, we need to compute only the odd sections. For the computation of $\bb{U}_{3}^\infty$ we need the knowledge of the Postnikov invariant $\kappa^4$ in the fiber sequence
\begin{equation}\label{eq:Postnikov_inf_unit}
K(\Z,3)\;\longrightarrow\; \bb{U}_{3}^\infty\;\stackrel{p_{3}}{\longrightarrow}\; \bb{U}_{2}^\infty\simeq \bb{U}_{1}^\infty\;\stackrel{\kappa^{4}}{\longrightarrow}\; K(\Z,4)\;.
\end{equation}
From its very definition we know that  $\kappa^{4}\in H^{4}(\bb{U}^\infty_{2}, \Z)\simeq H^{4}(\n{S}^1, \Z)=0$. The vanishing of $\kappa^{4}$  immediately yields 
\begin{equation}\label{post_dec_U_4-3}
\bb{U}_{4}^\infty\;\simeq\; \bb{U}_{3}^\infty\;\simeq\;K(\Z,1)\;\times\; K(\Z,3)\;.
\end{equation}
The next step  requires the computation of the Postnikov invariant $\kappa^5$ in the fiber sequence
\begin{equation}\label{eq:Postnikov_inf_unit2}
K(\Z,5)\;\longrightarrow\; \bb{U}_{5}^\infty\;\stackrel{p_{5}}{\longrightarrow}\; \bb{U}_{4}^\infty\simeq \bb{U}_{3}^\infty\;\stackrel{\kappa^{6}}{\longrightarrow}\; K(\Z,6)\;.
\end{equation}
In this case the Postnikov invariant is an element of 
$\kappa^{6}\in H^{6}(\bb{U}^\infty_{4}, \Z)\simeq H^{6}(\n{S}^1\times K(\Z,3) , \Z)$.
The knowledge of the non trivial cohomology $H^{k}(\n{S}^1, \Z)\simeq\Z$ if $k=0,1$ and the use of the 
K\"{u}nneth formula for cohomology provide
$$
H^{6}\big(\bb{U}^\infty_{4}, \Z\big)\;\simeq\;\Big(\Z\;\otimes_\Z\; H^{6}\big( K(\Z,3) , \Z\big)\Big)\;\oplus\;\Big(\Z\;\otimes_\Z\; H^{5}\big( K(\Z,3) , \Z\big)\Big)\;\simeq \Z\;\otimes_\Z\;\Z_2\;\simeq\Z_2\;.
$$
where we used the explicit results  $ H^{5}( K(\Z,3) , \Z)=0$ and $H^{6}( K(\Z,3) , \Z)\simeq\Z_2$ computed in \cite[Section 18]{bott-tu-82}. This is compatible with the more general result \cite[Lemma 5]{arlettaz-banaszak-95} which states that $\kappa^{6}$ is a (non-trivial) element of order $2$.
This implies that 
\begin{equation}\label{post_dec_U_6-5}
\bb{U}_{6}^\infty\;\simeq\; \bb{U}_{5}^\infty\;\simeq\;\Big(K(\Z,1)\;\times\; K(\Z,3)\Big)\;\times_{\kappa^6}\;  K(\Z,5)\;.
\end{equation}
where the last product is \virg{twisted} by the action of the non trivial class  $\kappa^6$. Evidently the construction becomes more and more involved when the degree of the Postnikov section increases. 
Let us consider now the implication of \eqref{post_dec_U_4-3} for the interpretation of the first two generators of $\n{U}(\infty)$. By construction the map $\alpha_4:\n{U}(\infty)\to \bb{U}_{4}^\infty$ is a $5$-equivalence, hence 
the \emph{Whitehead's second theorem}\footnote{\label{note_A1} More precisely the Whitehead's second theorem  \cite[Theorem 6.4.15]{arkowitz-11} (see also \cite[Chapter I, Section 8, Theorem 9]{spanier-66}) says that $\alpha_4$ induces a $5$-\emph{homology} equivalence (see \cite[Definition 6.4.10]{arkowitz-11}). At this point the proof that a $n$-{homology} equivalence implies a $n$-\emph{cohomology} equivalence is provided by the use of the  universal coefficient theorem as in \cite[Lemma 4.1]{peterson-59}.} assures that  $H^k(\n{U}(\infty),\Z)\simeq H^k(\bb{U}_{4}^\infty,\Z)$ for all $k\leqslant 4$. With the help of the K\"{u}nneth formula for cohomology and the explicit knowledge of cohomology of $K(\Z,1)$ and $K(\Z,3)$ one computes
$$
\begin{aligned}
H^0\big(\n{U}(\infty),\Z\big)\;&\simeq\;H^{0}\big( K(\Z,1) , \Z\big)\;\otimes_\Z\; H^{0}\big( K(\Z,3) , \Z\big)\; \simeq\;\Z\\
H^1\big(\n{U}(\infty),\Z\big)\;&\simeq\;H^{1}\big( K(\Z,1) , \Z\big)\;\otimes_\Z\; H^{0}\big( K(\Z,3) , \Z\big)\; \simeq\;H^{1}\big( K(\Z,1) , \Z\big)\;\simeq\;\Z\\
H^2\big(\n{U}(\infty),\Z\big)\;&\simeq\;0\\
H^3\big(\n{U}(\infty),\Z\big)\;&\simeq\;H^{0}\big( K(\Z,1) , \Z\big)\;\otimes_\Z\; H^{3}\big( K(\Z,3) , \Z\big)\;\simeq\;H^{3}\big( K(\Z,3) , \Z\big)\; \simeq\;\Z\\
H^4\big(\n{U}(\infty),\Z\big)\;&\simeq\;H^{1}\big( K(\Z,1) , \Z\big)\;\otimes_\Z\; H^{3}\big( K(\Z,3) , \Z\big)\; \simeq\;\Z\;.
\end{aligned}
$$
The above result shows that $\rr{w}_1$ and  $\rr{w}_2$ can be identified with the \emph{basic class} of $H^{1}( K(\Z,1) , \Z)$ and $H^{3}( K(\Z,3) , \Z)$, respectively (for a definition of basic class see \eg \cite[Definition 5.3.1]{arkowitz-11}). 
As a final comment we can observe that \eqref{post_dec_U_4-3} also describes the 3-rd and 4-th Postnikov sections of $\n{U}(m)$ for all $m\geqslant 2$ while \eqref{post_dec_U_6-5} describes the 5-rd and 6-th Postnikov sections of $\n{U}(m)$ for all $m\geqslant3$ (stable regime).
}\hfill $\blacktriangleleft$
\end{remark}

\medskip

%-----------------------------------%

\section{Fiber bundle identification of the classifying space}\label{sec:sub_ident}
The definition \eqref{eq:class_space1} says that the spaces $\chi_m(\C^n)$ are subspaces of $G_m(\C^n)\times \n{U}(n)$. 
These spaces can be also identified with the total spaces of  suitable fiber bundles. For this aim,
let us recall  a standard construction: For any Lie group $\n{G}$ and any principal
$\n{G}$-bundle $\pi: \bb{P} \to X$ we use the \emph{adjoint action} of $\n{G}$ on $\n{G}$ to get the associated
fiber bundle
$$
{\rm Ad}(\bb{P})\;:=\;\bb{P}\;\times_{\rm Ad}\;\n{G}\;=\;(\bb{P}\;\times\;\n{G})/\n{G}
$$
where $g\in\n{G}$ acts on $(p,h)\in\bb{P}\times\n{G}$ by $(p,h)\mapsto(p\cdot g^{-1},g\cdot h\cdot g^{-1})$. It is a well-known fact
that sections of ${\rm Ad}(\bb{P})\to X$ are in one to one correspondence
with automorphisms of $\bb{P}$ \cite[Chapter 7, Section 1]{husemoller-94}.

\medskip

Let us also recall that the 
{Grassmann manifold}  and its tautological  $\n{U}(m)$-frame bundle (Stiefel variety) $\bb{S}_m^n\to G_m(\C^n)$ can be realized as quotient spaces:
\begin{equation}\label{eq:grass_stirf_quot}
G_m(\C^n)\simeq\n{U}(n)/\big(\n{U}(m)\times \n{U}(n-m)\big)\;,\qquad\quad \bb{S}_m^n\simeq\n{U}(n)/ \n{U}(n-m)\;,
\end{equation}
where $\n{U}(m)$ and $\n{U}(n-m)$ act (on the right) on $\n{U}(n)$ through the inclusions of
$\C^m$ and $\C^{m-n}$ into $\C^n=\C^m\oplus\C^{n-m}$. Let $\Sigma\in G_m(\C^n)$ be a subspace of $\C^n$ of dimension $m$.
Given orthonormal basis $\mathbbm{v}:=({\rm v}_1,\ldots,{\rm v}_m)$ and $\mathbbm{w}:=({\rm w}_1,\ldots,{\rm w}_{n-m})$ of $\Sigma\subset\C^n$
and $\Sigma^\bot\subset\C^n$, respectively one can form an element $(\mathbbm{v},\mathbbm{w})\in\n{U}(n)$ by arraying the vectors as columns, \ie
\begin{equation}\label{eq:identi1}
(\mathbbm{v},\mathbbm{w})\;:=\;
\left(\begin{array}{cccccc}
| &  & | & | &  & | \\
{\rm v}_1 & \cdots & {\rm v}_m & {\rm w}_1 & \cdots & {\rm w}_{n-m} \\
| &  & | & | &  & |
\end{array}\right)\;.
\end{equation}
According to this notation $\mathbbm{v}$ and $\mathbbm{w}$ can be considered an $n\times m$ and an $n\times (m-n)$ matrices, respectively.
The pair $(\mathbbm{v},\mathbbm{w})$ provides a representative for $([\mathbbm{v}],[\mathbbm{w}])=\Sigma\in G_m(\C^n)$ according to the quotient description of the Grassmann manifold in \eqref{eq:grass_stirf_quot}. Here, the equivalence relations are naturally given by $\mathbbm{v}\sim \mathbbm{v}\cdot a$ and 
 $\mathbbm{w}\sim \mathbbm{w}\cdot b$ for some $a\in \n{U}(m)$ and $b\in \n{U}(n-m)$.
Similarly, we can use $(\mathbbm{v},[\mathbbm{w}])\in \bb{S}_m^n$ for a point in the Stiefel variety according to the
 quotient representation in \eqref{eq:grass_stirf_quot}
   and an element $c\in\n{U}(m)$ in the structure group can be represented by a block diagonal matrix
$$
\left(\begin{array}{c|c}c & 0 \\\hline 0 & \n{1}_{n-m}\end{array}\right)\;\subset\;\n{U}(n)\;.
$$
In this way points in ${\rm Ad}\big(\bb{S}_m^n\big)$ are given by equivalence classes $[(\mathbbm{v},\mathbbm{w}),c]$ with respect to the equivalence relation $((\mathbbm{v},\mathbbm{w}\cdot b),c)\sim ((\mathbbm{v}\cdot a^{-1},\mathbbm{w}),a\cdot c\cdot a^{-1})$ for some $a\in \n{U}(m)$ and $b\in \n{U}(n-m)$.
\medskip

The core of this appendix is the proof of the following identifications:

\begin{proposition}\label{prop:identif1}
There are natural homeomorphisms
$$
\chi_m(\C^n)\;\simeq\; {\rm Ad}\big(\bb{S}_m^n\big)\;,\qquad\text{and}\qquad \n{B}_\chi^m\;\simeq\; {\rm Ad}\big(\bb{S}_m^\infty\big)\;,
$$
where, as usual, $\bb{S}_m^\infty$ denotes the inductive limit obtained by the inclusions $\bb{S}_m^n\subset \bb{S}_m^{n+1}\subset\ldots\ $.
\end{proposition}

\medskip

\noindent
However, we start first with a technical preliminary result.
\begin{lemma}
For each pair of integers $1\leqslant m\leqslant n$ there is a bijective continuous map
\begin{equation}\label{eq:lem_identi0}
\vartheta\;:\;{\rm Ad}\big(\bb{S}_m^n\big)\;\longrightarrow \; \chi_m(\C^n)\;.
\end{equation}
\end{lemma}
\proof
Given an element $(\mathbbm{v},\mathbbm{w})\in\n{U}(n)$ as in \eqref{eq:identi1} and a $c\in \n{U}(m)$ there is a unique unitary matrix
$u(\mathbbm{v};\mathbbm{w};c)\in\n{U}(n)$ such that
$
u(\mathbbm{v};\mathbbm{w};c)\cdot (\mathbbm{v},\mathbbm{w})=(\mathbbm{v}\cdot c,\mathbbm{w})
$. Such a matrix is explicitly given by
$$
u(\mathbbm{v};\mathbbm{w};c)\;:=\;(\mathbbm{v},\mathbbm{w})\cdot \left(\begin{array}{c|c}c & 0 \\\hline 0 & \n{1}_{n-m}\end{array}\right)\cdot (\mathbbm{v},\mathbbm{w})^{-1}\;.
$$
By construction $u(\mathbbm{v};\mathbbm{w};c)$ preserves the $m$-dimensional subspace $([\mathbbm{v}],[\mathbbm{w}])=\Sigma\subset\C^n$ spanned by the columns of $\mathbbm{v}$. Therefore, the pair $(\Sigma,u):=(([\mathbbm{v}],[\mathbbm{w}]),u(\mathbbm{v};\mathbbm{w};c))$ provides a point in $\chi_m(\C^n)$. This allows to define the map
\begin{equation}\label{eq:lem_identi1}
\begin{aligned}
\vartheta\;:\;&\;\bb{S}_m^n\times\n{U}(m)&\;\longrightarrow \; &\ \ \chi_m(\C^n)&
\\
&\;\big((\mathbbm{v},[\mathbbm{w}]),c\big)&\;\longmapsto \; &\ \ (\Sigma,u)&.
\end{aligned}
\end{equation}
This map factors through the equivalence relation that defines ${\rm Ad}\big(\bb{S}_m^n\big)$ since $([\mathbbm{v}],[\mathbbm{w}])=([\mathbbm{v}\cdot a],[\mathbbm{w}])=\Sigma$ and 
$$
u\big(\mathbbm{v}\cdot a;\mathbbm{w}; a \cdot c\cdot a^{-1}\big)\;=\;u\big(\mathbbm{v};\mathbbm{w};c\big)\;=\;u\big(\mathbbm{v};\mathbbm{w}\cdot b;c\big)\;,\qquad\quad\forall\ a\in\n{U}(m),\ \ b\in\n{U}(n-m)
$$ 
Hence, the prescription
\eqref{eq:lem_identi1} defines a map like \eqref{eq:lem_identi0}. The map $\vartheta$ is also continuous. Indeed,
the topology of ${\rm Ad}\big(\bb{S}_m^n\big)$ is induced from $\n{U}(n)\times \n{U}(m)$ by a quotient,
and that of $\chi_m(\C^n)$ is induced from $G_m(\C^n)\times \n{U}(n)$ by an inclusion. Thus, to
prove that $\vartheta$ is continuous, it suffices to prove that the composition of the
following maps is continuous:
\beql{eq:lem_identi2}
\begin{diagram}
           \n{U}(n)\times \n{U}(m) &                                &   G_m(\C^n)\times \n{U}(n)       \\
                         \dTo^{\rm pr}                  &                                &  \uTo_{\imath} \\
  {\rm Ad}\big(\bb{S}_m^n\big)  &    \rTo_{\vartheta}    &   \chi_m(\C^n)   \\
 \end{diagram}\;.
\eeq
The natural projection ${\rm pr}$ and the inclusion $\imath$ are continuous by construction. Moreover, the composition of the three maps is easily computed to be $((\mathbbm{v},\mathbbm{w}),c)\mapsto (([\mathbbm{v}],[\mathbbm{w}]),u(\mathbbm{v};\mathbbm{w};c))$ which is evidently continuous. This, in turn, implies the continuity of $\vartheta$.

In order to prove that  $\vartheta$ is bijective we construct its inverse $\varrho$. Let us start with a point $(\Sigma,u)\in \chi_m(\C^n)$
and choose orthonormal basis $\mathbbm{v}$ and $\mathbbm{w}$ such that $([\mathbbm{v}],[\mathbbm{w}])=\Sigma$ as above. Since $u\in\n{U}(n)$ preserves $\Sigma$ there is a unique $c(\mathbbm{v};u)\in\n{U}(m)$ such that $u\cdot \mathbbm{v}=\mathbbm{v}\cdot c(\mathbbm{v};u)$. Such a matrix is explicitly given by
$$
c(\mathbbm{v};u)\;:=\;^t\overline{\mathbbm{v}} \cdot u\cdot\mathbbm{v}
$$
where $^t\overline{\mathbbm{v}}$ denotes the transpose of the matrix $\overline{\mathbbm{v}}$, complex conjugated of $\mathbbm{v}$. Now we define 
\begin{equation}\label{eq:lem_identi3}
\begin{aligned}
\varrho\;:\;&\;\chi_m(\C^n)&\;\longrightarrow \; &\ \ \ \  \ \ \ \ \ {\rm Ad}\big(\bb{S}_m^n\big)&
\\
&\;(\Sigma,u)&\;\longmapsto \; &\ \  \big((\mathbbm{v},[\mathbbm{w}]),c(\mathbbm{v};u)\big)&,
\end{aligned}
\end{equation}
where $(\mathbbm{v},[\mathbbm{w}])\in \n{U}(n)/ \n{U}(n-m)$ identifies an element of $\bb{S}_m^n$. The map $\varrho$ does not depend on the choice of a frame $\mathbbm{w}$ for $\Sigma^\bot$ (as denoted by the square brackets). Moreover, if $\mathbbm{v}'=\mathbbm{v}\cdot a$ with  $a\in\n{U}(m)$ is a new frame for  $\Sigma$, a direct computation shows $c(\mathbbm{v}';u)=a^{-1}\cdot c(\mathbbm{v};u)\cdot a$. These two facts prove that  
$\varrho$ really maps into ${\rm Ad}\big(\bb{S}_m^n\big)$. The proof that $\varrho=\vartheta^{-1}$ follows now from a direct, as well trivial, verification.
\qed

\medskip
\medskip

\proof[Proof of Proposition \ref{prop:identif1}]
Lemma \ref{eq:lem_identi0} proves the existence of a continuous bijection $\vartheta:{\rm Ad}\big(\bb{S}_m^n\big)\to\chi_m(\C^n)$.
Since ${\rm Ad}\big(\bb{S}_m^n\big)$ is compact (it is the quotient of a compact group) and $\chi_m(\C^n)$ is Hausdorff (it is the subspace of a Hausdorff space), $\vartheta$ turns out to be a homeomorphism (see \eg \cite[Chapter I, Section 8]{janich-84}).
This homeomorphism is compatible with taking the direct limits, so that it induces a homeomorphism
$\vartheta:{\rm Ad}\big(\bb{S}_m^\infty\big)\to\n{B}_\chi^m$.
\qed

\medskip

From the construction of the identifications in Proposition \ref{prop:identif1} it follows that the 
mapping $\chi_m(\C^n)\to G_m(\C^n)$ in \eqref{eq:class_space3} and $\n{B}_\chi^m\to G_m(\C^\infty)$ in \eqref{eq:class_space33}
agree with the bundle maps ${\rm Ad}\big(\bb{S}_m^n\big)\to G_m(\C^n)$ and ${\rm Ad}\big(\bb{S}_m^\infty\big)\to G_m(\C^\infty)$, respectively. Summarizing, one has the fiber sequence 
\begin{equation}\label{eq:fib_seq_m}
\n{U}(m)\;\longrightarrow\;{\rm Ad}\big(\bb{S}_m^\infty\big)\simeq\n{B}_\chi^m\;\stackrel{\pi}{\longrightarrow}\; G_m\big(\C^\infty\big)
\end{equation}
with projection $\pi$ given by \eqref{eq:class_space33}.

\medskip

When  $m=1$, just by exploiting the fact that $\n{U}(1)$ is abelian, one gets the following immediate consequence of  Proposition \ref{prop:identif1}.
\begin{corollary}\label{corol:m=1}
$$
\n{B}_\chi^1\;\simeq\;{\rm Ad}\big(\bb{S}_1^\infty\big)\;\simeq\;\C P^\infty\;\times\;\n{U}(1)\;.
$$
\end{corollary}
%

%------------------------------------------------------------------------------------------------------------------------%
%                                                                        bibliography
%------------------------------------------------------------------------------------------------------------------------%

\medskip
\medskip

%------------------------------------------------------------------------------------------------------------------------%

\begin{thebibliography} {[RMCPV]}
\frenchspacing \baselineskip=12 pt plus 1pt minus 1pt


%---A---%

\bibitem[AB]{arlettaz-banaszak-95}
{Arlettaz, D.; Banaszak, G.}: {\sl On the non-torsion elements in the algebraic $K$-theory of rings of integers}. 
J. reine angew. Math. {\bf 461}, 63-79 (1995)




\bibitem[AH]{atiyah-hopkins-04} {Atiyah, M.~F.; Hopkins M.~I.}: {\sl A variant of $K$-theory: $K_\pm$}. 
In: \emph{Topology, Geometry, and Quantum Field Theory} (Ed. U. Tilllmann).  Cambridge Univ. Press, pp. 05Ð17, 2004

\bibitem[AM]{ashcroft-mermin-76} {Ashcroft, N. W.;   Mermin N. D.}: 
{\em Solid State Physics}. 
Saunders College Pub., Philadelphia, 1976



\bibitem[Ark]{arkowitz-11} {Arkowitz, M.}: 
{\em Introduction to Homotopy Theory}. 
Springer, New York, 2011

\bibitem[Arl]{arlettaz-00}
{Arlettaz, D.}: {\sl Algebraic $K$-Theory of rings from a topological viewpoint}. 
Publ. Mat. {\bf 44}, 3-84 (2000)







\bibitem[At]{atiyah-67} {Atiyah, M.~F.}: {\em $K$-theory}. 
W. A. Benjamin, New York, 1967

\bibitem[AZ]{altland-zirnbauer-97}
{Altland, A.; Zirnbauer, M.:}: {\sl Non-standard symmetry classes in mesoscopic normal-superconducting hybrid structures}. 
Phys. Rev.B {\bf 55}, 1142-1161 (1997)


 
%---B---% 



 \bibitem[BCR]{bourne-carey-rennie-16}  Bourne,~C.;~Carey,~A.~L.;~Rennie,~A.: 
 {\sl A non-commutative framework for topological insulators}.
{Rev. Math. Phys.}~{\bf 28}, 1650004 (2016)
 



 \bibitem[Be]{bellissard-86} 
Bellissard,~J.~V.: {\sl $K$-Theory of $C^*$-algebras in Solid State Physics}.
In: {\em Statistical Mechanics and Field Theory, Mathematical Aspects'} (Eds. T. C. Dorlas, M. N. Hugenholtz and M. Winnink). Lecture Notes in Physics  {\bf 257}, 99-156, 1986 



\bibitem[BMKNZ]{bohm-mostafazadeh-koizumi-niu-zwanziger-03} {B\"{o}hm, A.; and Mostafazadeh, A.;  Koizumi, H.;  Niu, Q.;  Zwanziger, J.}: {\em The Geometric Phase in Quantum Systems}. 
Springer-Verlag, Berlin, 2003
 
 
 \bibitem[Bor1]{borel-53}  Borel,~A.: {\sl Sur la cohomologie des espaces fibres principaux et des espaces homogenes de groupes de Lie compacts}.
{Ann. of Math.}~{\bf 57}, 115-207 (1953)
 
 \bibitem[Bor2]{borel-55}  Borel,~A.: {\sl Topology of Lie groups and characteristic classes}.
{Bull. Am. Math. Soc.}~{\bf 61}, 397-433 (1955)
 
 

 

 
\bibitem[Bot1]{bott-58}  Bott,~R.: {\sl The space of loops on a Lie group}.
{Michigan Math. J.}~{\bf 5}, 35-61 (1958)
 
 
 
  \bibitem[Bot2]{bott-59}  Bott,~R.: {\sl The Stable Homotopy of the Classical Groups}.
{Ann. of Math.}~{\bf 70}, 313-337 (1959)

 
 
 \bibitem[Br]{brown-62}  Brown,~E.~H.~Jr.: {\sl Cohomology Theories.}
{Ann. of Math.}~{\bf 75}, 467-484 (1962)

 \bibitem[BT]{bott-tu-82} Bott,~R.; Tu,~L.~W.: {\em Differential Forms in Algebraic Topology}. 
Springer, Berlin, 1982
 
\bibitem[BuT]{budich-trauzettel-13}  Budich,~J.~C.; Trauzettel,~ B: {\sl From the adiabatic theorem of quantum mechanics to topological states of matter.
}
{Phys. Status Solidi RRL}~{\bf 7}, 109-129 (2013)



%---C---%

\bibitem[CDFG]{carpentier-delplace-fruchart-gawedzki-15}
{Carpentier, D.; Delplace, P.; Fruchart, M.;  Gaw\c{e}dzki, K.}:
{\sl Topological Index for Periodically Driven Time-Reversal Invariant 2D Systems},
  Phys. Rev. Lett. {\bf 114}, 106806 (2015)
  

\bibitem[CDFGT]{carpentier-delplace-fruchart-gawedzki-tauber-15} Carpentier, D.; Delplace, P.; Fruchart, M.;  Gaw\c{e}dzki, K.;  Tauber, C.: 
{\sl Construction and properties of a topological index for periodically driven time-reversal invariant 2D crystals}. 
{Nucl. Phys. B}~{\bf 896}, 779-834 (2015)

\bibitem[CJ]{chruscinski-jamiolkowski-04} {Chru\'{s}ci\'{n}ski, D.; Jamio{\l}kowski, A.}: {\em Geometric Phases in Classical and Quantum Mechanics}. 
Birkh\"{a}user, Basel, 2004



\bibitem[Cl]{clement-02} {Cl\'{e}ment, A.}: {\em Integral cohomology of finite Postnikov towers}. 
PhD Thesis. Universit\'{e} de Lausanne, 2002




%---D---%


\bibitem[DG1]{denittis-gomi-14} De~Nittis,~G.; Gomi,~K.: {\sl Classification of \virg{Real} Bloch-bundles: Topological Quantum Systems of type AI}.
J. Geom. Phys.~{\bf 86}, 303-338 (2014)


\bibitem[DG2]{denittis-gomi-14-gen} De~Nittis,~G.; Gomi,~K.: 
{\sl Classification of \virg{Quaternionic} Bloch-bundles: Topological Insulators of type AII}. 
Commun. Math. Phys. {\bf 339}, 1-55 (2015)



\bibitem[DG3]{denittis-gomi-15} De~Nittis,~G.; Gomi,~K.: 
{\sl Differential geometric invariants for time-reversal symmetric Bloch-bundles: the \virg{Real} case}. J. Math. Phys. {\bf 57}, 053506 (2016)


\bibitem[DG4]{denittis-gomi-17} De~Nittis,~G.; Gomi,~K.: 
{\sl The FKMM-invariant in low dimension}. Accepted for publication in Lett. Math. Phys.
E-print \texttt{arXiv:1702.04801} (2017)

\bibitem[DG5]{denittis-gomi-18} De~Nittis,~G.; Gomi,~K.: 
{\sl The cohomological nature of the Fu-Kane?Mele invariant}. J. Geometry Phys. {\bf 124}, 124-164 (2018)



\bibitem[DL]{denittis-landi-12} De~Nittis,~G.; Landi,~G.: {\sl Generalized TKNN equations}.
Adv. Theor. Math. Phys.~{\bf 16}, 505-547 (2012)



\bibitem[DK]{davis-kirk-01} Davis, J. F.; Kirk, P.: {\em Lecture Notes in Algebraic Topology}. 
AMS, Providence, 2001


%---E---%



%---F---%

\bibitem[FFG]{fomenko-fuchs-gutenmacher-86} Fomenko, A. T.; Fuchs, D. B.; Gutenmacher, V. L.: {\em Homotopic Topology}. 
Akad\'{e}miai Kiad\'{o}, Budapest, 1986

\bibitem[FM]{freed-moore-13} 
{Freed, D. S.; Moore, G. W.}: {\sl Twisted Equivariant Matter}. {Ann. Henri Poincar\'{e}}~{\bf 14}, 1927-2023 (2013)

\bibitem[FMP]{fiorenza-monaco-panati-14}  Fiorenza, D.; Monaco, D.; Panati, G.:  {\em $\Z_2$  invariants of topological insulators as geometric obstructions}.  
Commum. Math. Phys.  {\bf 343}, 1115-1157 (2016)


%---G---% 

\bibitem[GBVF]{gracia-varilly-figueroa-01}  Gracia-Bondia, J. M., Varilly, J. C.,  Figueroa,  H.:  {\em Elements of Noncommutative Geometry}.  Birkh\"{a}user, Boston, 2001



\bibitem[Go]{gomi-13} Gomi,~K.: {\sl A variant of K-theory and topological T-duality for Real circle bundles}.
Commun. Math. Phys. {\bf 334}, 923-975 (2015)

\bibitem[GP]{porta-graf-13} 
{Graf, G. M.; Porta, M.}: {\sl Bulk-Edge Correspondence for Two-Dimensional
Topological Insulators}. {Commun. Math. Phys.}~{\bf 324}, 851-895 (2013)

\bibitem[GS]{grossmann-schulz-baldes-16} Grossmann, J,; Schulz-Baldes H.: 
{\sl Index pairings in presence of symmetries with applications to topological insulators}.
Commun. Math. Phys. {\bf 343}, 477-513 (2016) 


%---H---% 

\bibitem[HK]{hasan-kane-10} Hasan,~M.~Z.; Kane,~C.~L.: {\sl Colloquium: Topological insulators}. 
{Rev. Mod. Phys.}~{\bf 82}, 3045-3067 (2010)



\bibitem[Hat]{hatcher-02} Hatcher,~A.: {\em  Algebraic Topology}. Cambridge University Press, Cambridge,  2002



\bibitem[Hu]{husemoller-94} Husemoller,~D.: {\em  Fibre bundles}. Springer-Verlag, New York, 1994

%---J---%

\bibitem[Ja]{janich-84} J\"{a}nich,~K.~W.: {\em  Topology}. Springer-Verlag, Berlin, 1984



\bibitem[JT]{james-thomas-65} James I.;  Thomas,~E.: 
{\sl  An approach to the enumeration problem for non-stable vector bundles}. 
{J. Math. Mech.}~{\bf 14}, 485-506 (1965)





%---K---%



\bibitem[Kar]{karoubi-97} 
Karoubi,~M.: {\em  $K$-Theory. An Introduction}.  Springer-Verlag, New York, 1978.



\bibitem[Kat]{kato-95} Kato, T.: {\em  Perturbation theory of linear operators.}
 Reprint of the 1980 edition, Springer, Berlin, 1995 



\bibitem[Ker]{kervaire-60}  Kervaire,~M.~A.:
{\sl Some nonstable homotopy groups of Lie groups}. 
{Illinois J. Math.}~{\bf 4}, 161-169 (1960)




\bibitem[Kel]{kellendonk-17}  Kellendonk~J.:
{\sl On the  $C^*$-Algebraic Approach to Topological Phases for Insulators}. 
{Ann. Henri Poincar\'{e}}~{\bf 18}, 2251-2300 (2017)



\bibitem[KG]{kennedy-guggenheim-15} Kennedy,~R.; Guggenheim,~C.: {\sl Homotopy Theory of Strong and Weak Topological Insulators}. 
Phys. Rev. B {\bf 91}, 245148 (2015)






\bibitem[Ki]{kitaev-09} Kitaev,~A.: 
{\sl Periodic table for topological insulators and superconductors}. 
{AIP Conf. Proc.}~{\bf 1134}, 22-30 (2009)


\bibitem[KM2]{kane-mele-05}  Kane,~C.~L.;  Mele,~E.~J.:
{\sl $\Z_2$ Topological Order and the Quantum Spin Hall Effect}. 
{Phys. Rev. Lett.}~{\bf 95}, 146802 (2005)




\bibitem[Ko]{kori-14}  Kori,~T.:
{\sl Extensions of current groups on $\n{S}^3$ and the adjoint representations}. 
{J. Math. Soc. Japan}~{\bf 66}, 819-838 (2014)



\bibitem[Kuc]{kuchment-93} Kuchment,~P.: {\em Floquet theory for partial differential equations}. Birkh\"{a}user, Boston,  1993

\bibitem[Kub]{kubota-17}
Kubota , Y.: {\sl Controlled Topological Phases and Bulk-edge Correspondence}
Commun. Math. Phys.
{\bf 349}, 493-525 (2017)

\bibitem[KZ]{kennedy-zirnbauer-15} Kennedy,~R.; Zirnbauer M.~R.: {\sl Bott periodicity for $\Z_2$ symmetric ground states of gapped free-fermion systems}. 
Commun. Math. Phys. {\bf 342}, 909-963 (2016)






%---L---%


\bibitem[Le]{lee-48} Lee, H. C.: 
{\sl On Clifford Algebras and Their Representations}. 
{Ann. of Math.}~{\bf 4}, 760-773 (1948)




\bibitem[LS]{lupton-smith-15} Lupton, G.; Smith, S. B.:
{\sl Gottlieb groups of function spaces}. 
Math. Proc. Cambridge Phil. Soc., 1-17 (2015)  

\bibitem[Lu]{Lundell-92} Lundell,~A.~T.: {\sl Concise Tables of James Numbers and Some Homotopy of Classical Lie Groups and Associated Homogeneous Spaces}. In: {\em Algebraic Topology Homotopy and Group Cohomology}. Lecture Notes in Mathematics vol. {\bf 1509}, pp. 250-272
Springer, Berlin, 1992
 





%---M---%



\bibitem[Mi]{milnor-62} 
{Milnor,~J.~W.}: {\sl On axiomatic homology theory}. {Pacific J. Math.}~{\bf 12}, 337-342 (1962)



\bibitem[MP]{may-ponto-12} May,~J.~P.;   Ponto,~K.: {\em More Concise Algebraic Topology: Localization, Completion, and Model Categories}. University of Chicago Press, 2012



\bibitem[MS]{milnor-stasheff-74} Milnor,~J.~W.;   Stasheff,~J.~D.: {\em Characteristic Classes}. Princeton University Press, 1974


\bibitem[MT]{monaco-tauber-17}
Monaco, D.; Tauber, C.: {\sl Gauge-theoretic invariants for topological insulators: a bridge between Berry, Wess-Zumino, and Fu-Kane-Mele}.
Lett. Math. Phys. (2017)


%---N---%


%---P---%



\bibitem[Pa]{palais-65} 
{Palais, R. S.}: {\sl On the homotopy type of certain groups of operators}. {Topology}~{\bf 3}, 271-279 (1965)



\bibitem[Per]{percy-10} 
{Percy, A.}: {\sl Some examples of relations between non-stable integral cohomology operations}. {Bull. Korean Math. Soc.}~{\bf 47}, 275-286 (2010)

\bibitem[Pet]{peterson-59} 
{Peterson, F. P.}: {\sl Some remarks on Chern classes}. {Ann. of Math.}~{\bf 69}, 414-420 (1959)



\bibitem[PS1]{prodan-schulz-baldes-14} Prodan,~E.; Schulz-Baldes~H.: {\sl Non-commutative odd Chern numbers and topological phases of disordered chiral systems}. 
J. Funct. Anal. {\bf 271}, 1150-1176 (2016) 

\bibitem[PS2]{prodan-schulz-baldes-book-16} Prodan,~E.; Schulz-Baldes~H.: {\em Bulk and Boundary Invariants for Complex Topological Insulators: From K-Theory to Physics}. 
Springer Series Mathematical Physics Studies, Springer Int. Pub., Switzerland, 2016








%---Q---%

\bibitem[QZ]{qi-zhang-11} 
{Qi,~X.-L.; Zhang,~S.-C.:}
 {\sl Topological insulators and superconductors}. {Rev. Mod. Phys.}~{\bf 83}, 1057-1110 (2011)





%---R---%

\bibitem[RSFL]{ryu-schnyder-furusaki-ludwig-10}
 Ryu,~S.; Schnyder,~A.~P.; Furusaki,~A.; Ludwig,~A.~W.~W.:
{\sl Topological insulators and superconductors: tenfold way and dimensional hierarchy}.
New J. Phys. {\bf 12}, 065010 (2010)




%---S---%



\bibitem[Sp]{spanier-66} Spanier, E. H.: {\em Algebraic Topology}. McGraw-Hill, New York, 1966





\bibitem[SRFL]{schnyder-ryu-furusaki-ludwig-08}
Schnyder,~A.~P.; Ryu,~S.; Furusaki,~A.; Ludwig,~A.~W.~W.:
{\sl Classification of topological insulators and superconductors in three spatial dimensions}.
Phys. Rev. {\bf B 78}, 195125 (2008)


\bibitem[St]{steenrod-51} Steenrod, N. E.: {\em The Topology of Fibre Bundles}. Princeton University Press, Princeton, 1951


\bibitem[ST]{stiepan-teufel-12}
Stiepan,~H-M.;~Teufel,~S.:
{\sl Semiclassical Approximations for Hamiltonians with Operator-Valued Symbols}.
Commun. Math. Phys. {\bf 320}, 821-849 (2012)



\bibitem[Sw]{swan-62}
Swan,~R.~G.:
{\sl Vector bundles and projective modules}.
Trans. Amer. Math. Soc. {\bf 105}, 264-277 (1962)


%---T---%




\bibitem[Te]{teufel-03} {Teufel, S.}: {\em Adiabatic Perturbation Theory in Quantum Dynamics}. 
 Lecture Notes in Mathematics {\bf 1821}, Springer-Verlag, 2003.




\bibitem[Th1]{thiang-14} Thiang,~G.~C.: {\sl On the $K$-theoretic classification of topological phases of matter}. Ann. Henri Poincar\'{e}, {\bf 17}, 757-794  (2016)

	

\bibitem[Th2]{thiang-15} Thiang,~G.~C.: {\sl A note on homotopic versus isomorphic topological phases}. E-print \texttt{arXiv:1412.4191} (2015)








%---V---%


 %---W---%
 
\bibitem[Wa]{wannier-37} Wannier,~G. H.: {\sl The structure of electronic excitation levels in insulating crystals}. 
Phys. Rev. {\bf 52}, 191-197 (1937)

\bibitem[Wi1]{witten-83} Witten,~E.: {\sl Current algebra, baryons, and quark confinement}. 
Nucl. Phys. B {\bf 223}, 433-444 (1983)

\bibitem[Wi2]{witten-98} Witten,~E.: {\sl $D$-branes and $K$-theory}. 
JHEP {\bf 12}, 019 (1998)





 
 
 \end{thebibliography}
\end{document}